\documentclass[preprint2]{aastex6}
\usepackage{txfonts}

\newcommand*{\unitstyle}[1]{\mathrm{#1}}
\newcommand*{\unitskip}{\,}                             %
\newcommand{\power}[2]{\ensuremath{{#1}^{#2}}}		    

\newcommand*{\meter}{\unitstyle{m}}

\newcommand*{\second}{\unitstyle{s}}

\newcommand*{\Kelvin}{\unitstyle{K}}
\newcommand*{\K}{\Kelvin}  

\newcommand*{\centi}{\unitstyle{c}}		
\newcommand*{\Mega}{\unitstyle{M}}		
\newcommand*{\Giga}{\unitstyle{G}}
\newcommand*{\cm}{\centi\meter}
\newcommand*{\gram}{\unitstyle{g}}

\newcommand*{\yr}{\unitstyle{yr}}
\newcommand*{\grampercc}{\gram\unitskip\power{\cm}{-3}} 
\newcommand*{\grampersquarecm}{\gram\unitskip\power{\cm}{-2}} 
\newcommand*{\columnunit}{\grampersquarecm}

\newcommand*{\MeVu}{\MeV\unitskip\unitstyle{u}^{-1}}

\newcommand*{\eV}{\unitstyle{eV}}               
\newcommand*{\MeV}{\Mega\eV}                  

\newcommand{\EE}[2]{\ensuremath{{#1}\times 10^{#2}}}
\newcommand{\density}[2]{$\rho={#1}\times 10^{#2}\,\grampercc$}
\newcommand{\logft}{\ensuremath{\log(ft)}}
\newcommand{\column}[2]{$y={#1}\times 10^{#2}\,\columnunit$}
\newcommand{\mue}[1]{$\mu_{\rm e}={#1}\,\MeV$}
\newcommand{\sn}[3]{$S_{\rm n}(^{#1}\mathrm{#2})=#3\,\MeV$}
\newcommand\iso[2]{\ensuremath{^{#2}{\rm #1}}}
\newcommand{\Urca}[4]{$^{#1}\mathrm{#2}\leftrightarrow {}^{#3}\mathrm{#4}$}
\newcommand*{\mdot}{\dot{m}}
\newcommand*{\mdotEdd}{\mdot_{\mathrm{Edd}}}

\newcommand*{\nuclei}[2]{\ensuremath{\mathrm{^{#1}#2}}}

\newcommand*{\magnesium}[1][24]{\nuclei{#1}{Mg}}
\newcommand*{\nitrogen}[1][14]{\nuclei{#1}{N}}
\newcommand*{\oxygen}[1][16]{\nuclei{#1}{O}}
\newcommand*{\potassium}[1][39]{\nuclei{#1}{K}}
\newcommand*{\calcium}[1][40]{\nuclei{#1}{Ca}}
\newcommand*{\chromium}[1][52]{\nuclei{#1}{Cr}}

\begin{document}
\title{Nuclear Reactions in the Crusts of Accreting Neutron Stars}

\email[Correspondence to: ]{schatz@nscl.msu.edu}

\author{
R. Lau\altaffilmark{1,2,3,4},
M. Beard\altaffilmark{3,5},
S. S. Gupta\altaffilmark{6}
H. Schatz\altaffilmark{1,2,3},
A.~V. Afanasjev\altaffilmark{7},
E. F. Brown\altaffilmark{1,2,3},
A. Deibel\altaffilmark{1,2,3,16},
L.~R. Gasques\altaffilmark{8},
G. W. Hitt\altaffilmark{15},
W.~R. Hix\altaffilmark{9,13},
L. Keek\altaffilmark{2,3,10},
P.  M{\"o}ller\altaffilmark{3,11},
P. S. Shternin\altaffilmark{12},
A. W. Steiner \altaffilmark{3,9,13},
M. Wiescher\altaffilmark{3,5},
Y. Xu\altaffilmark{14}
}

\altaffiltext{1}{National Superconducting Cyclotron Laboratory, Michigan State University, 640 South Shaw Lane, East Lansing, Michigan 48824, USA.}
\altaffiltext{2}{Department of Physics and Astronomy,Michigan State University, 567 Wilson Road, East Lansing, Michigan 48824, USA.}
\altaffiltext{3}{Joint Institute for Nuclear Astrophysics, Center for the Evolution of the Elements}
\altaffiltext{4}{Current Address:  Civil Engineering Department, Technological and Higher Education Institute of Hong Kong, 20A Tsing Yi Road, Tsing Yi Island, New Territories, Hong Kong.}
\altaffiltext{5}{Department of Physics, 225 Nieuwland Science Hall, University of Notre Dame, Notre Dame, Indiana 46556, USA.}
\altaffiltext{6}{Indian Institute of Technology Ropar, Nangal Road, Rupnagar (Ropar), Punjab 140 001, India.}
\altaffiltext{7}{Department of Physics and Astronomy, Mississippi State University, Mississippi State, MS 39762, USA}
\altaffiltext{8}{Departamento de Fisica Nuclear, Instituto de Fisica da Universidade de Sao Paulo, Caixa Postal 66318, 05315-970 Sao Paulo, Brazil.} 
\altaffiltext{9}{Physics Division, Oak Ridge National Laboratory, PO Box 2008, Oak Ridge, Tennessee 37831-6354, USA.}
\altaffiltext{10}{Current Address: Department of Astronomy, University of Maryland, College Park, MD 20742, USA.}
\altaffiltext{11}{Theoretical Division, MS B214, Los Alamos National Laboratory, Los Alamos, New Mexico 87545, USA.}
\altaffiltext{12}{Ioffe Institute, Politekhnicheskaya 26, Saint Petersburg, 194021, Russia}
\altaffiltext{13}{Department of Physics and Astronomy, University of Tennessee, 401 Nielsen Physics Building, 1408 Circle Drive, Knoxville, Tennessee 37996-1200, USA.}
\altaffiltext{14}{Extreme Light Infrastructure-Nuclear Physics, 077125 Magurele, Ilfov, Romania}
\altaffiltext{15}{Department of Physics and Engineering Science, Coastal Carolina University, P.O. Box 261954 Conway, SC 29528, USA}
\altaffiltext{16}{Current Address: Department of Astronomy, Indiana University, Bloomington, IN 47405, USA}

\begin{abstract}
X-ray observations of transiently accreting neutron stars during quiescence provide information about the structure of neutron star crusts and the properties of dense matter. Interpretation of the observational data requires an understanding of the nuclear reactions that heat and cool the crust during accretion, and define its non-equilibrium composition. We identify here in detail the typical nuclear reaction sequences down to a depth in the inner crust where the mass density is \density{2}{12} using a full nuclear reaction network for a range of initial compositions. The reaction sequences differ substantially from previous work. We find a robust reduction of crust impurity at the transition to the inner crust regardless of initial composition, though shell effects can delay the formation of a pure crust somewhat to densities beyond  \density{2}{12}.   This naturally explains the small inner crust impurity inferred from observations of a broad range of systems. The exception are initial compositions with A $\ge$ 102 nuclei, where the inner crust remains impure with an impurity parameter of $Q_{\rm imp} \approx 20$ due to the $N=82$ shell closure. In agreement with previous work we find that nuclear heating is relatively robust and independent of initial composition, while cooling via nuclear Urca cycles in the outer crust depends strongly on initial composition. 
This work forms a basis for future studies of the sensitivity of crust models to nuclear physics and provides profiles of composition for realistic crust models. 
\end{abstract}

\section{Introduction}

Approximately 190 Galactic X-ray sources are classified as low mass X-ray binaries \citep{Liu2007}, of which about 100 are confirmed to contain 
a neutron star accreting matter from a low mass ($< 1 M_\odot$)  companion star
at typical rates of $\lesssim 10^{-8}\,M_\odot\,\mathrm{yr^{-1}}$. 
Continued mass accretion pushes matter deeper into the crust; as the matter is compressed, the rising pressure and density induce nuclear reactions that generate heat, emit neutrinos, and increase neutron richness. Most low-mass X-ray binaries are expected to be older than 
$1\textrm{--}10\,\Mega\yr$, old enough for accretion to have replaced the entire crust of the neutron star. The accreted crust is never heated 
beyond $ \approx 1\,\Giga\K$ and  therefore differs fundamentally from the original crust, and that of isolated neutron stars, which form via annealing from a high temperature equilibrium \citep{bisnovatyui1979,Sato1979,Haensel1990}. Here we present reaction network calculations that delineate up to a density around \density{2}{12}, the full set of nuclear reactions that determine the composition and thermal profile of the accreted crust for a given set of astrophysical parameters. 

The properties of the accreted crust can be probed observationally in quasi-persistent transiently accreting neutron stars. These systems accrete continuously for years to decades, before accretion turns off and the source switches from outburst to quiescence. Despite the 2--5 order of magnitude drop in luminosity, modern X-ray telescopes can detect these systems in quiescence. The observed soft X-ray component is typically interpreted as thermal emission from the crust heated by nuclear reactions during the outburst \citep{Rutledge2002}, though there is some debate about the potential influence of residual accretion at a very low rate \citep[see, for example,][]{Parikh2017,Bernardini2013}. The time dependence of this thermal emission reflects the thermal profile of the neutron star crust and its thermal transport properties. For seven sources, the thermal emission in quiescence, following an outburst, has now been tracked observationally for many years \citep[see, for example, summaries in][]{Homan2014,Turlione2015,Waterhouse2016}. While there are large differences from source to source, in all cases a decrease in thermal emission over time is observed. This decrease in thermal emission is interpreted as the cooling of the heated crust \citep{Rutledge2002,Cackett2006,Shternin2007,Brown2009}. 

Constraints on the physics of neutron star crusts and dense matter in general have been derived from these observations through comparison with models that account for all relevant nuclear processes. Examples include the finding of a relatively high thermal conductivity indicating a relatively well-ordered lattice structure of the solid crust \citep{Cackett2006,Shternin2007,Brown2009} and constraints on its impurity \citep{Brown2009,Page2013,Turlione2015,Ootes2016,Merritt2016}; evidence for neutron superfluidity \citep{Shternin2007,Brown2009}; search for signatures of nuclear pasta \citep{Horowitz2015,Deibel2017}; possible signatures of chemical convection \citep{Degenaar2014, Medin2014}; constraints on surface gravity \citep{Deibel2015}; and evidence for a strong shallow heat source of unknown origin \citep{Brown2009,Degenaar2011,Page2013,Degenaar2013,Degenaar2015,Deibel2015,Turlione2015,Waterhouse2016,Merritt2016}. Heating and cooling from nuclear reactions in the crust also affects other regions of the neutron star. It influences explosive nuclear burning in regular X-ray bursts and rarer superbursts, which occur above the solid crust \citep{Cumming2006,Keek2008,Altamirano2012,Deibel2016,Meisel2017}, and it contributes towards heating of the neutron star core \citep{Brown1998,Cumming2017,Brown2018}. The latter effect can be used to constrain core neutrino emissivities and other core physics \citep{Brown1998,Colpi2001}. \citet{Cumming2017} recently used core heating constraints in connection with the transient light curve of KS1731-260 to place a lower limit on the core specific heat and concluded that the core could not be dominated by a quark color-flavor-locked phase. Observables related to crust nuclear reactions may not be limited to X-rays. \citet{Bildsten1998grav} and \citet{Ushomirsky2000} showed that density jumps induced by electron capture reactions in the crust, in combination with a temperature anisotropy, can lead to a mass quadrupole moment and significant gravitational wave emission that may balance the spin-up from the accretion torque and explain observed spin distributions \citep{Patruno2017}. 

The steady-state compositional profile of the outer layers of the neutron star is mapped out by the compositional changes of an accreted fluid element as it is incorporated deeper and deeper into the neutron star. These compositional changes are the result of a series of nuclear processes that occur with increasing density. Within hours of arrival on the neutron star, at around $\rho \approx 10^6$~g/cm$^3$,  hydrogen and helium burn into a broad range of heavier elements up to $Z \approx 48$. The reaction sequences are  the 3$\alpha$ reaction, the hot CNO cycles, the $\alpha$p-process, and the rapid proton capture process (rp-process) \citep{Wallace1981,Schatz1998}, and proceed either explosively in regular type I X-ray bursts \citep{Schatz2006a,Schatz2001,Fisker2008}, or in steady state \citep{Schatz1999}.  If the ashes contain significant amounts of carbon, explosive carbon burning in the ocean at $\rho \approx 10^8$~g/cm$^3$ may power the rare superbursts and transform the composition into elements around iron \citep{Schatz2003,Keek2011}. The ashes of these processes form the liquid ocean and eventually solidify around  $\rho \approx 10^9$~g/cm$^3$, setting the initial composition for the nuclear reactions in the solid crust. 

The nuclear reactions in the crust of accreting neutron stars, and the associated nuclear heating, were first calculated by \citet{bisnovatyui1979}, \citet{Sato1979} and later by \citet{Haensel1990}. They used a simplified model that assumed an initial composition of $^{56}$Fe, the presence of only a single species at a given depth, full 
$\beta$- and neutron equilibrium, zero temperature, and no shell structure. They found that electron capture reactions in the outer crust transform 
$^{56}$Fe stepwise into more neutron rich nuclei. Once the chain of nuclear reactions reaches the neutron drip line on the chart of nuclides (nuclei beyond the neutron drip line are neutron unbound with neutron separation energy $S_n < 0$), electron captures with neutron emission in the inner crust continue to transform nuclei to lower $Z$. The transition from the outer crust to the inner crust at around \density{6}{11}  is marked by the appearance of free neutrons, which coexist with nuclei. This location in the neutron star is commonly referred to as neutron drip. At \density{1.5}{12} 
density-induced (pycnonuclear) fusion reactions begin to fuse Ne ($Z=10$). The resulting heavy nuclei
are then again stepwise reduced in $Z$ by electron captures with neutron emission. This cycle repeats several times with 
increasing depth. \citet{Haensel2003,Haensel2008}
used the same model to investigate the fate of different initial isotopes, including $^{106}$Cd. \citet{Gupta2007} 
carried out the first reaction network calculation allowing the presence of an arbitrary mix of nuclei and including nuclear shell structure. 
They only considered electron capture reactions up to neutron drip, and demonstrated that heating can be 
substantially increased when taking into account electron capture into excited states.  \citet{Gupta2008} 
carried out a similar study including neutron captures and dissociations and following 
the electron captures just beyond neutron drip. They found that neutron reactions are not always in equilibrium, resulting 
in their superthreshold electron capture cascades (SEC), where a sequence of electron captures with neutron emission rapidly transform nuclei to lower Z, 
instead of the stepwise process found in simpler models. \citet{Steiner2012} developed a simple model similar to 
\citet{Haensel2003} but allowing for a multi-component plasma and a more realistic mass model. \citet{Schatz2014} used a full 
reaction network including $\beta$-decays to follow the crust composition in the outer crust, prior to neutron drip. They found a new type of 
neutron star crust reaction: nuclear Urca cycles with alternating electron captures and $\beta$-decays that cool the outer crust. This underlines the importance of using a full reaction network and allowing for the simultaneous presence of multiple species of nuclei.

In this work, we carry out the first full reaction network calculation of the compositional changes in accreted neutron star crusts through neutron drip and into the first pycnonuclear fusion reactions. We follow a broad range of individual reactions and also account for nuclear shell structure. This provides the full picture of nuclear transformations governing the transition from the outer crust to the inner crust. There are a number of 
open questions that we aim to address: (1) What are the nuclear reaction sequences in the neutron star crust for a realistic multicomponent composition when allowing for branchings and competition between different types of rates? (2) Is the crust evolving towards equilibrium, once free neutrons are available for neutron capture reactions to produce heavier elements, as suggested by \citet{Jones2005}, or are previous predictions of an evolution towards lighter elements correct? (3) How does the crust impurity as characterized by the breadth of nuclear composition evolve from the outer to the inner crust?  Is the inner crust impurity influenced by nuclear burning at the surface, and therefore likely different from system to system? (4) Can nuclear reactions provide more heating than previously assumed, alleviating, at least in some sources, the need for an exotic additional heat source? \citet{Horowitz2008} proposed that heat released by shallower fusion reactions of lighter nuclei may explain some of the additional heating. 

\section{Model}
\label{Sec:Model}

\begin{figure}
\plotone{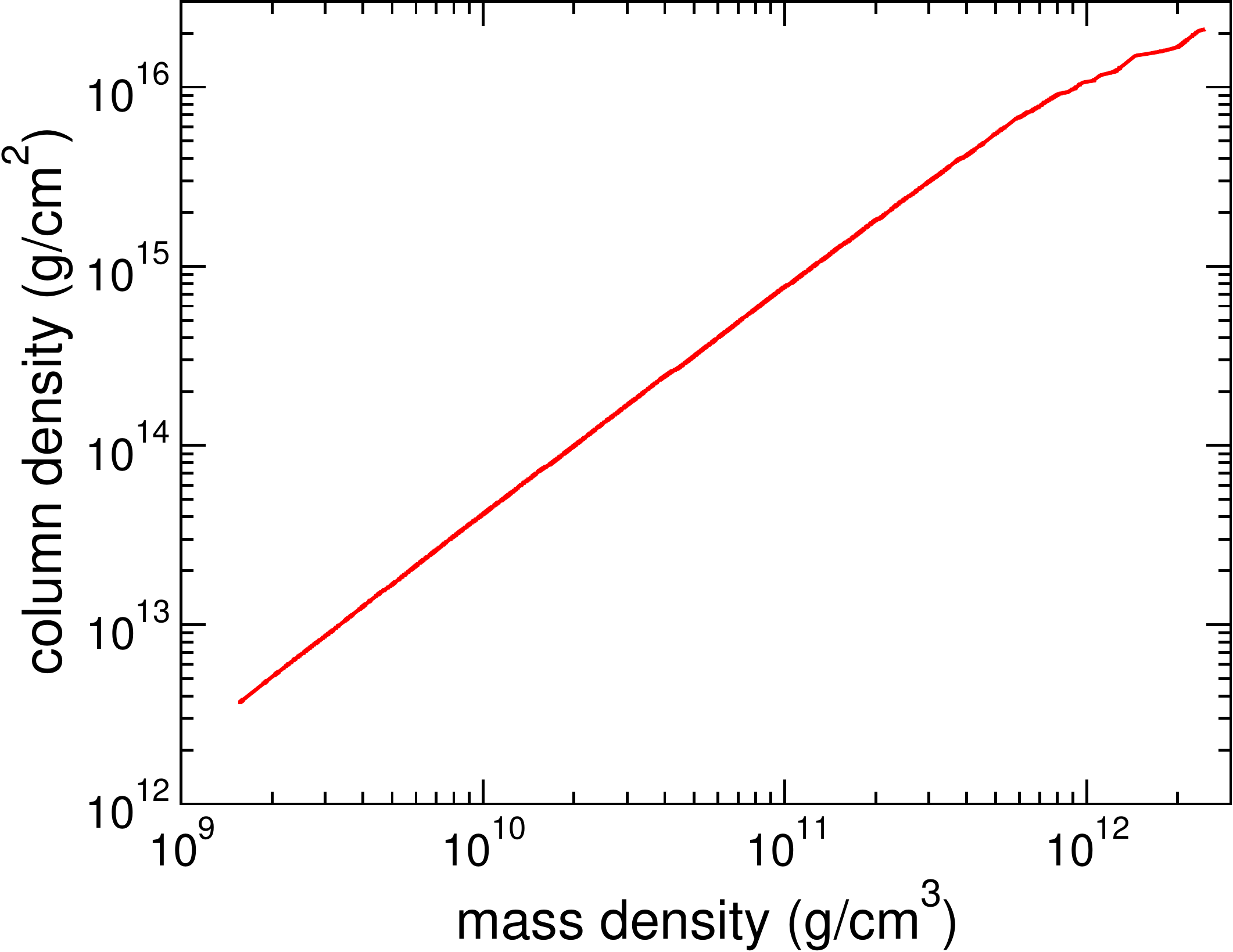}
\caption{\label{Fig:Ycol} Column density as a function of mass density for extreme burst ashes. The change in slope around \density{6}{11} indicates the change of the dominant pressure source from electrons to neutrons.}
\end{figure}

The crust model used here is similar to that in \citet{Gupta2007,Schatz2014}. The crust is modeled as a plane-parallel slab in a local Newtonian frame with constant gravity $g$.
We follow the compositional changes of an accreted fluid element induced by the increasing pressure $P=\mdot gt$, with local accretion rate $\mdot$ and time $t$,
to determine the steady state composition of the crust.  Time is therefore a measure of depth throughout this work. The mass density is 
calculated using an equation of state $P=P(T,\rho,Y_i)$ with temperature $T$, and nuclear abundances $Y_i$ (including the neutron abundance) as described in \citet{Gupta2007}. The pressure
of the free neutrons is computed using a zero-temperature compressible
liquid-drop model \citep{Mackie1977}. 
Fig.~\ref{Fig:Ycol} shows the resulting column density $y=\mdot t$ as a function of mass density $\rho$.  An accreted fluid element takes about 24,000 yr to reach the end of our calculation around \column{2}{16}. 

In order to track the time evolution of the nuclear abundances, $Y_i$, an implicitly solved nuclear reaction network is used that includes 
electron captures, $\beta$-decays, neutron capture, neutron dissociation, and pycnonuclear fusion reactions. The nuclear heat $dQ$ deposited in a time step $\Delta t$ is obtained as $dQ=\sum \Delta_i dY_i - \mu_e dY_e - \mu_n dY_n - \epsilon_\nu -dW_L$ with atomic mass excesses $\Delta_i$, electron chemical potential (without rest mass) $\mu_e$, neutron chemical potential (without rest mass)  $\mu_n$, electron fraction $Y_e$, neutron abundance $Y_n$, neutrino energy losses from electron captures and beta decays $\epsilon_\nu$, and lattice energy $W_L$ \citep{Chamel2008}.

\subsection{Astrophysical Parameters}
Unless otherwise
stated, we use $\mdot = 0.3\mdotEdd$ in the rest frame at the surface, with the local Eddington accretion rate $\mdotEdd=\EE{8.8}{4}\,\grampersquarecm\,\second^{-1}$, and $g=\EE{1.85}{14}\,\cm\,\second^{-2}$. This accretion rate is in the range for mixed H/He bursts powered by the rp-process as well as superbursts and is therefore appropriate for the initial compositions explored in this work. The calculation starts at a density of \density{1.4}{9}. The temperature is treated as a free parameter and set to $T=0.5$~GK throughout the crust. This corresponds closely to the temperature profile used in \citet{Gupta2007}. This approach is suitable for identifying the typical nuclear reactions, independent of specific temperature profiles that vary from system to system and with time, and depend on a number of additional parameters outside of our model (see discussion below). Temperature is not expected to dramatically alter reaction sequences as $kT \ll \mu_{\rm e}$ everywhere and  $kT \ll \mu_{\rm n}$ everywhere except for a very narrow layer at neutron drip. Pycnonuclear fusion reaction rates are not temperature sensitive either. The one nuclear process that is strongly temperature dependent is the strength of nuclear Urca cooling in the outer crust \citep{Schatz2014}. Choosing a relatively high constant temperature allows us to clearly identify critical Urca cooling pairs with their intrinsic strengths that may play a role in limiting crustal heating. 

\subsection{Nuclear Physics Input}
Nuclear masses are among the most important input parameters. We use the Atomic Mass Evaluation AME12 \citep{AME12} for experimental masses closer to stability. For the majority of nuclei for which masses are experimentally unknown, we employ the FRDM \citep{Moller1995} mass model. We do not mix experimental and theoretical masses to calculate reaction Q-values of interest here, such as electron capture thresholds or neutron separation energies. 

In general, mass models for isolated nuclei such as the FRDM are not applicable in the inner crust, where interactions with the free neutrons result in significant modifications of masses and nuclear structure. However, the goal of this work is to calculate the nuclear reactions in the outer crust and the transition from the outer crust into the inner crust. We stop the calculations in the outermost region of the inner crust below \density{2.7}{12} before such modifications become significant. While the neutron mass fraction at the end of our calculation is $X_n=0.52$, the neutron Fermi energy is 1.4~MeV, and the free neutron density is only \EE{9}{-4}~fm$^{-3}$, less than 1\% of the neutron density inside a nucleus.  For these conditions, the \citet{Mackie1977} mass model, which includes the impact of free neutrons on the surface energy, but neglects shell structure, shows an average correction of neutron separation energies of only 200~keV (maximum 300 keV), well within mass model uncertainties. Interactions with the free neutron gas in the inner crust also affect the single particle level structure of nuclei. In particular, with increasing free neutron density the shell structure of spherical nuclei is expected to be modified, and for sufficiently large densities effectively disappears \citep{Negele1973}. This would affect the shell correction term in the FRDM, which turns out to be important in our work. Fig. 5 in \citet{Negele1973} shows that modifications of single particle levels are expected to only set in at baryon densities in excess of $n_{\rm b}=$\EE{2}{36}, above the maximum $n_{\rm b}=$\EE{1.6}{36} in our calculations.

Electron capture rates and $\beta^-$-decay rates are determined from strength
functions calculated in a model based on wave functions in a
deformed folded-Yukawa single-particle potential with residual pairing and
Gamow-Teller interactions. They are solved for in a quasi-particle
random-phase approximation (QRPA). The original theory was based on
a deformed  oscillator single-particle potential \citep{Krumlinde1984}.
To obtain greater global predictive power a folded-Yukawa single-particle
model has been used later instead 
\citep{Moller1990,Moller1997}. That is the model used here and we refer to it as QRPA-fY.
Only allowed Gamow-Teller transitions are considered. Parent nuclei are assumed to be in their 
ground state, which is a reasonable assumption for the low temperatures ($T < 1$~GK) encountered in neutron star crusts. Weak interaction thresholds are corrected for lattice energy changes following \citet{Chamel2008}. 
The weak reaction rates are then calculated for each  time step using nuclear masses and a fast phase space approximation \citep{Becerril2006,Gupta2007}. Neutron emission is 
determined individually for each transition from the parent ground state to a daughter state with excitation energy $E_x$. We 
make the simplifying assumption that the highest number of emitted neutrons that is energetically possible will occur in all 
cases, similar to the approach of calculating branchings for $\beta$-delayed neutron emission in \citet{Moller1997}. To take into account 
Pauli blocking due to the positive neutron chemical potential, we impose the additional condition of $(E_x - S_n)/N > \mu_n$ with 
neutron separation energy $S_n$ and number of emitted neutrons $N$. 

Neutron capture rates were computed with the TALYS statistical model code as part of a systematic effort to create a reaction rate database for nucleosynthesis 
studies, using the same atomic masses used to calculate the weak interaction rates \citep{Xu2013}. 
These neutron capture rates and the rates of the reverse reactions were corrected to account for plasma screening of photons and neutron degeneracy following \citet{Shternin2012}. Pycnonuclear fusion rates were calculated from the S-factors of \citet{Beard2010,Afanasjev2012} using the formalism described in \citet{Yakovlev2006} in the uniformly mixed multicomponent plasma approximation. We implement a total of 4844 pycnonuclear fusion rates from Be to Si.

\section{Results}

We performed calculations of the compositional evolution in the accreted crust for different initial compositions. The initial composition is determined by the nuclear ashes of thermonuclear burning near the neutron star surface, which is expected to vary from system to system depending on companion star composition, accretion rate, and neutron star mass. We use here four sets of ashes: a pure $^{56}$Fe ash composition to facilitate comparison with previous work by \citet{Haensel1990}, predictions for the ashes of extremely hydrogen rich X-ray bursts powered by an extended rp-process \citep{Schatz2001}, predictions for the ashes of a realistic mixed hydrogen and helium burst with a moderate rp-process expected to power GS 1826-238 \citep{Woosley2004a,Cyburt2016}, and predictions for a superburst powered by explosive carbon burning \citep{Keek2012}.

\subsection{Reaction sequence for initial $^{56}$Fe composition}
\label{Sec:Fe56}

\begin{figure}
\plotone{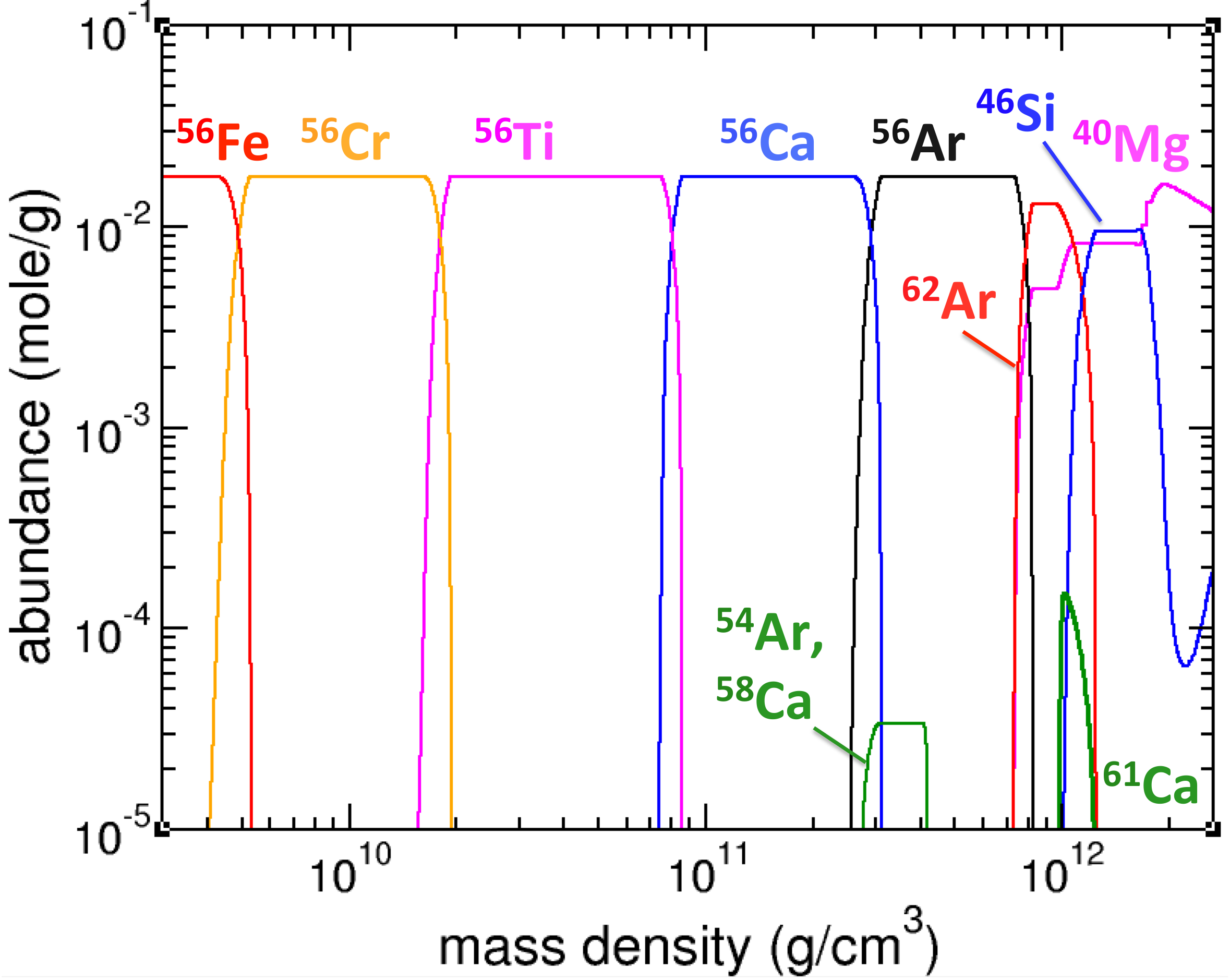}
\caption{\label{Fig:Fe56_Comp} Abundance as a function of density of the most important nuclides for initial $^{56}$Fe burst ashes.}
\end{figure}

We begin by discussing in detail the reaction sequences for an initial composition of pure $^{56}$Fe.  Our calculation follows the compositional change in an accreted fluid element as density and therefore electron chemical potential $\mu_e$ are slowly rising. The evolution of the main composition as a function of depth is shown in Fig.~\ref{Fig:Fe56_Comp} and the major compositional transitions are listed in Tab.~\ref{Tab:56Fe_Transitions}. 

\begin{deluxetable}{crrrl}
  \tablecaption{\label{Tab:56Fe_Transitions} Major compositional transitions for initial $^{56}$Fe 
}
\tablewidth{0pt}
\tablehead{
\colhead{Transition} & \colhead{$P$\tablenotemark{a}} & \colhead{$\rho$\tablenotemark{b}} & \colhead{$\mu_e\tablenotemark{c}$} & \colhead{$X_n$\tablenotemark{d}} 
}
\startdata
$^{56}$Fe$\rightarrow^{56}$Cr & \EE{3.4}{27} & \EE{4.9}{9} & 6.2 &  $<$10$^{-25}$\\
$^{56}$Cr$\rightarrow^{56}$Ti & \EE{1.7}{28}  & \EE{1.8}{10} & 9.6 & $<$10$^{-25}$\\
$^{56}$Ti$\rightarrow^{56}$Ca & \EE{1.1}{29} & \EE{8.1}{10} & 15.6 & $<$10$^{-25}$\\
$^{56}$Ca$\rightarrow^{56}$Ar,$^{54}$Ar,$^{58}$Ca & \EE{5.5}{29} & \EE{2.9}{11} & 23.3 &  \EE{1.2}{-18}\\
$^{56}$Ar,$^{54}$Ar,$^{58}$Ca$\rightarrow^{56}$Ar & \EE{8.3}{29} & \EE{4.2}{11} & 25.9 & \EE{7.2}{-20}\\
$^{56}$Ar$\rightarrow^{40}$Mg,$^{62}$Ar & \EE{1.8}{30} & \EE{7.8}{11} & 31.6 & \EE{5.4}{-8}\\
$^{40}$Mg,$^{62}$Ar$\rightarrow^{40}$Mg,$^{48}$Si & \EE{2.3}{30} & \EE{1.1}{12} & 33.5 & 0.13\\
$^{40}$Mg,$^{48}$Si$\rightarrow^{40}$Mg & \EE{4.2}{30} & \EE{2.8}{12} & 37.1 & 0.54
\enddata
\tablenotetext{a}{Pressure in dyne/cm$^2$}
\tablenotetext{b}{Mass density in g/cm$^3$}
\tablenotetext{c}{Electron chemical potential (without rest mass) in MeV}
\tablenotetext{d}{Neutron abundance}
\end{deluxetable}

\begin{figure}[ht]
\plotone{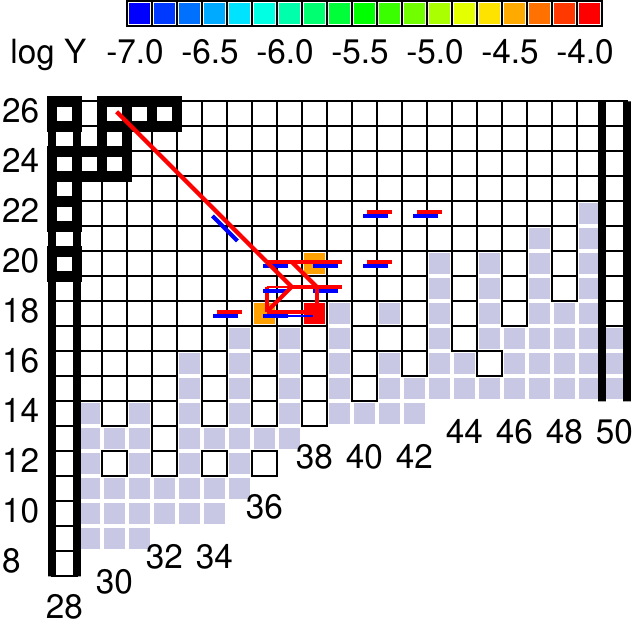}
\caption{Integrated reaction flows on the chart of nuclides for the initial electron capture sequence on $^{56}$Fe down to a depth where  \density{3.5}{11} (\column{3.6}{15}). Rows are labelled on the left with charge number $Z$, columns at the bottom with neutron number $N$.  The isotope colors indicate final abundances $Y$ in mol/g  at the end of the integration time period (see legend). Abundances $\log Y > -4$ are colored red, abundances $\log Y < -7$ are uncolored. The thick black squares mark stable nuclei, the grey squares neutron unbound nuclei included in the network, and the medium thick vertical lines the magic neutron numbers. Shown are flows that lead to lower $Z$ or higher $N$ (red lines) and flows that lead to higher $Z$ and lower $N$ (blue lines). Thick lines indicate flows above 10$^{-6}$~mol/g, thin lines flows between 10$^{-8}$~mol/g and 10$^{-6}$~mol/g. The reaction path splits, leading to a multi-component layer.
\label{FigFluxFe56Step1}}
\end{figure}

The initial reaction sequence up to $\mu_e=15.6$ MeV and \density{8.1}{10} takes 
$2.3 \times 10^{10}$~s and is
characterized by a series of three, two step, electron capture (EC) reactions $^{56}$Fe(2EC)$^{56}$Cr, $^{56}$Cr(2EC)$^{56}$Ti, and
$^{56}$Ti(2EC)$^{56}$Ca described already in \citet{Haensel1990} (see Fig.~\ref{FigFluxFe56Step1}). These electron captures proceed in steps of two because of the 
odd-even staggering of the electron capture thresholds. During the last sequence, ($\gamma$,n) reactions release small amounts of  neutrons that get recaptured but do not appreciably change the reaction flows.  For most EC transitions, the inverse process, $\beta^-$-decay, is blocked, as the decay feeds primarily excited daughter states, which reduces the energy of the emitted electrons resulting in effective Fermi blocking. The exception is $^{56}$Ti(EC)$^{56}$Sc where $\beta^-$-decay of $^{56}$Sc does occur, leading to a  $^{56}$Ti - $^{56}$Sc EC/$\beta^-$ Urca cycle \citep{Schatz2014,Meisel2015}.  However, as discussed in \citet{Meisel2015}, the cycle is weak because of the fast $^{56}$Sc(EC)$^{56}$Ca reaction for the nuclear physics inputs used here, and thus does not affect nuclear energy generation. 

At $\mu_e=23.3$ MeV and \density{2.9}{11}, the destruction of $^{56}$Ca by electron capture occurs. However, this step proceeds entirely differently owing to the rising significance of free neutrons (see Fig.~\ref{FigFluxFe56Step1}). These neutrons are released in the second step of the two step electron capture sequence, which proceeds as $^{56}$Ca(EC)$^{56}$K(EC,2n)$^{54}$Ar. The neutron separation energy of $^{56}$Ar is sufficiently low for most of the EC transitions from $^{56}$K to proceed to neutron-unbound states leading to the emission of neutrons. 
The released neutrons are recaptured by the most abundant nucleus, which is still $^{56}$Ca, leading to a neutron capture sequence to $^{58}$Ca. The reaction path therefore splits into two branches leading to $^{54}$Ar and $^{58}$Ca, respectively. However, branchings between electron capture and neutron capture at $^{57}$Ca and $^{56}$K divert some of the reaction flow to $^{56}$Ar via $^{57}$Ca(EC)$^{57}$K(EC,n)$^{56}$Ar and $^{56}$K(n,$\gamma$)$^{57}$K(EC,n)$^{56}$Ar, respectively. The result is a three nuclide composition, dominated by $^{56}$Ar, but with admixtures of $^{58}$Ca and $^{54}$Ar at about 0.2\% mass fraction each.

\begin{figure}[ht]
\plotone{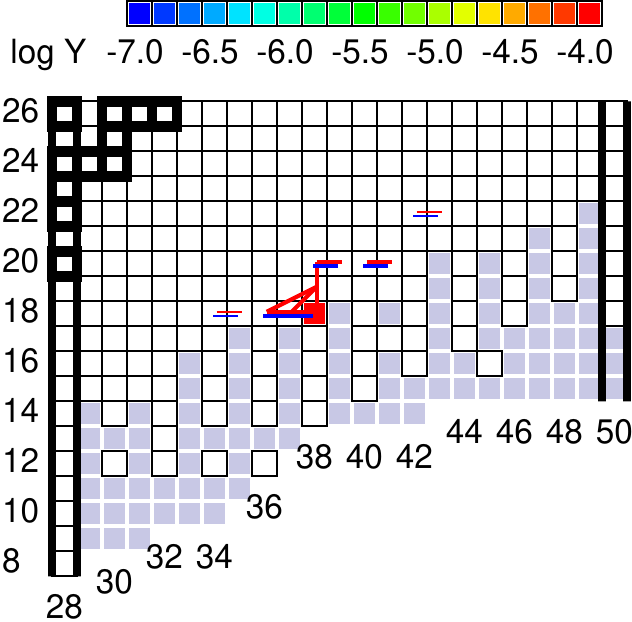}
\caption{Integrated reaction flows for initial $^{56}$Fe ashes from \density{3.6}{11} (\column{3.7}{15})  to \density{4.6}{11} (\column{5.0}{15}).   See Fig.~\ref{FigFluxFe56Step1} for details.
\label{FigFluxFe56Step2}}
\end{figure}

This admixture is, however, short lived, as at $\mu_e=25.9$ MeV and \density{4.2}{11}, $^{58}$Ca and $^{54}$Ar are converted into $^{56}$Ar (Fig.~\ref{FigFluxFe56Step2}). The destruction of $^{58}$Ca proceeds via $^{58}$Ca(EC,1n)$^{57}$K(EC,1n,2n,3n), resulting in a range of Ar isotopes, which, together with the already existing $^{54}$Ar, are quickly transformed into $^{56}$Ar by neutron capture. At this point, the crust is rather pure and mainly composed of $^{56}$Ar. 

\begin{figure}[ht]
\plotone{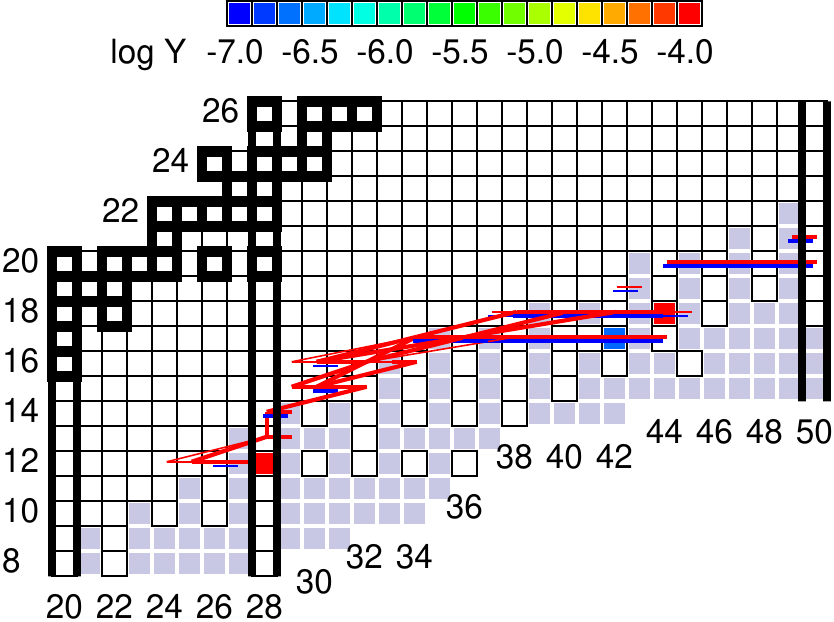}
\caption{Integrated reaction flows for initial $^{56}$Fe ashes from \density{7.0}{11}  (\column{9.0}{15})  to \density{9.4}{11} (\column{1.2}{16}).  See Fig.~\ref{FigFluxFe56Step1} for details.
\label{FigFluxFe56Step3}}
\end{figure}

At $\mu_e$=31.6 MeV and \density{7.8}{11}, $^{56}$Ar is destroyed by the first previously termed superthreshold electron capture cascade (SEC) \citep{Gupta2008} (see Fig.~\ref{FigFluxFe56Step3}). This reaction sequence
occurs when the neutron emission following an electron capture leads to a nucleus with $|Q_{EC}| \ll \mu_e$, which therefore immediately captures electrons again and so on. 
In this particular case, an SEC leading from $^{56}$Ar all the way to $^{40}$Mg is established. The detailed reaction sequence is shown in Fig.~\ref{FigFluxFe56Step3} and is characterized 
by electron captures with the emission of mostly 4--5 neutrons. The released neutrons are recaptured by $^{56}$Ar, which is still the most abundant nuclide. 
This leads again to a split of the reaction path into the SEC from
$^{56}$Ar to $^{40}$Mg  and 
a sequence of neutron captures from $^{56}$Ar to $^{62}$Ar. In the initial phase of the SEC, there is a significant abundance buildup of $^{50}$S produced by neutron capture
from the SEC path, and to a lesser extent of $^{42}$Si. However, with only a slight rise of $\mu_e$, electron capture quickly destroys these isotopes and they are converted into $^{40}$Mg as well. The end result is a layer that consists primarily of $^{62}$Ar (80\% mass fraction) and $^{40}$Mg (20\% mass fraction). There is a small admixture of $^{59}$Cl (10$^{-5}$ mass fraction). The free neutron abundance is significantly increased to 4.8$\times$10$^{-5}$. 

\begin{figure*}[ht]
\plotone{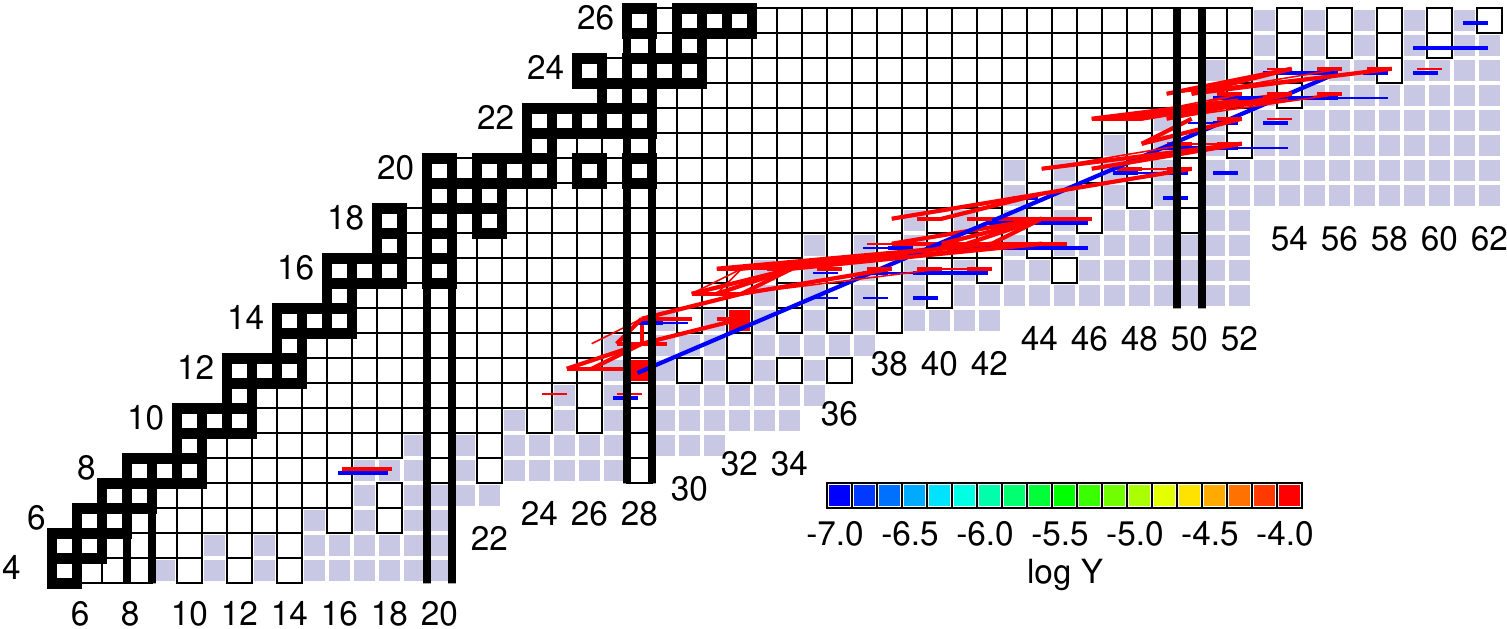}
\caption{Integrated reaction flows for initial $^{56}$Fe ashes from \density{9.7}{11} (\column{1.2}{16}) to \density{1.7}{12} (\column{1.9}{16}).  See Fig.~\ref{FigFluxFe56Step1} for details.
\label{FigFluxFe56Step4}}
\end{figure*}

\begin{figure}[ht]
\plotone{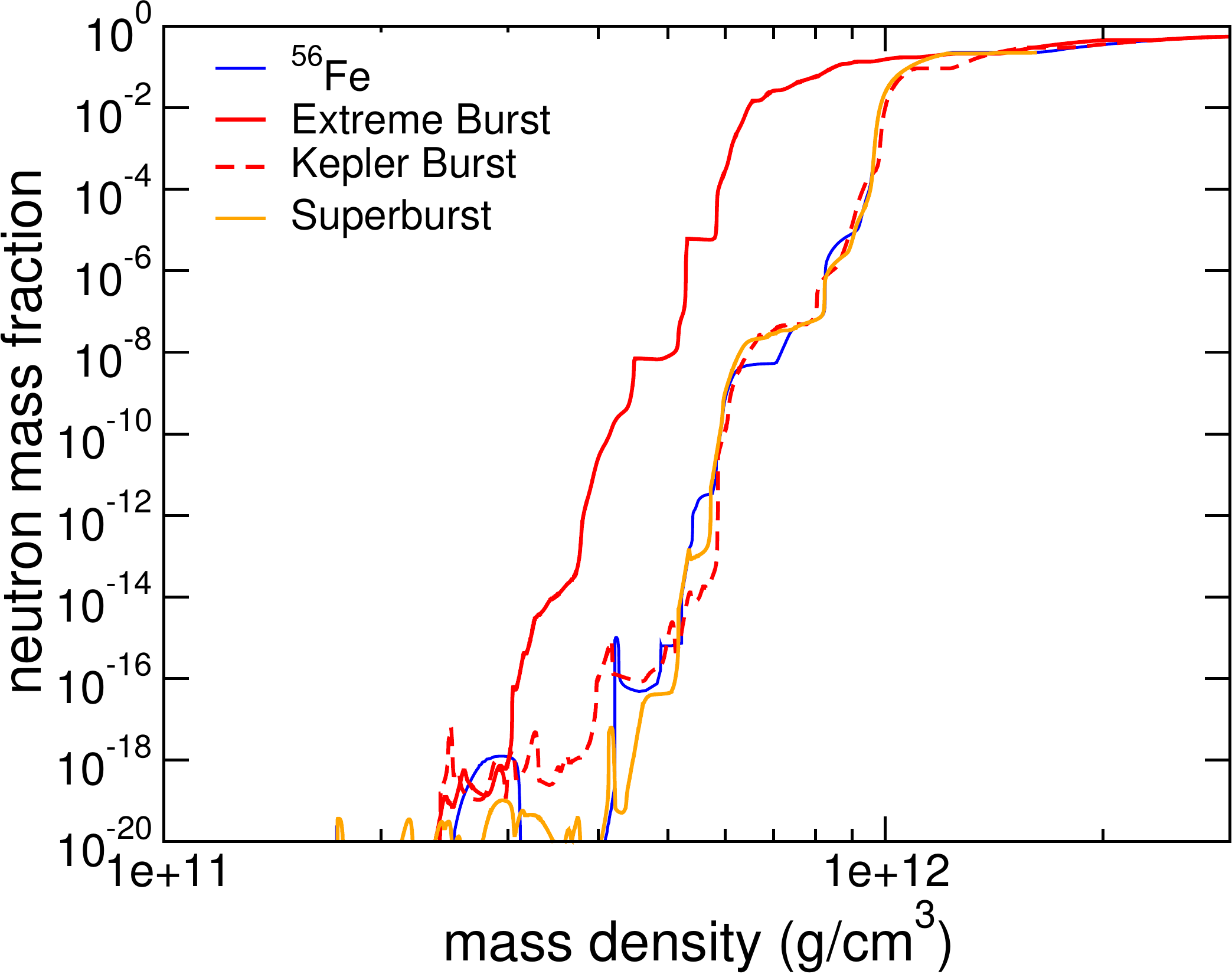}
\caption{Neutron mass fraction $Y_{\rm n}$ as a function of density for pure $^{56}$Fe ashes (solid blue), extreme burst ashes (solid red), KEPLER burst ashes (dashed red), and superburst ashes (solid orange). Neutrons become degenerate for $Y_{\rm n} \gtrsim 10^{-4}$. 
\label{FigYn}}
\end{figure}

At $\mu_e$=33.5 MeV and \density{1.1}{12} $^{62}$Ar is destroyed by an SEC and converted into $^{48}$Si and $^{40}$Mg (Fig.~\ref{FigFluxFe56Step4}). The initial reaction sequence proceeds via $^{62}$Ar(EC,3n)$^{59}$Cl(2n,$\gamma$)$^{61}$Cl(EC,11n)$^{50}$S (Fig.~\ref{FigFluxFe56Step4}).
At $^{50}$S, four destruction paths carry significant flow: (EC,3n), (EC,4n)$^{46}$P(n,$\gamma$), (EC,5n)$^{45}$P(2n,$\gamma$), and (4n,$\gamma$)$^{54}$S(EC,7n) all leading to $^{47}$P, which then 
undergoes a (EC,5n) reaction leading to $^{42}$Si ($N=28$). At this point the neutron abundance has reached $Y_n=0.23$ (Fig.~\ref{FigYn}) and this considerable neutron density results in a significant neutron capture branch that drives reaction flow out of $N=28$ into $^{48}$Si. However, $^{42}$Si(EC,n) is not negligible and results in an additional build up of $^{40}$Mg via
$^{42}$Si(EC,n)$^{41}$Al(EC,3n and 4n)$^{37,38}$Mg(2-3n,$\gamma$)$^{40}$Mg. As a consequence,  $^{40}$Mg increases significantly in abundance. After the destruction of $^{62}$Ar is complete, the layer consists of $^{46}$Si (44\% mass fraction), $^{40}$Mg (32\% mass fraction) and neutrons (23\% mass fraction).

The destruction of $^{62}$Ar coincides with the onset of a weak reaction flow through the first pycnonuclear fusion reaction,
 $^{40}$Mg$+^{40}$Mg$\rightarrow^{80}$Cr (Fig.~\ref{FigFluxFe56Step4}). The fusion reaction is immediately 
followed by a rapid SEC sequence leading back to $^{40}$Mg and establishing a pycnonuclear fusion-SEC cycle 
(Fig.~\ref{FigFluxFe56Step4}). The net effect of the cycle is a  $^{40}$Mg$+^{40}$Mg$\rightarrow^{40}$Mg$ + 40$n reaction resulting in the conversion of $^{40}$Mg into neutrons with increasing depth. $^{46}$Si is the only significant bottle neck in the cycle besides $^{40}$Mg and maintains a significant, roughly constant abundance while reactions produce and destroy the isotope. 

\begin{figure*}[ht]
\plotone{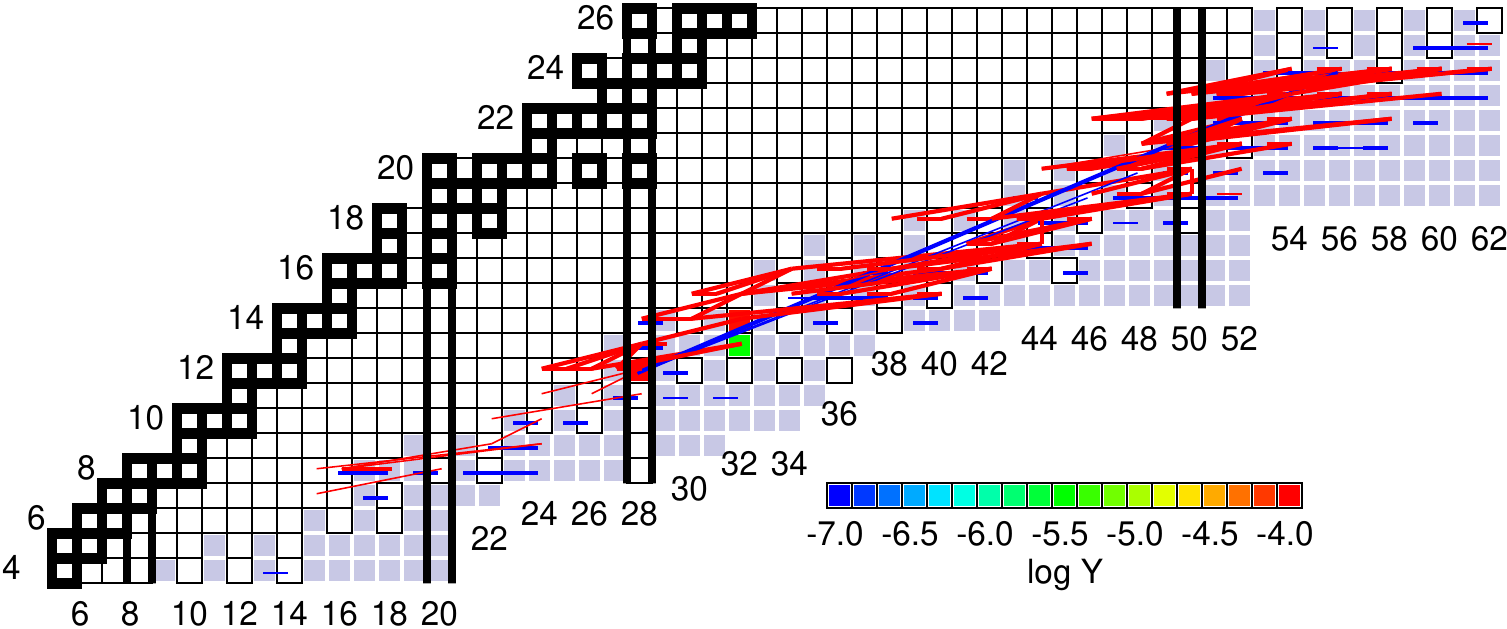}
\caption{Integrated reaction flows for initial $^{56}$Fe ashes from \density{1.7}{12} (\column{1.9}{16}) to \density{2.4}{12} (\column{2.1}{16}). 
 See Fig.~\ref{FigFluxFe56Step1} for details.
\label{FigFluxFe56Step5}}
\end{figure*}

At $\mu_e=37.1\,\MeV$ and \density{1.7}{12}, $^{46}$Si begins to be depleted significantly. However, because of its location in the fusion-SEC cycle, its abundance is initially not dropping to zero but is merely reduced to about $10^{-4}$. At this depth, $^{40}$Mg electron capture begins to initiate a SEC sequence towards lighter nuclei (Fig.~\ref{FigFluxFe56Step5}). This SEC sequence ends at $^{25}$N where the pycnonuclear fusion reaction $^{25}$N$+^{40}$Mg$\rightarrow^{65}$K (not visible in Fig.~\ref{FigFluxFe56Step4} because the flow is too weak) dominates over further electron capture. The resulting $^{65}$K is immediately destroyed by a reaction sequence that merges into the SEC of the $^{40}$Mg+$^{40}$Mg main pycnonuclear fusion-SEC cycle. An additional branching occurs at $^{28}$O, where EC is comparable to $^{28}$O$+^{40}$Mg$\rightarrow^{68}$Ca fusion. Again, the resulting $^{68}$Ca does not accumulate but merges into the main SEC.  EC on $^{40}$Mg therefore effectively leads to a branching of the reaction flow into a two pycnonuclear fusion sub-cycles. The effect of all these cycles is the same $^{40}$Mg$+^{40}$Mg$\rightarrow^{40}$Mg$ + 40$n net reaction converting one $^{40}$Mg nucleus into neutrons on each full loop. The first significant depletion of $^{40}$Mg (Fig.~\ref{Fig:Fe56_Comp}) marks therefore the onset of significant pycnonuclear fusion. 

At a slightly larger depth, at $\mu_e$=37.2 MeV, $\rho=2.4 \times 10^{12}$~g/cm$^3$, $P=3.9 \times 10^{30}$~dyne/cm$^2$ and $y=2.1 \times 10^{16}$~g/cm$^2$ abundance builds up at the edge of our network and we stop the calculation. The neutron mass fraction has reached 46\%. The calculations indicate, though, that the next step is the conversion of $N=28$ $^{40}$Mg into $^{44}$Mg as a consequence of the increasing neutron density. Deeper fusion-SEC cycles develop then starting on $^{44}$Mg. While $^{44}$Mg is not magic, it is the preferred nucleus at these higher neutron densities, because of the jump in neutron binding from Na to Mg predicted by the FRDM mass model leading to an extension of the neutron drip line by 8 isotopes  (see, for example, Fig.~\ref{FigFluxFe56Step5}). 

\subsection{Reaction sequence for extreme rp-process ashes}
\label{Sec:XRB}

\begin{figure}[ht]
\plotone{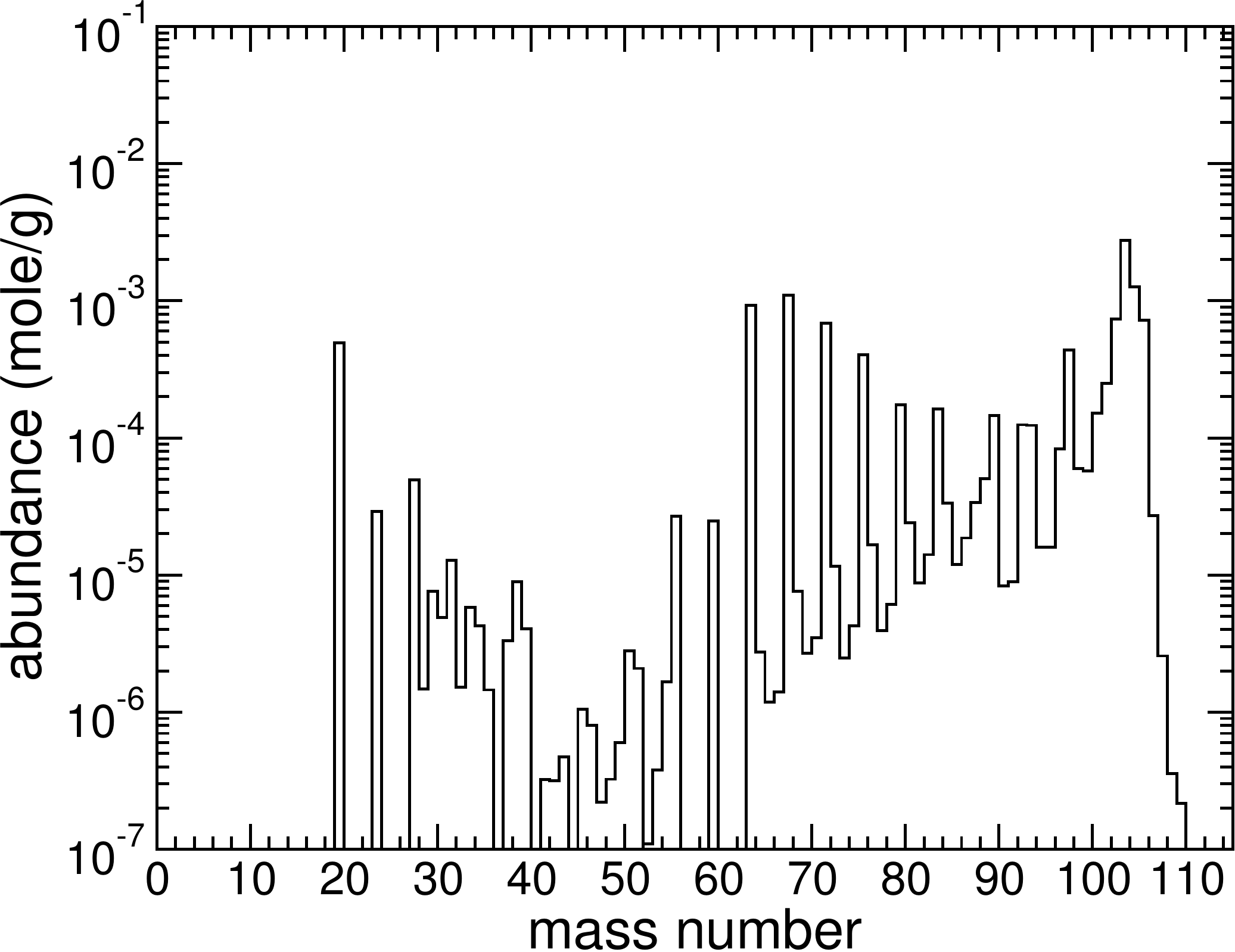}
\caption{Initial composition set by an X-ray burst with an extreme rp-process, summed by mass number. 
\label{FigInitcompXRB}}
\end{figure}

For X-ray bursters that do not exhibit superburst burning, the ashes of the rp-process
is the appropriate initial composition for the crust processes. We use here the composition calculated by the X-ray burst model of \citet{Schatz2001}, which has ignition conditions that correspond to systems with high accretion rate and low metallicity, resulting in a relatively large amount of hydrogen (mass fraction $X=0.66$) at ignition (see Fig.~\ref{FigInitcompXRB}). While such bursts would be rare in nature, the model serves as a useful tool to explore the consequences of a maximally extended rp-process that reaches the Sn-Sb-Te cycle. We ignore elements lighter than neon  assuming they are destroyed by residual helium burning and other thermonuclear fusion processes near the surface. 

\begin{figure*}[ht]
\plotone{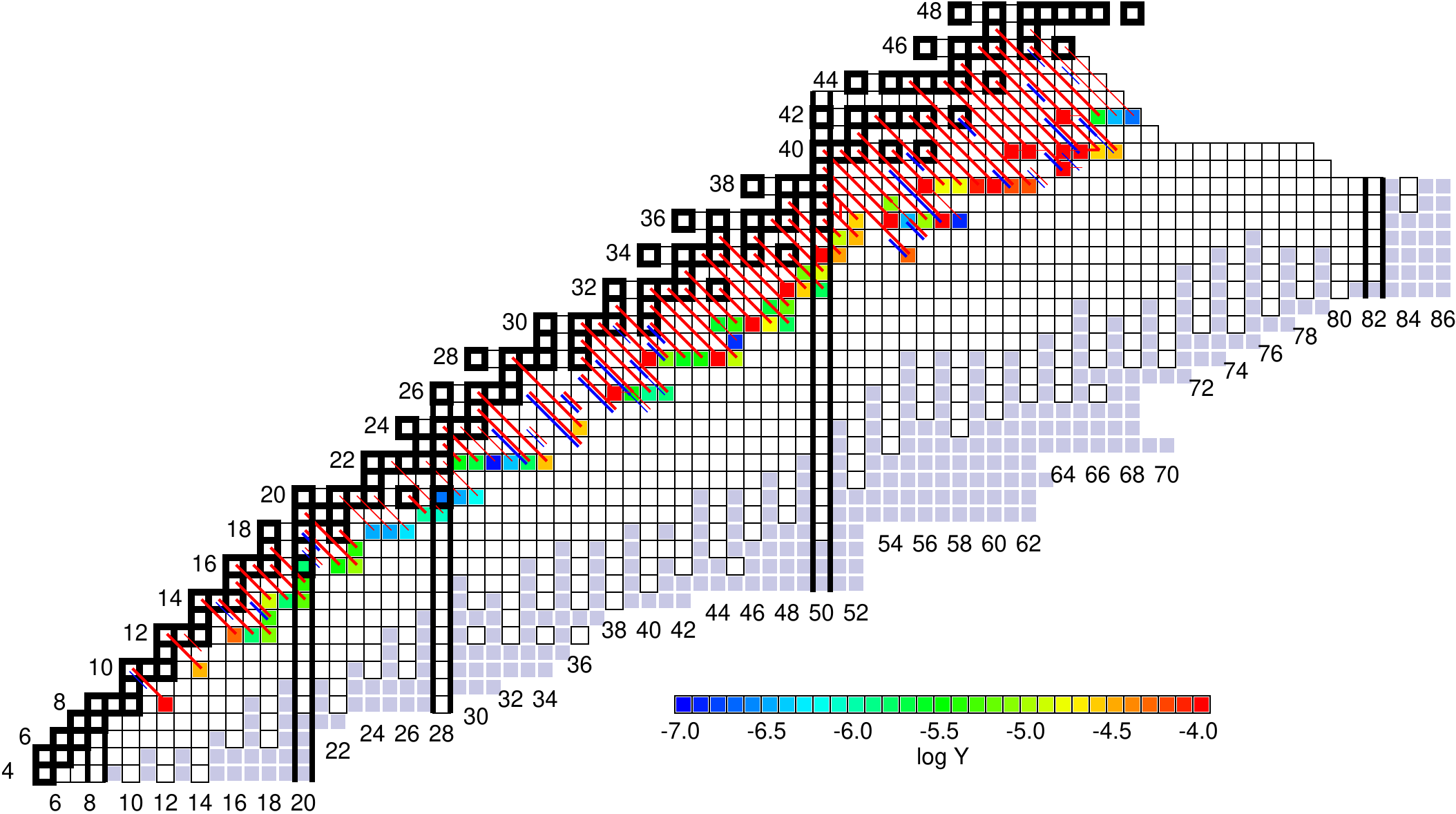}
\caption{Integrated reaction flows and final composition for extreme rp-process ashes down to a depth where 
\density{2.50}{10} (\column{1.32}{14}).  See Fig.~\ref{FigFluxFe56Step1} for details.
\label{FigFluxXRBStep1}}
\end{figure*}

\begin{figure*}[ht]
\plotone{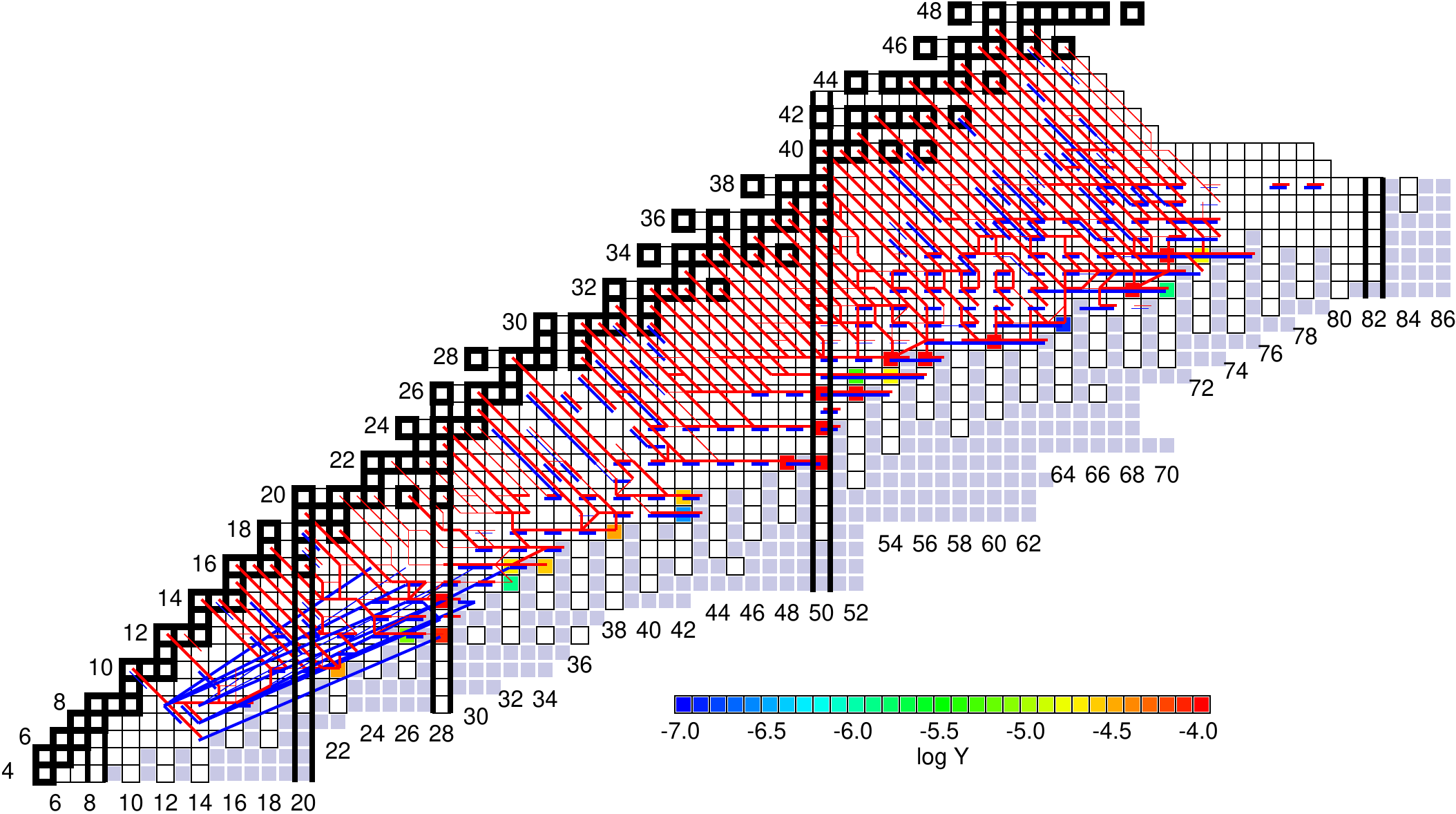}
\caption{Integrated reaction flows and final composition for extreme rp-process ashes down to  a depth where
\density{3.77}{11} (\column{3.96}{15}).  See Fig.~\ref{FigFluxFe56Step1} for details.
\label{FigFluxXRBStep123}}
\end{figure*}

The initial compositional  evolution is characterized by sequences of electron capture reactions along chains of constant mass number (Fig.~\ref{FigFluxXRBStep1}). In this shallow region, the original composition as a function of mass number is preserved and simply pushed to more neutron rich nuclei. In some cases, $\beta^-$ decay is not completely blocked creating a local nuclear Urca cycle \citep{Schatz2014} where both, EC and $\beta^-$ decay occur between a pair of nuclei. This can lead to significant neutrino cooling at the depth where the cycle forms, especially at high temperatures. Such Urca cycles do not occur for all EC transitions. They require a strong ground state to ground state transition (or a transition to a very low lying state with excitation energy $E_x \ll kT$) and an effective blocking of the subsequent EC reaction that would otherwise drain the cycle. EC - $\beta$-decay pairs therefore occur predominantly in odd $A$ chains, though there are a few exceptions, for example in the $A=96$ and $A=98$ chains. The complete set of relevant Urca pairs can be identified in Fig.~\ref{FigFluxXRBStep123} using the $\beta^-$ flow as an indicator of the Urca cycling strength. Tab.~\ref{TabXRBUrca} lists the most important Urca cycling pairs. 

\begin{deluxetable}{ccc}
  \tablecaption{\label{TabXRBUrca} Strongest Urca pairs for extreme rp-process ashes}
\tablewidth{0pt}
\tablehead{
\colhead{Urca pair} & \colhead{$\rho$ (g/cm$^3$)} & \colhead{relative flow\tablenotemark{a}}
}
\startdata
\iso{Zr}{105} - \iso{Y}{105} & \EE{3.15}{10} &  1.00\\
\iso{Sr}{103} - \iso{Rb}{103} & \EE{6.40}{10} &  0.42\\
\iso{Zr}{103} - \iso{Y}{103} & \EE{2.19}{10} &  0.31\\
\iso{Y}{105} - \iso{Sr}{105} & \EE{4.40}{10} &  0.14\\
\iso{Nb}{105} - \iso{Zr}{105} & \EE{1.64}{10} &  0.08\\
\iso{Kr}{ 98} - \iso{Br}{ 98} & \EE{1.16}{11} &  0.08\\
\iso{Kr}{ 96} - \iso{Br}{ 96} & \EE{8.38}{10} &  0.06\\
\iso{Mg}{ 31} - \iso{Na}{ 31} & \EE{8.72}{10} &  0.06\\
\iso{Sr}{ 93} - \iso{Rb}{ 93} & \EE{1.10}{10} &  0.04\\
\iso{Rb}{ 93} - \iso{Kr}{ 93} & \EE{1.60}{10} &  0.03\\
\iso{Sr}{ 99} - \iso{Rb}{ 99} & \EE{3.80}{10} &  0.02\\
\iso{Rb}{ 97} - \iso{Kr}{ 97} & \EE{3.58}{10} &  0.02\\
\iso{Rb}{ 99} - \iso{Kr}{ 99} & \EE{4.99}{10} &  0.02\\
\iso{Kr}{ 97} - \iso{Br}{ 97} & \EE{6.40}{10} &  0.01\\
\iso{Mg}{ 29} - \iso{Na}{ 29} & \EE{5.58}{10} &  0.01\\
\iso{Ti}{ 55} - \iso{Sc}{ 55} & \EE{3.94}{10} &  0.01\\
\iso{Al}{ 31} - \iso{Mg}{ 31} & \EE{3.96}{10} &  0.01\\
\iso{Y}{ 93} - \iso{Sr}{ 93} & \EE{2.03}{ 9} &  0.01\\
\iso{O}{ 21} - \iso{N}{ 21} & \EE{1.22}{11} &  0.01\\
\iso{Rb}{103} - \iso{Kr}{103} & \EE{7.74}{10} &  0.01\\
\enddata
\tablenotetext{a}{Time integrated reaction flow relative to the strongest Urca pair.}
\end{deluxetable}

At around \mue{10.3} and \density{2.36}{10} the first neutrons are created. These neutrons are immediately recaptured by nuclei in other mass chains. The further evolution is therefore characterized by a combination of EC reactions that drive the composition more neutron rich, and neutron capture reactions that deplete some mass chains, and enhance others.  Fig.~\ref{FigFluxXRBStep1} shows the reaction sequences up to the point where neutrons first appear. The first neutrons are created by 
$^{88}$Rb(EC,n)$^{87}$Kr, which competes with $^{88}$Rb(EC)$^{88}$Kr. The reason that neutron emission can occur relatively close to stability is that EC on $^{88}$Rb is predicted to occur through a relatively high lying excited state in $^{88}$Kr at $E_x=$6.9~MeV. This is a consequence of the proximity of the $N=50$ neutron shell closure ($^{88}$Rb has $N=51$) resulting in spherically shaped nuclei and EC strength distributions that are concentrated in a few states \citep{Schatz2014}. This highly excited state is sufficiently close to \sn{88}{Kr}{7.0}  to lead to a significant neutron emission branch. The released neutrons are readily recaptured by nuclei in other mass chains, primarily at higher mass numbers where neutron capture rates tend to be higher. In this case, neutron capture is dominated by $^{105}$Zr, with smaller capture branches on $^{105}$Y, $^{130,104,106}$Zr, and $^{108}$Mo, which are the most abundant high $A$ nuclei present at this time (Fig.~\ref{FigFluxXRBStep1}). Because of the rapid recapture of the released neutrons, the free neutron abundance stays negligibly small. The chief result of the early neutron release is therefore not the appearance of free neutrons, but changes in the composition as function of mass number (see discussion in Section \ref{Sec:FreeN}). 

The first fusion reactions are initiated relatively early at \density{1.2}{11} and \mue{17.5} (Fig.~\ref{FigFluxXRBStep123}). Previously, $^{20}$O produced by EC processes from the initially abundant $^{20}$Ne, has been partially converted by neutron capture into $^{21}$O.  As soon as $^{21}$O undergoes an EC transition to $^{21}$N, $^{21}$N$+^{21}$O and $^{21}$N$+^{21}$N fusion reactions occur. 
Slightly deeper 
at \density{1.3}{11} and $\mu_e$=18.0~MeV, two EC transitions on the remaining $^{20}$O produce $^{20}$C, triggering $^{20}$C$+^{20}$O, $^{20}$C$+^{21}$N, and $^{20}$C$+^{20}$C fusion reactions. Other fusion reaction combinations, including reactions involving $^{22}$O produced by neutron capture from $^{21}$O, also occur but are an order of magnitude weaker. At somewhat higher \density{1.92}{11} and \mue{20.4}, the pycnonuclear fusion of oxygen becomes possible. $^{24}$O, produced by electron captures from the initially present $^{24}$Mg, with some contribution from $^{21}$O neutron captures, is destroyed via $^{24}$O$+^{24}$O. With this reaction, all oxygen is destroyed, leaving neon as the lightest element present in the crust. 

At \density{3.78}{11} and \mue{25.2} nuclei in most reaction chains have reached the neutron drip line (Fig.~\ref{FigFluxXRBStep123}). The free neutron abundance is still low with $Y_{\rm n}$=\EE{1.1}{-13}. This is sufficient however to drive the composition into a (n,$\gamma$)-($\gamma$,n) equilibrium within each isotopic chain. Note that the neutron Fermi energy $E_{\rm F n} \ll kT$ at this stage. Overall, neutron capture reactions have significantly altered the composition as a function of mass number. In particular, abundances in most odd $A$ chains have been drastically reduced, the only remaining odd $A$ nucleus with a significant abundance is $^{89}$Cu (Section \ref{Sec:FreeN}). 

 \begin{figure*}[ht]
\plotone{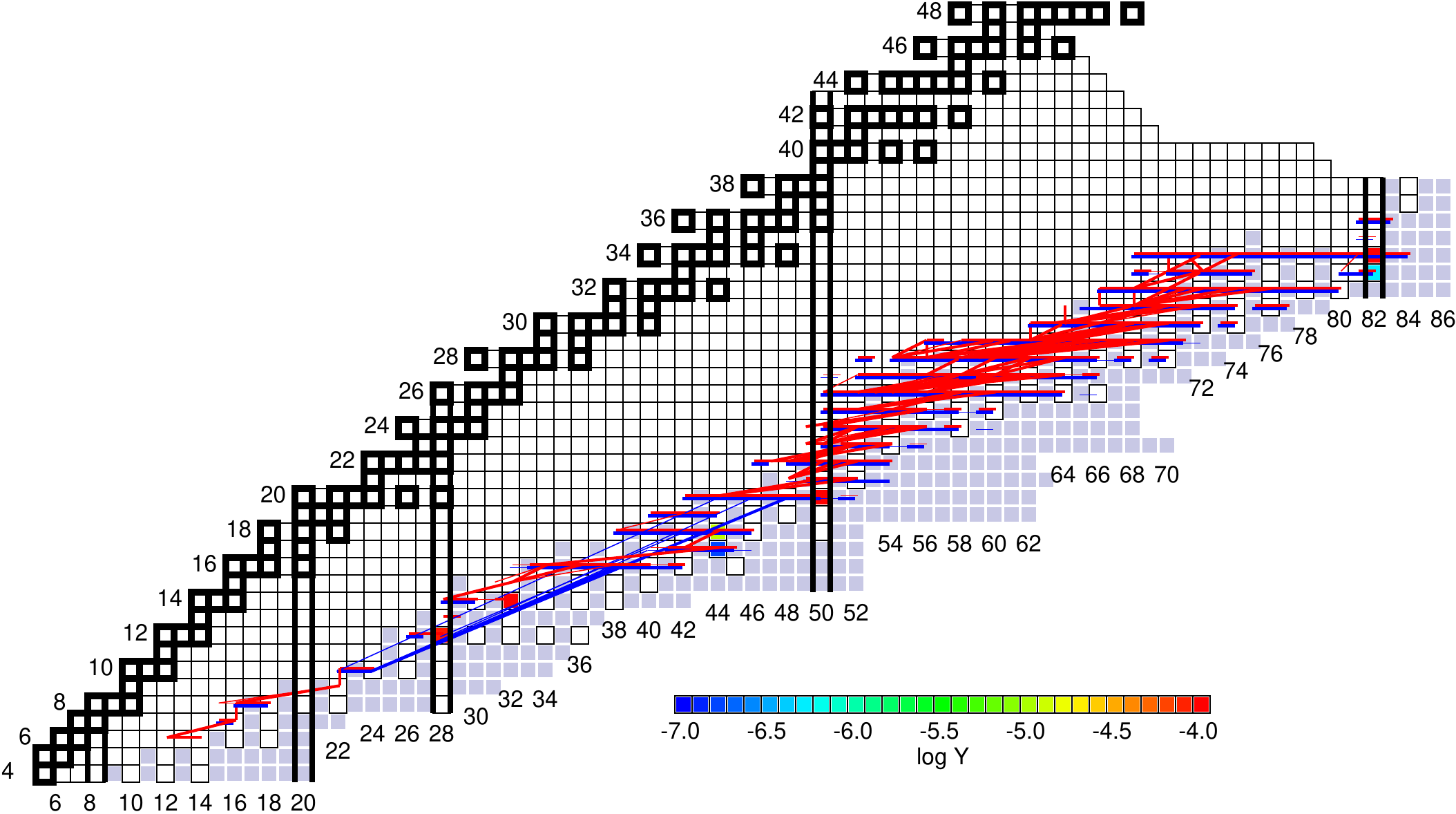}
\caption{Integrated reaction flows and final composition for extreme rp-process ashes starting at  
\density{3.77}{11} (\column{3.96}{15}) and ending at \density{1.28}{12} (\column{1.27}{16}).  See Fig.~\ref{FigFluxFe56Step1} for details.
\label{FigFluxXRBStep4}}
\end{figure*}

Beyond \density{3.78}{11} and \mue{25.2}, SEC chains begin to play a role and rapidly convert nuclei along the neutron drip line into lighter species until a particularly strongly bound nucleus with a large EC threshold is reached (Fig.~\ref{FigFluxXRBStep4}). The associated release of neutrons leads to a drastic increase of the free neutron abundance, marking the location of neutron drip. 

Neutron captures also drive the abundance in the neon isotopic chain, predominantly originating from initial $^{28}$Si in the burst ashes, into $^{32}$Ne and $^{34}$Ne. At \density{7.7}{11}, pycnonuclear fusion reactions set in and destroy $^{32}$Ne and $^{34}$Ne. The most important reactions are $^{34}$Ne$+^{34}$Ne, $^{32}$Ne$+^{34}$Ne, $^{34}$Ne$+^{24}$O, and $^{34}$Ne$+^{20}$C. $^{24}$O is produced via $^{32}$Ne(EC,n)$^{31}$Na(EC,8n)$^{23}$O(n,$\gamma$)$^{24}$O. 
$^{20}$C is produced from $^{24}$O via $^{24}$O(EC,2n)$^{22}$N(n,$\gamma$)$^{23}$N(EC,5n)$^{18}$C(2n,$\gamma$)$^{20}$C.    

All these processes are essentially completed at \density{1.28}{12} and \mue{33.6}, at which point the composition is concentrated in a few nuclei at $N=82$ ($^{116}$Se, abundance $Y$=\EE{4.4}{-4}), at $N=50$ ($^{70}$Ca $Y$=\EE{1.0}{-2}, near $N=28$ ($^{46}$Si, $Y$=\EE{2.6}{-4}), and 
at $N=28$ ($^{40}$Mg, $Y$=\EE{8.4}{-5}). The neutron abundance has reached $Y_{\rm n}=0.21$ (Fig.~\ref{FigYn}) and $E_{\rm F n}$ has reached 0.62~MeV, exceeding $kT \approx 40$~keV resulting in degenerate neutrons. The abundance accumulated at the three locations where the neutron drip line intersects the neutron numbers $N=28, 50$, and 82 can be mapped to different mass ranges in the initial composition. $N=82$ is mostly produced from initial $A \ge 106$ nuclei, with some contribution from $A=102-105$. The initial $A \ge 106$ abundance is \EE{7.5}{-4}, already larger than the final $N=82$ abundance. The main branch points that govern leakage to lighter nuclei for $A \ge 106$ material are $^{106}$Kr and $^{104}$Se. At $^{106}$Kr, neutron capture moves material towards $N=82$, while EC feeds $^{104}$Se via a (EC)($\gamma$,n)(EC,n) sequence. At $^{104}$Se, (EC,n) moves material ultimately to $N=50$, while neutron capture feeds $N=82$. The final $N=82$ abundance may be increased by a small contribution from $A=102-105$. The key branchings are $^{102}$Se and again $^{104}$Se, where in each case the EC branch moves the abundance towards $N=50$.

$A \le 56$ nuclei are mostly converted into nuclei near the $N=28$ region. $A=56$ is the borderline case, and the reaction sequence is similar to the pure $^{56}$Fe case discussed in section \ref{Sec:Fe56}. An isolated exception is the initial $^{28}$Si abundance which in part ends up in the $N=50$ region due to fusion reactions: $^{28}$Si is converted into neon isotopes via ECs. At $^{28}$Ne neutron capture competes with $^{28}$Ne(EC,n). The $^{28}$Ne(EC,n) branch leads to fluorine and then oxygen, which then fuses into nuclei in the sulfur region. However, a significant fraction of the initial $^{28}$Si abundance is processed through the $^{28}$Ne(n,$\gamma$) branch  leading to neutron rich neon isotopes, which then fuse into calcium, and ultimately end up in $^{70}$Ca ($N=50$). Initial elements lighter than silicon fuse into nuclei below calcium and are therefore converted into $N \approx 28$ nuclei. Initial elements heavier than silicon but lighter than iron end up as magnesium or silicon isotopes, which do not fuse until much greater depths, when nuclei are sufficiently neutron rich for SECs to prevent accumulation at $N=50$.

 \begin{figure*}[ht]
\plotone{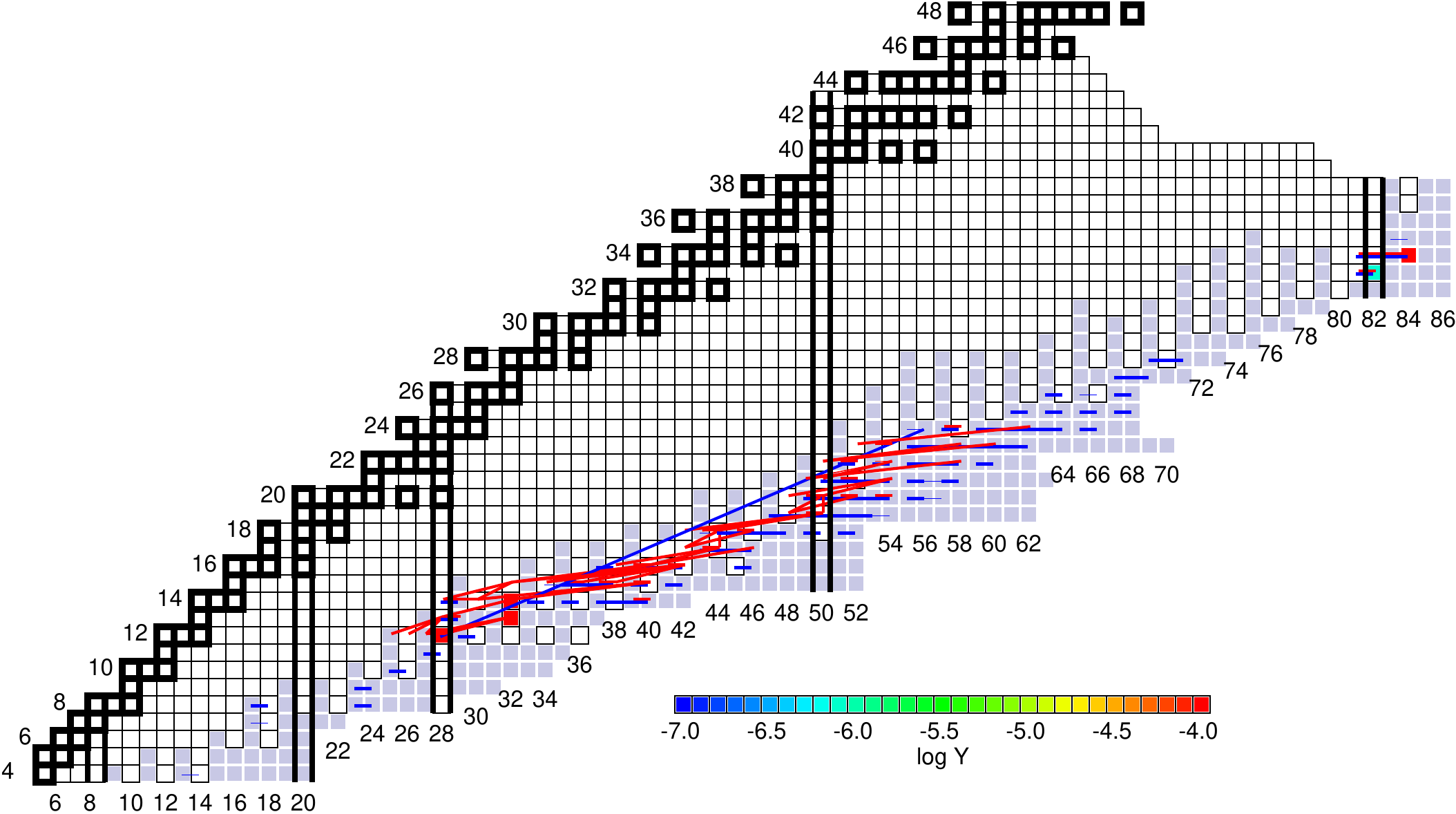}
\caption{Integrated reaction flows and final composition for extreme rp-process ashes starting at  
\density{1.76}{12} (\column{1.58}{16}), and ending at \density{2.49}{12} (\column{2.11}{16}).  See Fig.~\ref{FigFluxFe56Step1} for details.
\label{FigFluxXRBStep5}}
\end{figure*}

We continue the simulation beyond \density{1.25}{12} and \mue{33.6} to \density{2.5}{12} and \mue{37}, at which point the increasing neutron density drives the composition towards the edge of our reaction network. At \density{2}{12} and \mue{35.1} $N=50$ $^{70}$Ca is destroyed by a SEC and converted into nuclei in the $N=28$ region, leaving only the $N=28$ and $N=82$ regions with significant abundance. We also see the onset of significant $^{40}$Mg$+^{40}$Mg fusion, resulting in a similar fusion-SEC cycle as discussed in section \ref{Sec:Fe56}. 

 \begin{figure}[ht]
\plotone{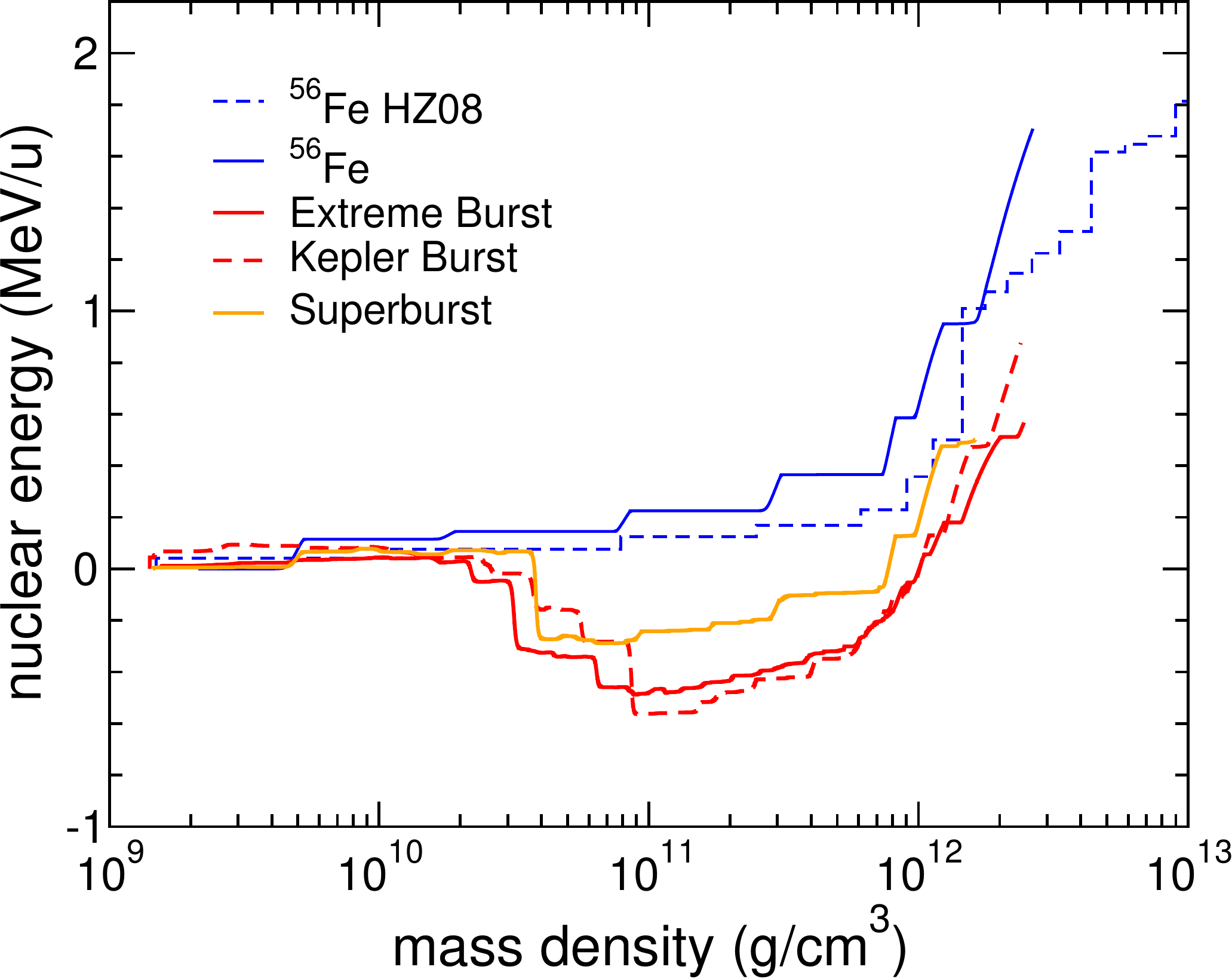}
\caption{Integrated nuclear energy release as a function of mass density for pure $^{56}$Fe ashes (solid blue), extreme burst ashes (solid red), KEPLER burst ashes (dashed red), and superburst ashes (solid orange). The nuclear energy release obtained by \citet{Haensel2008} for pure $^{56}$Fe ashes is shown for comparison (dashed blue).
\label{FigEnergy}}
\end{figure}

Fig.~\ref{FigEnergy} shows the calculated time integrated nuclear energy production. The various drops indicate significant cooling from nuclear Urca pairs in the outer crust. Indeed, the location of the top three pairs listed in Tab.~\ref{TabXRBUrca} coincides with the major drops visible in Fig.~\ref{FigEnergy} around \density{2.2}{10}, \density{3.2}{10}, and \density{6.4}{10}.

\subsection{Reaction sequence for initial KEPLER X-ray burst ashes}

\begin{figure}[ht]
\plotone{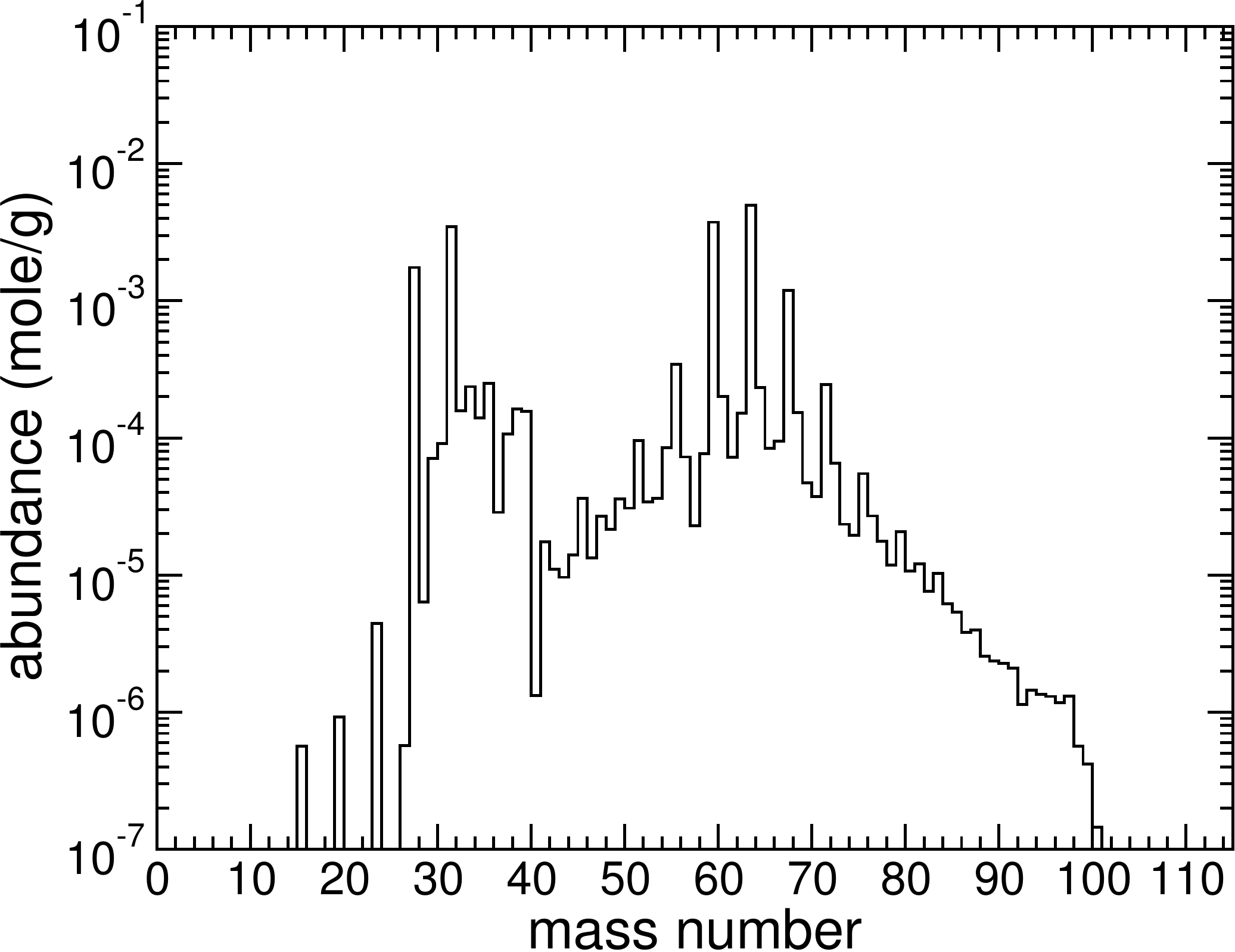}
\caption{Initial composition set by the ashes of an X-ray burst modeled with KEPLER, summed by mass number. 
\label{FigInitcompKEPLER}}
\end{figure}

A more typical estimate of the final composition of mixed H/He bursts is provided by calculations using the 1D multi-zone code KEPLER. We use the model described in more detail in \citet{Cyburt2016}, which was shown to reproduce the observed light curve features of GS1826-24 reasonably well \citep{Heger2007}. The final composition entering the crust is calculated by averaging over the deeper layers in the accreted material after a sequence of about 14 bursts, excluding the bottom layers that are produced by the atypical first burst (see \citet{Cyburt2016} for details). Fig.~\ref{FigInitcompKEPLER} shows the initial composition as a function of mass number. The main difference to the extreme rp-process discussed in section \ref{Sec:XRB} is the reduced amount of heavier nuclei beyond $A=72$. This is due to increased CNO hydrogen burning in between bursts that leads to a lower hydrogen abundance at ignition, and a lower ignition depth that leads to lower peak temperature and a less extended $\alpha$p-process. Both of these effects result in a lower hydrogen to seed ratio and a shorter rp-process (see Eq. 13 in \citet{Schatz1999}). 

\begin{figure*}[ht]
\plotone{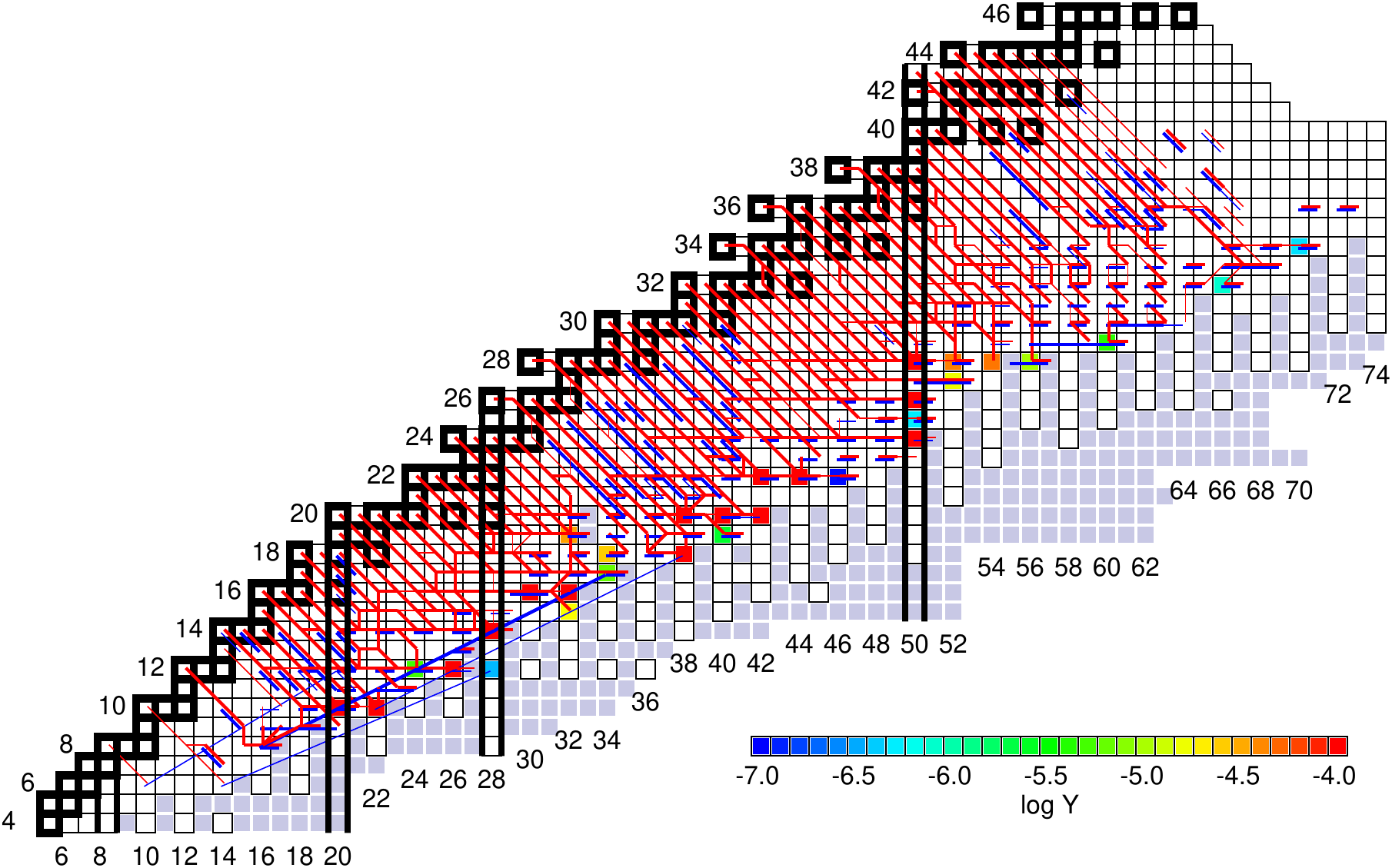}
\caption{Integrated reaction flows and final composition for KEPLER X-ray burst ashes down to  a depth where
\density{3.28}{11} (\column{3.43}{15}).  See Fig.~\ref{FigFluxFe56Step1} for details.
\label{FigFluxKEPLERPhase1}}
\end{figure*}

\begin{figure*}[ht]
\plotone{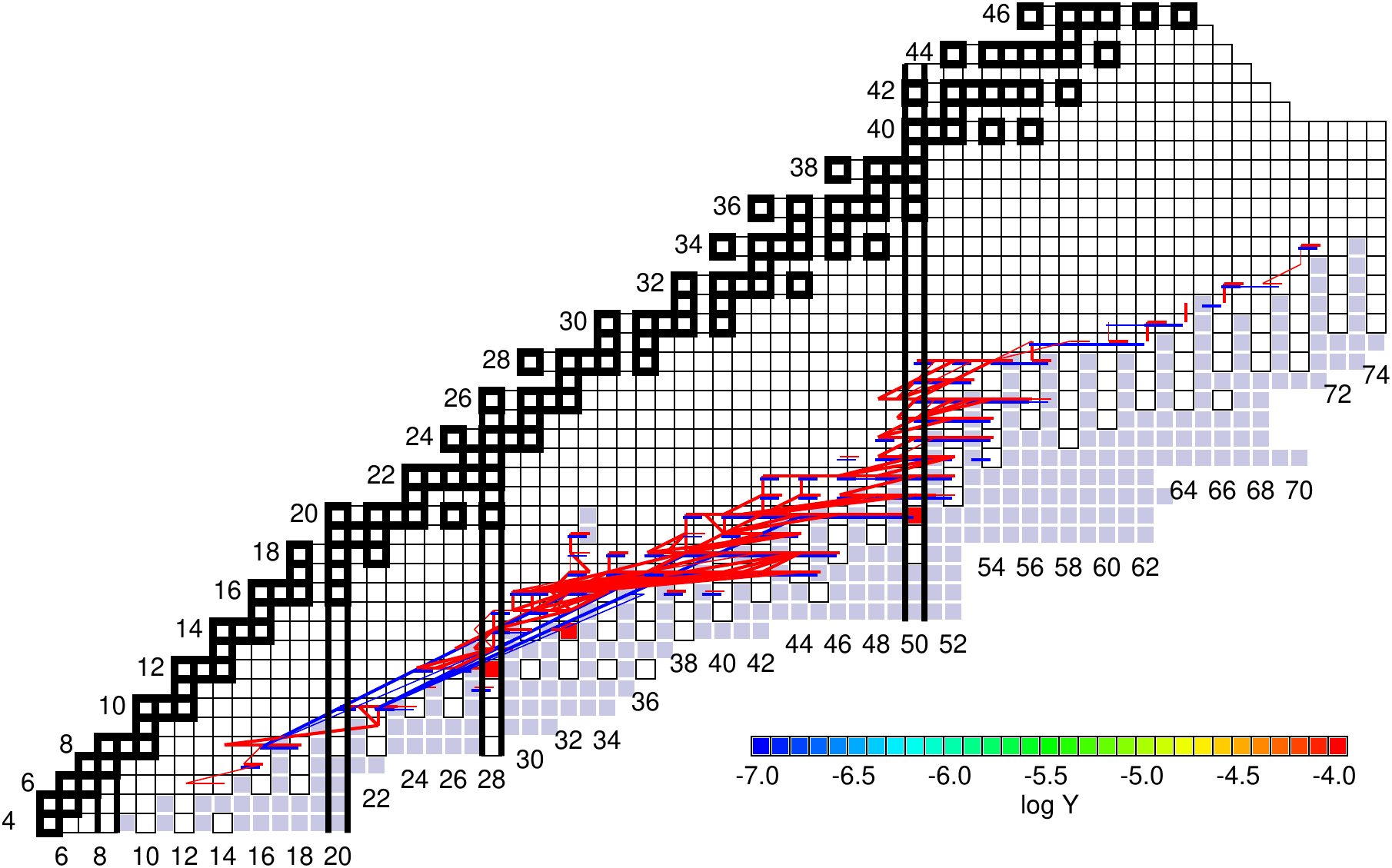}
\caption{Integrated reaction flows and final composition for KEPLER X-ray burst ashes starting at  
\density{3.28}{11} (\column{3.43}{15}), and ending at \density{1.10}{12} (\column{1.27}{16}).  See Fig.~\ref{FigFluxFe56Step1} for details.
\label{FigFluxKEPLERPhase2}}
\end{figure*}

The evolution of the composition with increasing depth is overall very similar to the extreme rp-process ashes case, though the different mass chains have different relative abundances due to the different initial composition (Fig.~\ref{FigFluxKEPLERPhase1}). The main global difference is a shift of the appearance of significant amounts of free neutrons towards higher densities by about 50\%, which is more in line with the calculation for pure $^{56}$Fe ashes (Fig.~\ref{FigYn}). This is due to the much reduced abundance of heavy ($A>72$) nuclei that tend to reach the neutron drip line at shallower depth and lead to an early release of neutrons in the case of the extreme rp-process ashes. The deeper onset of neutron drip has some consequences for the further evolution of the composition. In particular, when the heavier mass chains reach the neutron drip line, the neutron abundance is lower, and the equilibrium nucleus is therefore closer to stability where electron capture thresholds are lower. SEC cascades therefore set in earlier, compared to the case of extreme rp-process ashes where nuclei are pushed to more neutron rich isotopes that require higher $\mu_{\rm e}$ to capture electrons (Fig.~\ref{FigFluxKEPLERPhase2}). This, together with the much lower initial abundance of $A>103$ nuclei leads to a negligible production of $N=82$ nuclei.  The final composition after SECs have converted the composition into nuclei at or near closed shells or with large single particle energy level gaps is $^{70}$Ca ($N=50$, $Y=$\EE{8.5}{-3}), $^{46}$Si ($N=32$, $Y=$\EE{1.24}{-3}), and $^{40}$Mg ($N=28$, $Y=$\EE{6.4}{-3}) with a neutron abundance of $Y_{\rm n}=0.092$.

Again we can map these final abundances to the distribution by mass number of the initial composition. As in the case of the extreme rp-process ashes, $A \approx 56$ is the approximate dividing line between material ending up in $N=50$ and near $N=28$. However, because of the lower free neutron abundance nuclei tend to be less neutron rich and have lower EC thresholds. Therefore, in the case of the KEPLER X-ray burst ashes, there is some leakage from the $A=58-60$ mass chains towards the $N=28$ region, primarily due to branch points where neutron capture competes with EC such as $^{59}$Ca(EC),$^{60}$Ca(EC,n), and $^{62}$Ca(EC,3n). While the abundance that remains in the Ca isotopic chain is ultimately converted into $N=50$ $^{70}$Ca, any leakage to lighter elements feeds the $N=28$ region. As in the case of the extreme rp-process ashes, the initial $^{28}$Si abundance is fed into the Ca isotopic chain and converted to $^{70}$Ca. Indeed, the sum of the initial abundances of $^{28}$Si and $A > 60$ is  \EE{9.5}{-3}~mol/g exceed the produced $N=50$ abundance slightly, indicating that most of the other nuclei end up near $N=28$. 

At depths beyond \density{1.10}{12} (\column{1.27}{16}) the evolution is essentially the same as in the case of the extreme rp-process ashes (see section \ref{Sec:XRB}). The nuclear energy release is overall similar (Fig.~\ref{FigEnergy}), though there is somewhat stronger nuclear Urca cooling, and a slightly higher nuclear energy generation. The dominant Urca pairs are listed in 
Tab. \ref{TabKEPLERUrca}. Compared to the extreme rp-process ashes there are two additional important Urca pairs in lighter mass chains, $^{33}$Al-$^{33}$Mg and $^{65}$Fe-$^{65}$Mn. The lack of shallower cooling from heavier nuclei is more than offset by the very strong cooling from $^{31}$Mg-$^{31}$Na at around \density{1}{11} (Fig.~\ref{FigEnergy}). This is due to the much larger initial abundance of $A=31$ nuclei (\EE{9.1}{-5} vs \EE{4.9}{-6}). 

\begin{deluxetable}{ccc}
  \tablecaption{\label{TabKEPLERUrca} Strongest Urca pairs for KEPLER X-ray burst ashes}
\tablewidth{0pt}
\tablehead{
\colhead{Urca pair} & \colhead{$\rho$ (g/cm$^3$)} & \colhead{relative flow\tablenotemark{a}}
}
\startdata
\iso{Mg}{ 31} - \iso{Na}{ 31} & \EE{8.60}{10} &  1.00\\
\iso{Al}{ 33} - \iso{Mg}{ 33} & \EE{5.62}{10} &  0.47\\
\iso{Ti}{ 55} - \iso{Sc}{ 55} & \EE{3.72}{10} &  0.40\\
\iso{Al}{ 31} - \iso{Mg}{ 31} & \EE{3.75}{10} &  0.14\\
\iso{Fe}{ 65} - \iso{Mn}{ 65} & \EE{2.59}{10} &  0.11\\
\iso{Cr}{ 59} - \iso{V}{ 59} & \EE{2.42}{10} &  0.09\\
\iso{Mg}{ 29} - \iso{Na}{ 29} & \EE{5.32}{10} &  0.07\\
\iso{Cr}{ 57} - \iso{V}{ 57} & \EE{1.36}{10} &  0.05\\
\iso{V}{ 57} - \iso{Ti}{ 57} & \EE{2.62}{10} &  0.04\\
\iso{Fe}{ 63} - \iso{Mn}{ 63} & \EE{1.61}{10} &  0.04\\
\iso{Mn}{ 59} - \iso{Cr}{ 59} & \EE{1.08}{10} &  0.03\\
\iso{Cr}{ 63} - \iso{V}{ 63} & \EE{4.56}{10} &  0.03\\
\iso{Ni}{ 65} - \iso{Co}{ 65} & \EE{5.27}{ 9} &  0.03\\
\iso{Co}{ 65} - \iso{Fe}{ 65} & \EE{1.21}{10} &  0.02\\
\iso{V}{ 59} - \iso{Ti}{ 59} & \EE{3.39}{10} &  0.02\\
\iso{Fe}{ 61} - \iso{Mn}{ 61} & \EE{9.09}{ 9} &  0.02\\
\iso{Fe}{ 59} - \iso{Mn}{ 59} & \EE{3.48}{ 9} &  0.01\\
\iso{Ti}{ 57} - \iso{Sc}{ 57} & \EE{5.78}{10} &  0.01\\
\iso{V}{ 55} - \iso{Ti}{ 55} & \EE{9.90}{ 9} &  0.01\\
\enddata
\tablenotetext{a}{Time integrated reaction flow relative to strongest Urca pair.}
\end{deluxetable}

\subsection{Reaction sequence for initial superburst ashes}

\begin{figure}[ht]
\plotone{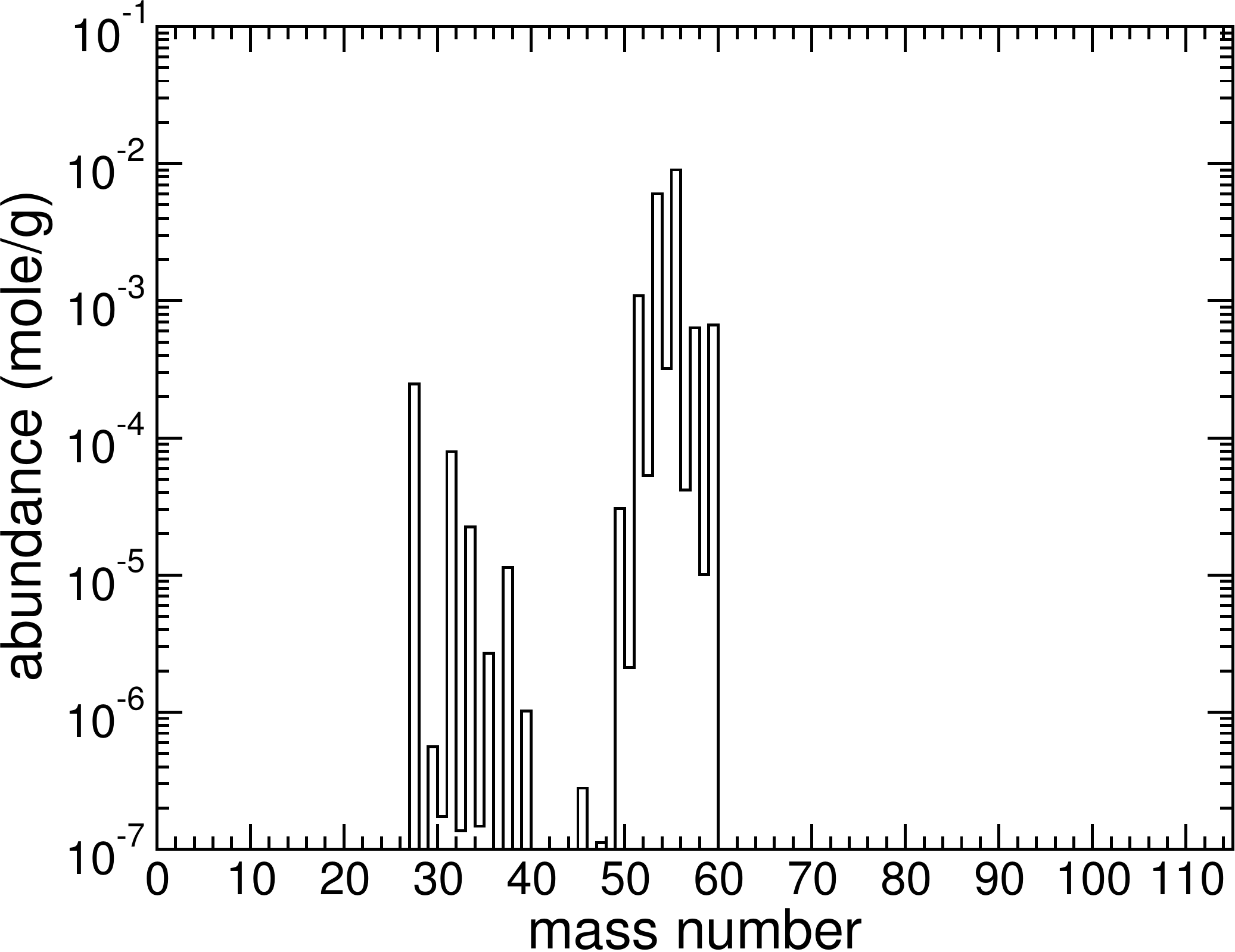}
\caption{Initial composition set by the ashes of a superburst, summed by mass number. 
\label{FigInitcompSB}}
\end{figure}

\begin{figure*}[ht]
\plotone{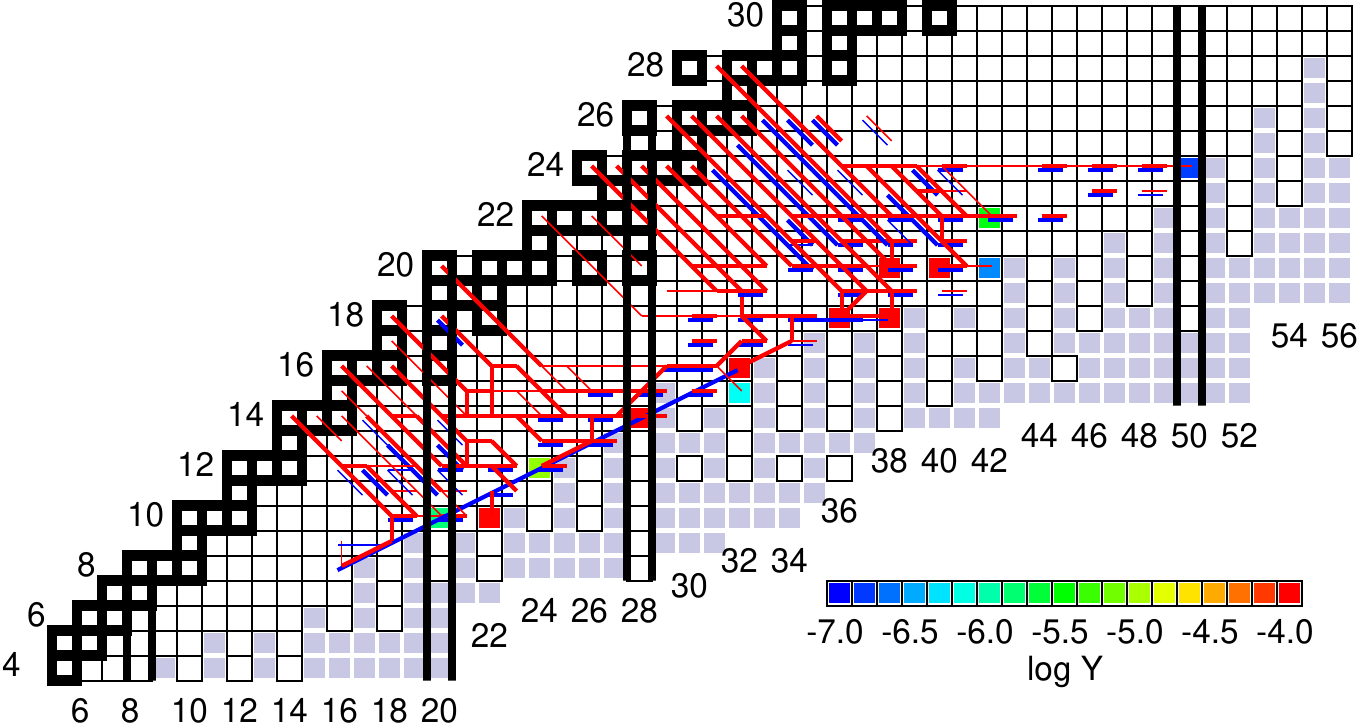}
\caption{Integrated reaction flows and final composition for superburst ashes down to  a depth where
\density{3.39}{11} (\column{3.43}{15}).  See Fig.~\ref{FigFluxFe56Step1} for details.
\label{FigFluxSBPhase1}}
\end{figure*}

\begin{figure*}[ht]
\plotone{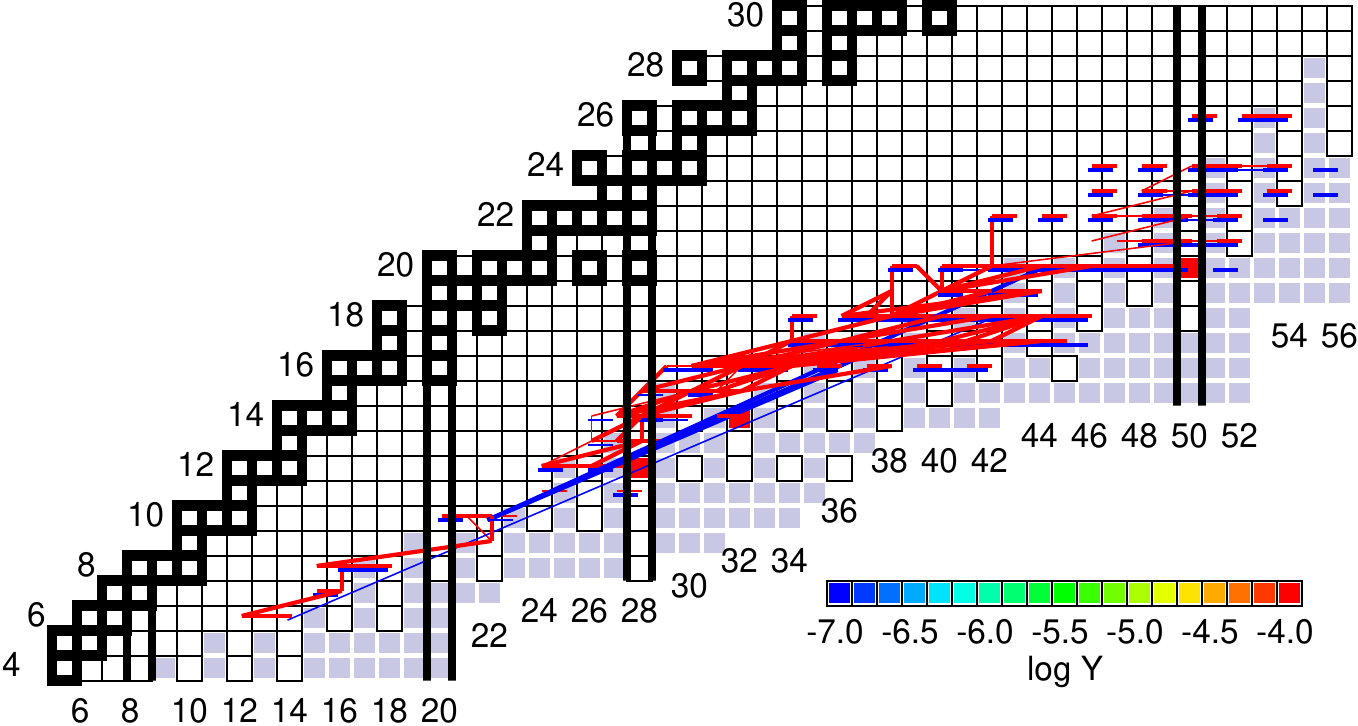}
\caption{Integrated reaction flows and final composition for superburst ashes starting at  
\density{3.39}{11} (\column{3.43}{15}), and ending at \density{1.23}{12} (\column{1.30}{16}).  See Fig.~\ref{FigFluxFe56Step1} for details.
\label{FigFluxSBPhase2}}
\end{figure*}

In some X-ray bursting systems, rare superbursts may further modify the composition at a depth around \density{1}{9} \citep{Cumming2006}. For systems that regularly exhibit superbursts, the ashes of superburst burning is the appropriate initial composition for nuclear processes in the crust. We use the final composition produced in a superburst model calculated with the KEPLER code \citep{Keek2011}. This composition is shown in Fig.~\ref{FigInitcompSB}.  The sequence of EC and neutron captures leading to the neutron drip line is shown in Fig.~\ref{FigFluxSBPhase1} and, for the mass chains with significant abundance, is very similar to the result with the KEPLER X-ray burst ashes. Fig.~\ref{FigFluxSBPhase2} shows the reaction sequences and final composition when the composition is consolidated to a few nuclei that are particularly strongly bound due to shell effects. One difference to the calculation with the KEPLER X-ray burst ashes is that this point is reached at a slightly higher density (\density{1.23}{12} instead of \density{1.1}{12}). The reason is that $Y_{\rm e}$ is smaller due to the higher neutron abundance and therefore a higher density is required to achieve \mue{33.7}, needed to destroy $^{62}$Ar. 

The final composition shown in Fig.~\ref{FigFluxSBPhase2} has been determined at the same $\mu_{\rm e}$ as Fig.~\ref{FigFluxKEPLERPhase2} and demonstrates that indeed the same nuclei are populated. Because this composition is reached at a greater depth, it is much closer to the onset of pycnonuclear fusion of $^{40}$Mg. However, the relative population of $^{40}$Mg ($Y$=\EE{9.1}{-3}), $^{46}$Si ($Y$=\EE{8.7}{-3}), and $^{70}$Ca ($Y$=\EE{3.2}{-4}) is different because of the different initial composition. Most of the abundance is concentrated around $N=28$ with only a smaller $N=50$ contribution from $^{70}$Ca. Because the initial composition has only a small amount of $A > 60$ nuclei ($Y$=\EE{8}{-8} ~mol/g), the only contribution to $N=50$ comes from parts of the initial $A=60$ and $A=28$ abundances (see discussion in section \ref{Sec:XRB}). Again the competition of neutron capture and EC at $^{60}$Ca and $^{28}$Ne, respectively, is critical in determining the relative distribution of $N=28$ and $N=50$ nuclei in the inner crust. 

In crusts with initial superburst ashes, Urca cooling is comparable to the case of extreme rp-process ashes (Fig.~\ref{FigEnergy}), however, cooling comes almost exclusively from the $^{55}$Sc-$^{55}$Ti pair. All other Urca cooling pairs are at least a factor of 50 weaker. Heating is significantly higher than for either of the X-ray burst ashes cases.


\section{Discussion}

\subsection{Reaction Sequences} 
\label{Sec:Sequence}

The reaction sequences obtained here with a full reaction network differ substantially from previous work, especially when compared to models that consider only a single species at a given time such as \citet{Haensel1990,Haensel2008}. One important difference are the EC/$\beta$ Urca pairs already discussed in \citet{Schatz2014}, which can occur when both the parent and the daughter nucleus of an EC are present, and the $\beta^-$ decay is not fully blocked. 

In the case of the initial $^{56}$Fe ashes, we agree with \citet{Haensel1990} that the composition reaches the neutron drip line with the destruction of $^{56}$Ar at around 
\density{7.8}{11}. However, taking into account the finite time needed for the transition and the change in neutron density during the transition, we find that the reaction flow branches into an EC sequence and a neutron capture sequence unlike \citet{Haensel1990}. This leads to the appearance of more than one species in the composition. Also, unlike \cite{Haensel1990}, we confirm that beyond neutron drip, nuclei are converted rapidly via the superthreshold electron capture cascades (SECs) found in \citet{Gupta2008} into much lighter nuclei. For example, $^{56}$Ar is converted into $^{40}$Mg in a single step so that $^{40}$Mg is already produced at the time of $^{56}$Ar destruction at \density{7.8}{11} and $\mu_e$=31.6 MeV. This is in contrast to  \citet{Haensel1990}, where, after $^{56}$Ar destruction at \density{6.1}{11}, several EC reactions at stepwise increasing $\mu_{\rm e}$ have to occur before $^{40}$Mg is produced at \density{1.1}{12}. In addition, we find that the reaction sequence branches at $^{42}$Si leading to the additional production of $^{48}$Si via neutron captures, resulting in a two component composition of $^{40}$Mg and $^{48}$Si. 

The onset of pycnonuclear fusion also differs. In \citet{Haensel1990}, the first fusion is $^{34}$Ne$+^{34}$Ne, triggered by EC on $^{40}$Mg at \density{1.46}{12} and $\mu_e=34.3$~MeV. We find that, due to our mass model that includes shell effects, the threshold for $^{40}$Mg(EC) is higher so that $^{40}$Mg(EC) occurs deeper at  \density{1.8}{12}. As our calculation allows for the presence of multiple nuclear species, we find that the lighter nuclides produced by the ensuing SEC chains preferably fuse with the still abundant $^{40}$Mg, rather with themselves, leading to the occurrence of fusion between unlike nuclides. Furthermore, the SEC chains on $^{40}$Mg are faster, leading to lighter nuclides before fusion sets in. A branching at $^{28}$O, where EC and fusion competes, leads to the creation of two major species undergoing fusion, resulting in two major fusion reactions, $^{25}$N$+^{40}$Mg$\rightarrow^{65}$K  and $^{28}$O$+^{40}$Mg$\rightarrow^{68}$Ca. In addition, at the larger depth where these reactions are triggered,  $^{40}$Mg$+^{40}$Mg$\rightarrow^{80}$Cr becomes significant as well. In summary, instead of a single fusion reaction between like species, three fusion reactions occur simultaneously, two of them between very different species. 

A fundamental difference here is that we do not find a large abundance buildup of the fusion reaction products as found in \citet{Haensel1990,Haensel2008}. Instead, the reaction product is immediately recycled via an SEC. Fusion reactions therefore lead to fusion-SEC cycles. In the case of $^{40}$Mg, for example, the resulting net reaction of all fusion-SEC cycles is $^{40}$Mg$+^{40}$Mg$\rightarrow ^{40}$Mg+40n. The fusion-SEC cycles slowly convert $^{40}$Mg into neutrons, until the increasing neutron density shifts the composition to more neutron rich nuclei. A single fusion reaction effectively only destroys a single $^{40}$Mg nucleus instead of two, allowing for more fusion reactions at a shallower depth. A fusion induced cycle for the destruction of $^{40}$Mg has been described in \citet{Steiner2012}, who used a Quasi Statistical Equilibrium model that allows for the presence of multiple species. He found a cycle that starts at $^{40}$Mg with a SEC to $^{22}$C, followed by $^{22}$C+$^{22}$C$\rightarrow^{44}$Mg($\gamma$,4n)$^{40}$Mg. However, taking into account the finite speed of the nuclear reactions, we find that $^{40}$Mg and the lighter nuclides produced by a SEC coexist leading to asymmetric fusion reactions. Our model also tracks individual reaction channels and can therefore resolve branchings between competing reactions. This also broadens the range of fusion reactions. 

The reaction sequences starting with broader initial composition distributions are of similar type, characterized by 4 phases - EC chains without neutrons, EC chains with neutron-induced reactions, SECs at neutron drip, and pycnonuclear fusion. As soon as neutrons are released, the evolution in a given isobaric EC chain starts to depend on what happens in other chains. Initial abundances of lighter species such as $^{20}$Ne, $^{24}$Mg, or $^{28}$Si are transformed into even lighter nuclei, which then undergo pycnonuclear fusion prior to neutron drip. This has already been suggested by \citet{Horowitz2008}. We confirm that these reactions occur at a depth around \density{1}{11} but the types of fusion reactions differ significantly from previous predictions. The most dominant fusion reactions around this depth are $^{21}{\rm N}+^{21}$O, $^{21}$N$+^{21}$N, $^{20}$C$+^{20}$O, $^{20}$C$+^{21}$N,  $^{20}$C$+^{20}$C, and $^{24}$O$+^{24}$O, but many weaker reactions occur. Ne fusion sets in at slightly higher \density{4}{11} via $^{34}$Ne$+^{34}$Ne, $^{32}$Ne$+^{34}$Ne, $^{34}$Ne$+^{24}$O, and $^{34}$Ne$+^{20}$C.  At greater depths we find the pycnonuclear fusion-SEC cycles:
\begin{eqnarray*}
\magnesium[40]+\magnesium[40]\rightarrow\chromium[80] \rightarrow\magnesium[40]+40 \mathrm{n}\\
\nitrogen[25]+\magnesium[40]\rightarrow\potassium[65]\rightarrow\magnesium[40]+25\mathrm{n}\\
\oxygen[28]+\magnesium[40]\rightarrow\calcium[68]\rightarrow\magnesium[40]+28\mathrm{n}
\end{eqnarray*}

Here the fusion reaction rates determine how rapidly nuclei are converted into free neutrons.

\subsubsection{Urca Cooling}
Nuclear Urca cooling has been discussed in detail in \citet{Schatz2014,Deibel2016}. With our model temperature of 0.5~GK, we can identify the strongest Urca pairs for the different initial compositions investigated here. For superburst ashes, \Urca{55}{Ti}{55}{Sc} is the only strong Urca cooling pair, owing to the limited mass range of the nuclei in the burst ashes. The strongest Urca pair identified in \citet{Schatz2014} for superburst ashes, \Urca{56}{Ti}{56}{Sc} has been shown to be ineffective as a consequence of newly measured masses and newly calculated transition strengths from shell model calculations \citep{Meisel2015}. 

The situation for the \Urca{55}{Ti}{55}{Sc} pair is less clear. One of the key prerequisites for a strong Urca cycle in an odd $A$ chain is a strong allowed ground state to ground state (or within a few 10 keV of the ground state) EC and $\beta^-$ transition. Experimental studies of the ground state of $^{55}$Ti indicate a spin and parity of 1/2$^-$ \citep{Maierbeck2009}. The ground state of $^{55}$Sc is expected to be 7/2$^-$, based on systematics. Such a large spin difference would preclude a fast ground state to ground state transition. \citet{Crawford2010} therefore assume that the significant missing strength that they observed in a study of the $^{55}$Sc $\beta^-$ decay is not due to a ground state transition, but due to a sizable $\beta$-delayed neutron emission branch. As low lying excited states are not expected in these isotopes, one would have to conclude that, in contrast to the QRPA-fY predictions used here, the \Urca{55}{Ti}{55}{Sc} Urca pair is not effective, and that therefore there no strong Urca cooling pair exists in crusts composed of superburst ashes. Nevertheless, an experimental confirmation of the absence of a strong ground state to ground state transition in the $\beta^-$ decay of $^{55}$Sc, or direct evidence of a ground state to ground state $\beta$-delayed neutron emission branch, would be desirable to clarify whether Urca cooling can play a role in accreting neutron stars with superbursts. 

For ashes from regular X-ray bursts that produce a wider range of nuclei, additional strong nuclear Urca pairs can be populated. For KEPLER X-ray burst ashes, the dominant cooling comes from the 
\Urca{31}{Mg}{31}{Na} pair (Tab.~\ref{TabKEPLERUrca}). This pair had also been identified in \citet{Schatz2014} when using the FRDM mass model (see below). Experimental data 
indeed indicate a strong ground state to ground state transition, although the experimentally derived  $\logft=4.9(2)$ indicates a roughly a factor of 4 slower transition than what is used in our model \citep{Klotz1993,GUILLEMAUD1984}. Another very strong Urca pair for KEPLER X-ray burst ashes is \Urca{33}{Al}{33}{Mg}. The experimentally derived \logft\ value of 5.2 is very close to the QRPA-fY prediction (5.0) \citep{Tripathi2008} and would confirm a very strong Urca pair. However, there is some debate about the experimental interpretation, in particular about the parity of the $^{33}$Al ground state \citep{Yordanov2010}.

For our calculation with the ashes of an extreme X-ray burst, the production of $A>100$ nuclei opens up a number of additional possible Urca cooling pairs (Tab.~\ref{TabXRBUrca}), the strongest of which had already been identified in \citet{Schatz2014}. Indeed, while \Urca{31}{Mg}{31}{Na} is also important, Urca cooling is largely dominated by the $A=103$ and $A=105$ chains. Not much is known experimentally about the relevant nuclei $^{103,105}$Sr, $^{103,105}$Y, $^{103,105}$Zr, and $^{105}$Nb. Data on ground state to ground state $\beta^-$ transitions, and the masses of $^{103,105}$Sr and $^{105}$Y remain to be determined to put the existence and strengths of these Urca cooling pairs on solid experimental footing. 

The strongest Urca cooling pairs identified here are located at depths in the range of \density{3.5}{9} to \density{1.1}{11}. Pairs at shallower depths are considerably weaker, though they may still be important in limiting the strong shallow heating that is indicated by observations of cooling transients and superburst ignition depths \citep{Deibel2016,Meisel2017}. Urca cooling at greater depths is largely precluded by the onset of neutron emission and capture reactions, that tend to deplete odd $A$ mass chains (see below), and prevent the coexistence of parent and daughter nuclides once the drip line has been reached. 

\subsubsection{Appearance of free neutrons} \label{Sec:FreeN}
We find that free neutrons start to play a role long before the composition reaches the neutron drip line, the traditional point where free neutrons appear. This early release of neutrons stems from EC reactions that populate neutron unbound excited states ($E_x > S_n$), which then decay by neutron emission. There are two basic mechanisms for EC to populate high lying excited states. First, the EC threshold of a particular reaction may be increased by the excitation energy of the lowest lying daughter state for an allowed transition. If this state is above the neutron separation energy, neutron emission will occur. Also, once the transition proceeds at threshold, $\mu_e$ can be higher than the threshold of the subsequent EC reaction, leading again to the population of excited daughter states that may be above the neutron separation energy. These effects can occur in even and odd $A$ chains. An example is $^{88}$Rb(EC,n)$^{87}$Kr discussed in section \ref{Sec:XRB}, where the lowest lying EC transition is predicted to go to a 6.9 MeV state in $^{88}$Kr, close to \sn{88}{Kr}{7.02}. This leads to neutron emission relatively close to stability. It will be important to explore how lower lying forbidden transitions not included in the QRPA-fY calculations may reduce this effect.

The second mechanism to release neutrons prior to reaching the neutron drip line is the odd-even staggering of $Q_{\rm EC}$ in even $A$ chains, 
$\Delta Q_{\rm EC} = | Q_{\rm EC, even-even} - Q_{\rm EC, odd-odd} | $. In these chains, an EC reaction on an even-even nucleus is immediately followed by an EC reaction on an odd-odd nucleus \citep{Haensel1990}, where excited states up to $\Delta Q_{\rm EC}$ can be populated \citep{Gupta2007}. Neutron emission is possible if $\Delta Q_{\rm EC} > S_{\rm n}$.
 $\Delta Q_{\rm EC}$ depends strongly on the mass model \citep{Meisel2015}. For typical values of 3~MeV neutron release would only start closer to the drip line at $S_{\rm n} \approx 3$~MeV. However, the FRDM mass model predicts significantly larger $\Delta Q_{\rm EC}$ in some cases.
 
\begin{figure}[ht]
\plotone{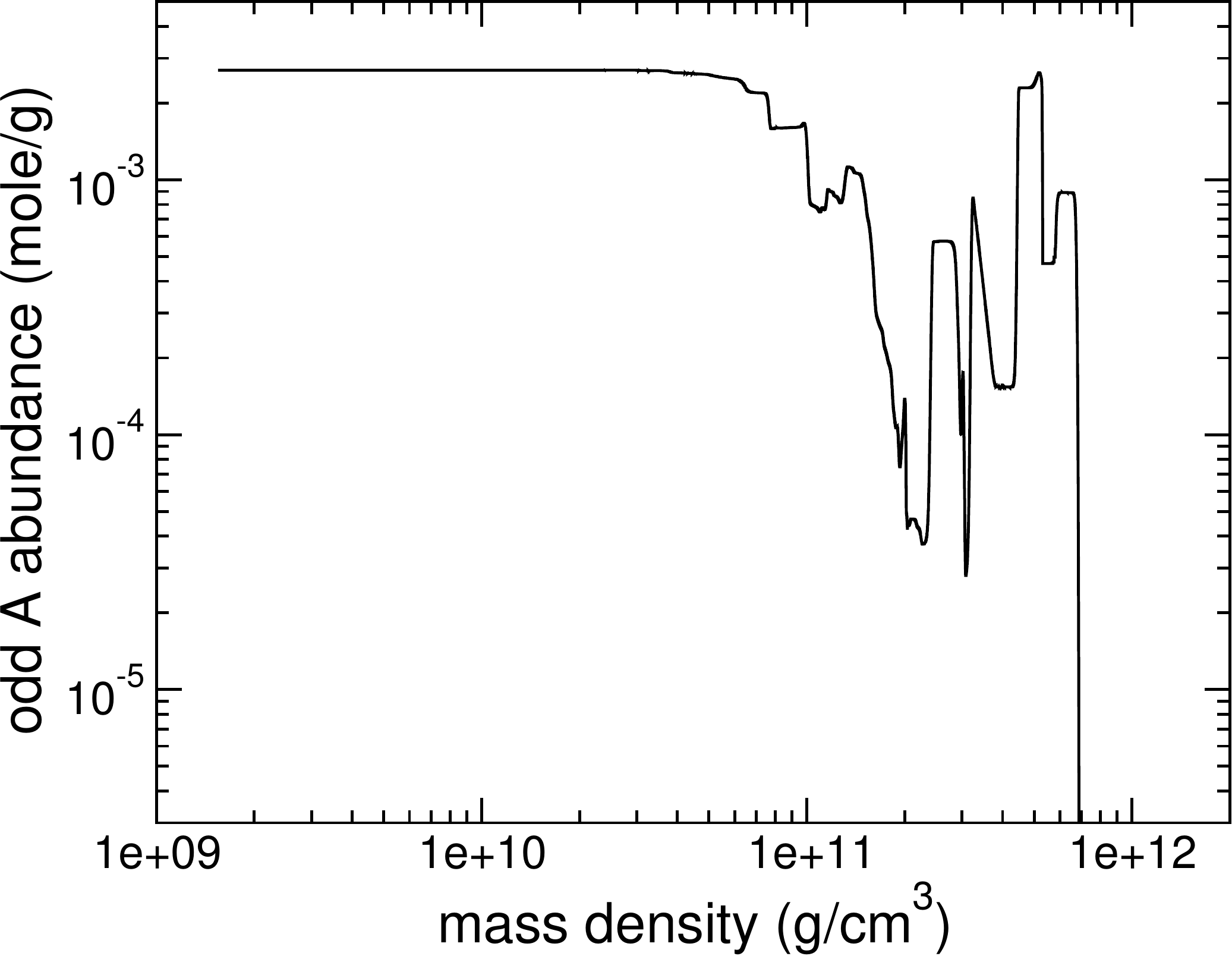}
\caption{Summed nuclear abundance in odd $A$ mass chains as a function of density for extreme burst ashes. 
\label{FigOddA}}
\end{figure}

The early release of neutrons does not lead to a buildup of a large free neutron abundance. Instead, the released neutrons are recaptured by other nuclei present at the same depth. This is a feature of the multi-component composition of the outer crust. Nuclei with the largest abundance and largest neutron capture cross sections will dominate the neutron absorption. Interestingly this tends to lead to the depletion of odd $A$ chains, starting as early as at \density{4}{10} (see Fig.~\ref{FigOddA}). As the odd $A$ mass chains tend to have most of the nuclear Urca pairs, the early release of neutrons strongly limits Urca cooling in the deeper regions of the outer crust. 

\begin{figure}[ht]
\plotone{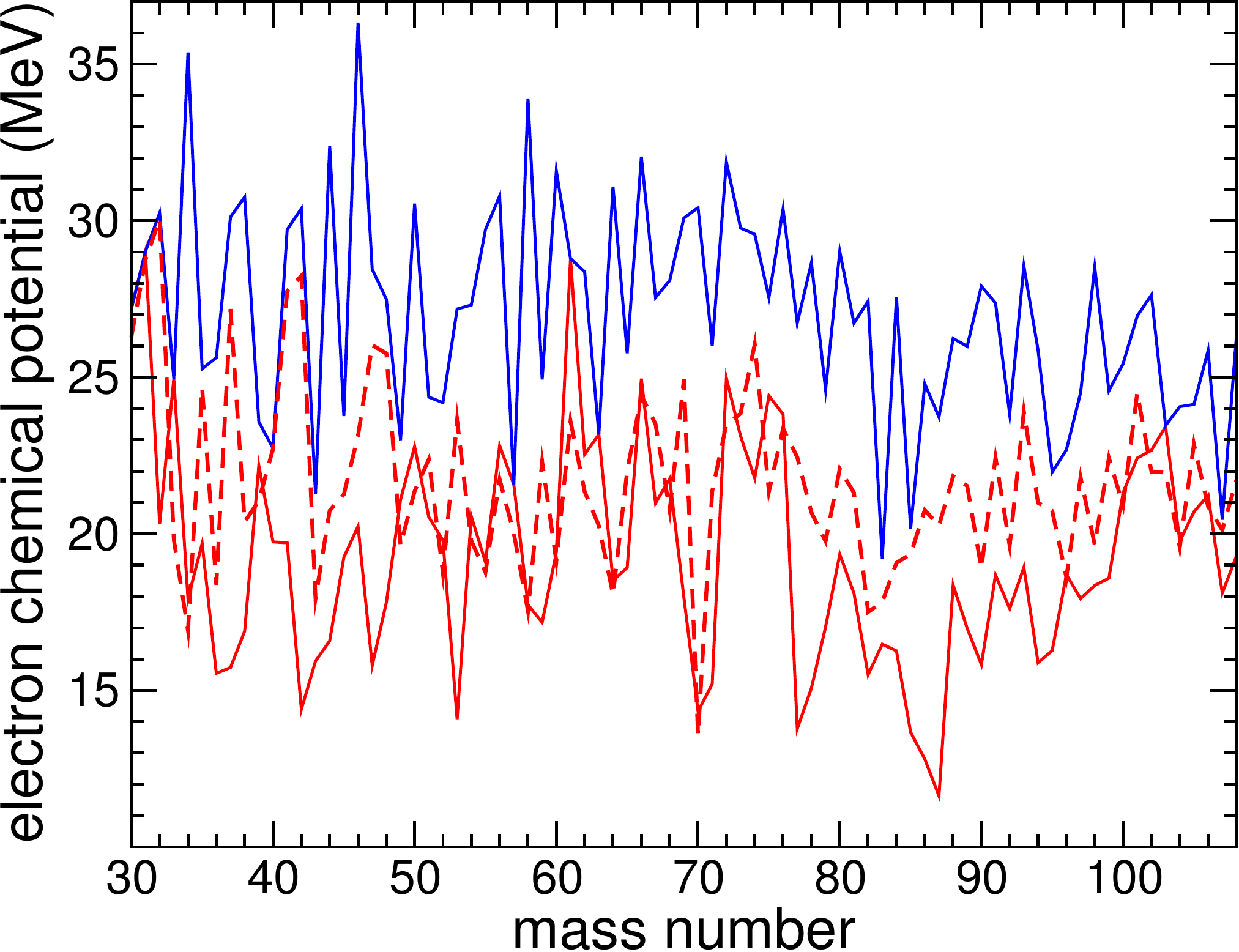}
\caption{Estimated minimum $\mu_{\rm e}$ for neutron release following an electron capture for each mass chain as a function of mass number. Shown are estimates obtained when taking into account all transitions into excited states (red, solid), estimates obtained when using only ground state to ground state EC thresholds but taking into account transitions to excited states for a subsequent transition in even mass chains  (red, dashed), and estimates obtained when neglecting transitions into excited states entirely limiting neutron release to reaching the neutron drip line (blue, solid). The estimates are solely based on nuclear mass differences and strength functions. Lattice energy and finite temperature corrections are neglected. 
\label{FigNeutronRelease}}
\end{figure}

The $\mu_{\rm e}$ required for pre-drip line neutron release varies greatly from mass chain to mass chain. To illustrate this point, we provide a simple estimate for the minimum $\mu_{\rm e}$ for neutron release in each mass chain, based on nuclear mass differences ($\mu_{\rm e} > |Q_{\rm EC}|+S_{\rm n}$) (Fig.~\ref{FigNeutronRelease}). EC transitions are assumed to proceed when $\mu_{\rm e} > |Q_{\rm EC}| + E_{\rm x0}$, with $E_{\rm x0}$ being the daughter excitation energy of the lowest lying EC transition.  This simple estimate neglects lattice energy and finite temperature corrections, which depend on overall composition and astrophysical parameters and are included in the full network calculation. Clearly, the depth of early neutron release depends strongly on the mass chain and therefore on the initial composition created by thermonuclear burning on the neutron star surface. While transitions into excited states move the release of neutrons to shallower depths, on average by 6~MeV in $\mu_{\rm e}$ (red solid line in Fig.~\ref{FigNeutronRelease}), the odd-even staggering of $Q_{\rm EC}$ alone leads to significant neutron release (red dashed line in Fig.~\ref{FigNeutronRelease}) prior to reaching the neutron drip line (blue line in Fig.~\ref{FigNeutronRelease}). All curves in Fig.~\ref{FigNeutronRelease} show a pronounced variation in $\mu_{\rm e}$ from mass chain to mass chain of up to about 10~MeV. Therefore, regardless of the detailed transition energies and odd-even staggering, there will be a transition region between the outer and inner crust where some mass chains release neutrons, and others capture them. The characteristics of the compositional evolution in this region will depend sensitively on the composition of the X-ray burst ashes. 

\subsubsection{Superthreshold Electron Capture Cascades}
In agreement with \citet{Gupta2008} we find that EC and neutron emission sequences at the neutron drip line proceed not in single steps but in a rapid sequence spanning many isotopic chains. Once the neutron drip line is reached, the composition is therefore rapidly converted into lower $Z$ nuclei. We also find that similar rapid sequences of EC and neutron emission drive the products of pycnonuclear fusion instantly back to the originating nucleus, leading to pycnonuclear fusion-SEC cycles (section \ref{Sec:Sequence}).

In the SEC mechanism, EC with neutron emission drives the composition away from the neutron drip line towards lower EC thresholds. EC reactions can then become faster than neutron capture reactions and another EC reaction follows immediately, before neutron capture can restore (n,$\gamma$)-($\gamma$,n) equilibrium in the isotopic chain. If the subsequent EC reaction again leads to neutron emission, the sequence can repeat many times, greatly accelerating the conversion of heavier elements into lighter ones. 

An example is the SEC sequence shown in Fig.~\ref{FigFluxFe56Step3} for the initial $^{56}$Fe composition. For the neutron density and temperature at the location shown, the dominant (n,$\gamma$)-($\gamma$,n) equilibrium abundance in the $Z=14,17,18$ isotopic chains would be $N=34, 44$, and 44, respectively. Yet, EC occurs significantly closer to stability at $N=28$, 34--36, and  38--42, respectively. Clearly the reaction sequence is off equilibrium and the competition of EC and neutron capture rates, as well as the number of neutrons emitted following an EC are important. On the other hand, in the $Z=15$, and 16 isotopic chains, neutron captures drive the composition back to equilibrium following the EC-induced neutron emission.  Nevertheless the cascade continues, as the EC thresholds of the equilibrium nuclei are lower than the current $\mu_{\rm e}$. Fig.~\ref{FigFluxXRBStep4} shows a similar mix of EC reactions off equilibrium (immediately following a preceding EC reaction) and in equilibrium (EC reactions followed by neutron capture). The reaction sequences driving the composition to lower $Z$ can therefore be quite complex and depend on $Q_{\rm EC}$, $S_{\rm n}$, the feeding of states above $S_{\rm n}$ in EC transitions, and neutron capture rates. 

\subsubsection{Shell effects and the composition beyond neutron drip}
The FRDM mass model includes shell effects that have a significant impact on the composition in the inner crust. The term "shell effect" refers here to the occurrence of large energy gaps in neutron or proton single particle levels in some nuclei. We emphasize that this is not limited to closed shell nuclei, where the term ``shell" refers to the highly degenerate set of levels that occur only in nuclei with spherical shape \citep{Mayer1949,Mayer1950a,Mayer1950b}. Rather, this also includes gaps in the single particle levels that occur in deformed nuclei where the level degeneracy is lifted and shells in the traditional sense therefore do not exist \citep{Nilsson1955,Mottelson1959}. In fact, the vast majority of the nuclei of relevance to this study are deformed. 

These energy gaps give rise to additional nuclear binding and thus can, as we show, prevent the formation of a single species composition at neutron drip. Instead, depending on the initial composition,  abundance peaks form where the $N=28, 50$, and 82 shell effects coincide with the neutron drip line. This can have a strong impact on thermal conductivity in the inner crust (section \ref{Sec:Imp}). We find that initial $A \le 56$ nuclei (but not $A=28$) end up feeding $N \approx 28$, initial $60 < A \le 105$ and $A=28$ nuclei feed $N=50$, and initial $A \ge 106$ nuclei feed mostly $N=82$. $A=56$--60 and $A=$102--106 are borderline areas, where branchings lead to feeding of two different final mass regions. The exact split depends on neutron abundance and competing reaction rates. 

It is therefore important to understand the composition of the X-ray burst ashes, in particular the amount of  $^{28}$Si and $A \ge 102$ nuclei.  \citet{Cyburt2016} show that $^{28}$Si synthesis in X-ray bursts depends on helium burning reactions such as $^{12}$C($\alpha$,$\gamma$) and $^{24}$Mg($\alpha$,$\gamma$). They also show that the mass fraction of $^{28}$Si in the burst ashes can exceed 10\%, depending on the value of the uncertain $^{15}$O($\alpha$,$\gamma$) reaction rate. 

The amount of $A \ge 102$ nuclei produced in X-ray bursts is still an open question. \citet{Schatz2001} used a one zone model with ignition conditions assuming low accreted metallicity ($Z=10^{-3}$) and a relatively high accretion rate ($\dot{m}=0.3 \dot{m}_{\rm Edd}$ with Eddington accretion rate $\dot{m}_{\rm Edd}=8.8 \times 10^4$~g/cm$^2$/s ) to explore the maximum possible extent of an rp-process. They indeed find significant production of $A \ge 102$ nuclei in bursts that exhibit long $\approx$200~s tails. \citet{Woosley2004} confirmed this result with a multi-zone X-ray burst model. Their model zM assumes similar system parameters and their first burst indeed produces more than 30\% mass fraction of $A=104$ with a light curve extending about 200-250 s before cooling exponentially. However, they also find that subsequent bursts are influenced by the ashes from previous bursts, resulting in a more moderate rp-process that produces only negligible amounts of $A \ge 102$ material and a more rapidly cooling light curve.  In contrast \citet{Jose2010} use a different model but similar system parameters and find that while bursts after the first burst become somewhat shorter, lasting about 200 s, they still do produce large amounts ($> 10$\% mass fraction) $A \ge 102$ material.

Another important question is whether the shell effects for nuclei near the neutron drip line predicted by the FRDM exist. In particular, the production of $N=82$ nuclei is strongly facilitated by the interplay of predicted masses and spherical shell closure induced shape changes of neutron rich nuclei around $Z=38$ and $N=70$--82. An increase in $S_{\rm n}$ with increasing neutron number in this mass region leads to a jump of the (n,$\gamma$)-($\gamma$,n) equilibrium nucleus to $N=82$. This is the same effect that leads to an underproduction of nuclei below $A=130$ in the rapid neutron capture process \citep{Kratz1993}. 

It has been pointed out that calculations based on some self-consistent Hartree-Fock Bogolubov and Relativistic Mean Field models predict a weakening of the spherical shell gaps far from stability (see, for example, \citet{Sorlin2008,Afanasjev2015} or \citet{Chen1995} in the context of the r-process). Experimental evidence indeed indicates that the $N=28$ spherical shell closure disappears with decreasing $Z$ because strong deformation sets in already at sulfur ($Z=16$) and silicon ($Z=14$) isotopes \citep{GLASMACHER1997,Bastin2007,Meisel2015Ar}. $^{40}$Mg, which plays a critical role in our model, has been discovered experimentally \citep{Baumann2007}. 
First nuclear structure studies confirm the presence of deformation \citep{Crawford2014}. However, this does not necessarily mean that shell effects as defined in this work do not occur. Indeed, the FRDM mass model predicts strong deformation of $^{40}$Mg in agreement with experiment, but nevertheless also predicts increased binding because of a large deformed N=28 single particle energy level gap for the predicted oblate deformation. Mass measurements of $^{40,41,42}$Mg that will be come possible at future rare isotope facilities will be needed to confirm the predicted trends in neutron separation energy. 

For the relevant $N=50$ and $N=82$ nuclei $^{70}$Ca ($N=50$) and $^{116}$Se ($N=82$) the FRDM predicts spherical shell closures. However, these nuclei are currently out of experimental reach and neither has been observed in laboratory experiments. The most proton deficient $N=50$ nucleus studied so far is $^{78}$Ni. Measurements of $\beta$-decay half-lives of $^{78}$Ni and nearby isotopes indicate strong spherical shell closures at $Z=28$ and $N=50$ \citep{Xu2014}. For  $N=82$, recent studies of $^{128}$Pd indicate a robust spherical shell closure for $Z=46$ \citep{Watanabe13}. This is in contrast to earlier work that provided evidence for a weakening of the spherical shell gap already at $^{130}$Cd \citep{Dillmann2003}. Shell model calculations  \citep{Taprogge2014} and covariant density functional theory \citep{Afanasjev2015} predict a gradual weakening of the $N=82$ spherical shell gap towards $Z=40$, though the gap is predicted to remain significant. This remains to
be confirmed experimentally. 

\subsection{Heating and Cooling}

Heating and cooling by nuclear reactions in the crust links the nuclear processes identified in this work with observables. Fig.~\ref{FigEnergy} shows that the considerably different nuclear processes obtained with a full reaction network and a realistic initial multi-component composition lead to differences in the heating and cooling of the neutron star crust compared to simplified single component equilibrium calculations \citep{Haensel2008}. In particular, for all types of realistic burst ashes, Urca cooling is significant at the 0.5~GK temperature investigated here, and would likely lead to a cooler crust in a self-consistent model. As expected, the location and strength of Urca cooling depends sensitively on the initial composition. 

 \begin{figure}[ht]
\plotone{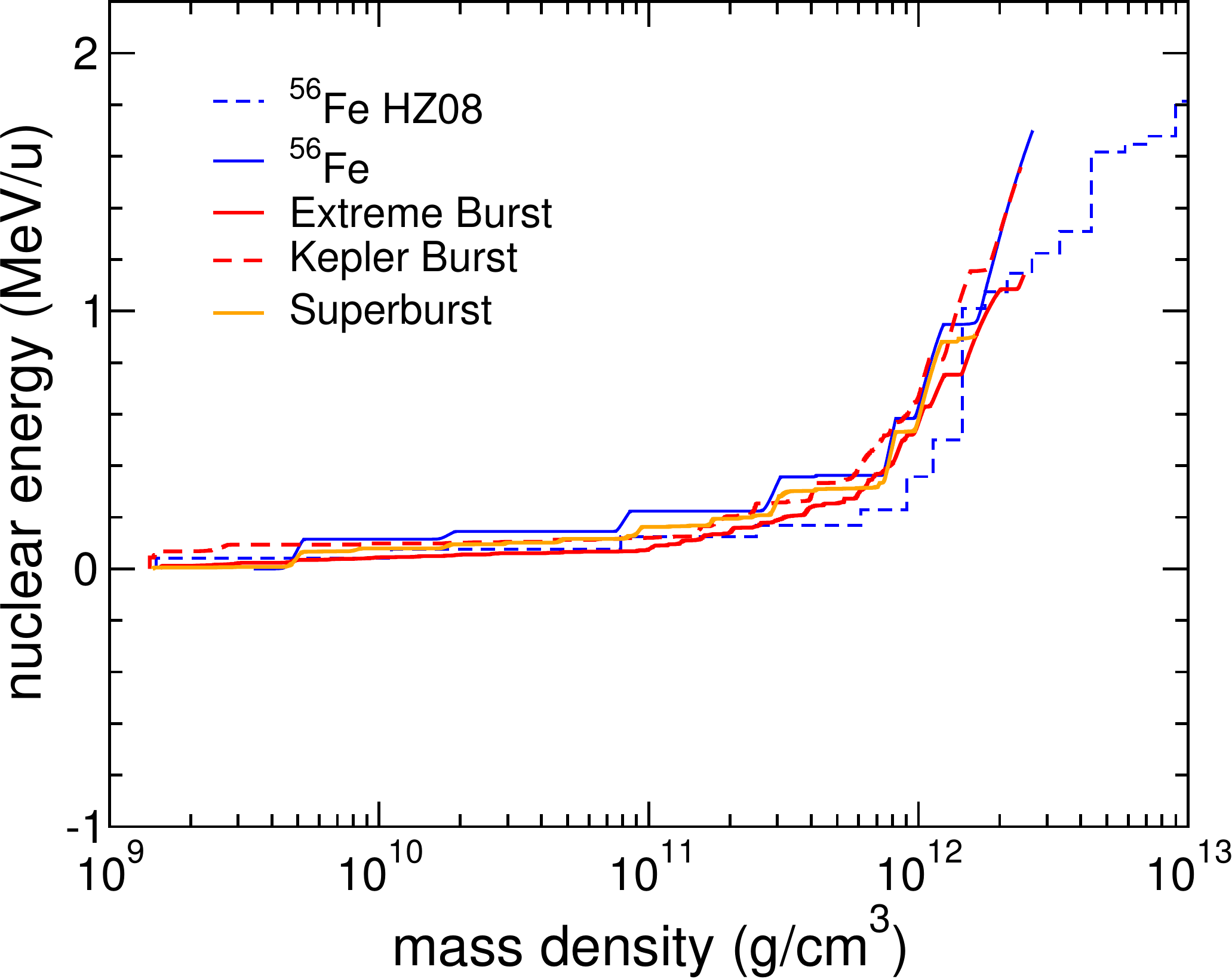}
\caption{Integrated nuclear energy release during episodes with heat deposition as a function of mass density for pure $^{56}$Fe ashes (solid blue), extreme burst ashes (solid red), KEPLER burst ashes (dashed red), and superburst ashes (solid orange). Any heating during cooling episodes is not included. The nuclear energy release obtained by \citet{Haensel2008} for pure $^{56}$Fe ashes is shown for comparison (dashed blue).
\label{FigHeat}}
\end{figure}

There are also significant differences in heating between the models. This is shown in Fig.~\ref{FigHeat}, where we only integrate over segments of positive slope in Fig.~\ref{FigEnergy}. This provides a lower limit of the heating, as we neglect any heating during a cooling episode. On one hand, all our calculations with a full reaction network show significantly more heating at shallower depths than the \citet{Haensel2008} (HZ08) estimate. At around \density{1.3}{12}, the integrated difference has accumulated to about $0.5\,\MeVu$ (though this is a lower limit). This is in part due to our inclusion of transitions into excited states in the first step of the two step electron capture sequences in even $A$ chains. These transitions not only reduce neutrino emission (as considered in HZ08), but also increase the electron capture energy thresholds and thus the total energy release in the sequence \citep{Gupta2007}. On the other hand, the calculations with realistic burst ashes and a full reaction network show remarkable similarity, despite the significant variations in initial compositions. Differences in the thermal structure for different realistic burst ashes will therefore predominantly arise from differences in Urca cooling, not from differences in heating. 

The much shallower onset of fusion reactions in the models with realistic ashes, around \density{1.2-7.7}{11}, compared to \density{1.1}{12} for pure $^{56}$Fe ashes, contributes to an increased heat deposition at shallower depths. This is due to two effects. First, lighter nuclei in the initial composition tend to be converted more rapidly into the low $Z$ species that can undergo fusion reactions. Second, the superthreshold electron capture cascade (SEC) effect creates lighter nuclei earlier.  \citet{Horowitz2008} pointed out the potential importance of fusion of lighter nuclei at shallower depths. Indeed such reactions can deposit of the order of $\epsilon=0.7\textrm{--}0.9\,\MeVu$ of heat \citep{Horowitz2008}, provided they would make up 100\% of the composition. However, the mass fraction $X$ of $A \le 28$ nuclei in the initial composition is only 0.7\%, 5\%, and 1\% for superburst, KEPLER, and extreme burst ashes, respectively. The associated heating $\epsilon X$ is therefore rather small, $0.005\textrm{--}0.05\,\MeVu$, and comparable to electron capture heating in the more abundant mass chains.  

Despite these differences in the distribution of heat deposition, the total heat deposited is remarkably similar for all our models, at least down to a depth where \density{1.6}{12}. At that depth, total heat deposition is $1.1\,\MeVu$, $0.96\,\MeVu$, $0.88\,\MeVu$, $1.2\,\MeVu$ and $0.9\,\MeVu$ for HZ08, pure Fe ashes, extreme burst ashes, KEPLER burst ashes, and superburst ashes, respectively. Note however, that the latter three cases are lower limits, as some heat may be released in regions with net cooling. Our results confirm with a full reaction network the robustness of heating (but not Urca cooling) in respect to initial composition found in previous work using simplified approaches \citep{Haensel2008} or reaction networks without pycnonuclear fusion \citep{Gupta2008}.

\subsection{Impurity}
\label{Sec:Imp}

\begin{figure}[ht]
\plotone{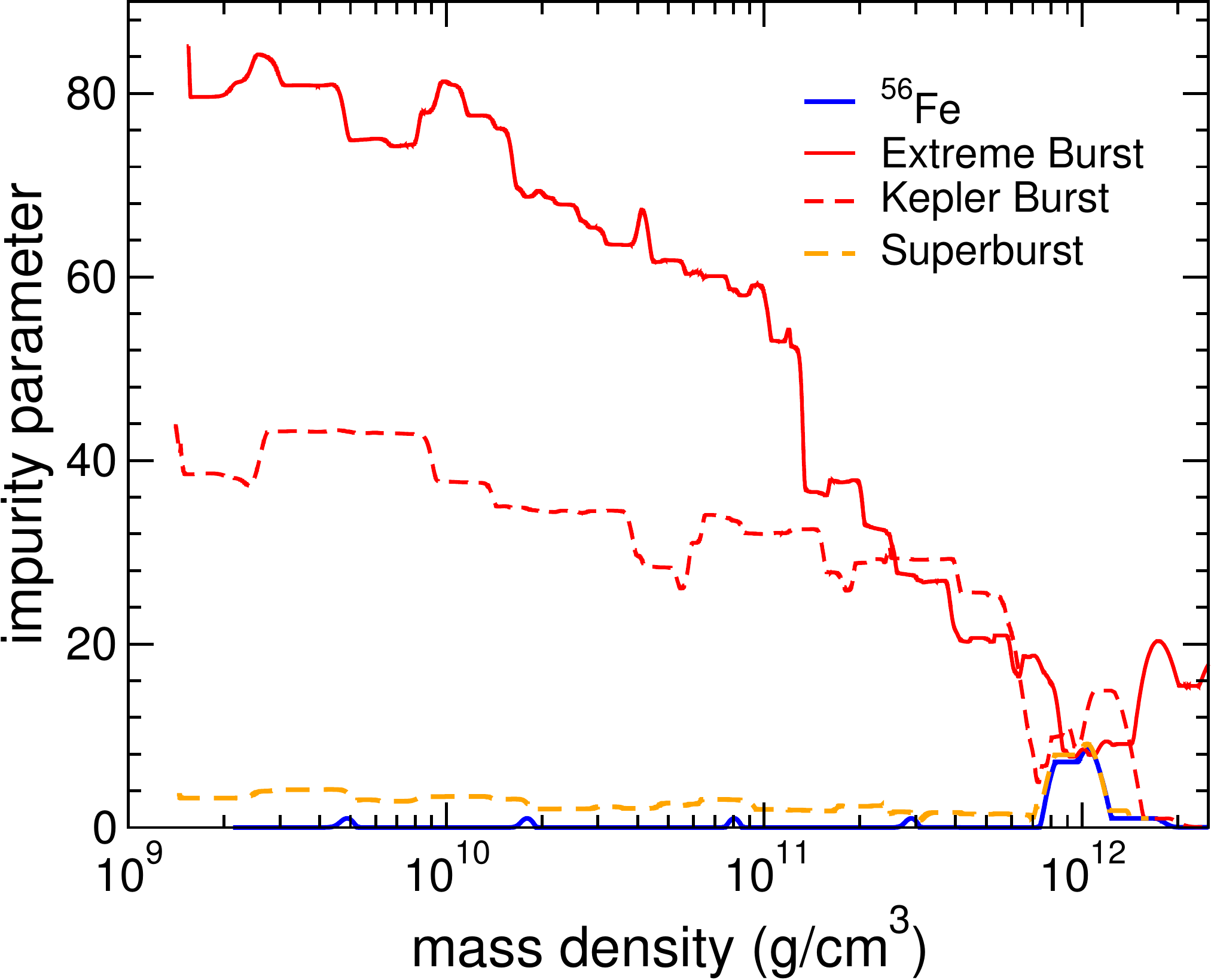}
\caption{Impurity parameter $Q_{\rm imp}$ as a function of mass density for pure $^{56}$Fe ashes (solid blue), extreme burst ashes (solid red), KEPLER burst ashes (dashed red), and superburst ashes (dashed orange).
\label{FigImpurity}}
\end{figure}

We are now in the position to predict the impurity parameter $Q_{\rm imp}=\sum_i Y_i (Z_i - \langle Z \rangle )^2 / \sum_i Y_i$ with average charge number $\langle Z \rangle$ and abundances $Y_i$ (excluding neutrons) as a function of depth. $Q_{\rm imp}$ is important as it determines the thermal conductivity of the crust due to electron impurity scattering. Fig.~\ref{FigImpurity} shows impurity parameters for the various initial compositions as a function of density. The extreme X-ray burst ashes exhibit the broadest range of isotopes and has the largest $Q_{\rm imp} \approx 80$. The rp-process in the more typical KEPLER burst produces much fewer $Z=30-46$ nuclei resulting in a lower initial $Q_{\rm imp} \approx 40$. Superbursts drive the composition into nuclear statistical equilibrium, resulting in much less diverse ashes with a much smaller initial $Q_{\rm imp} \approx 4$. Down to a depth where \density{1}{10}, $Q_{\rm imp}$ stays rather constant. At greater depth it begins to decrease substantially because heavier nuclei tend to electron capture more, reducing their $Z$ faster, and because the early release of neutrons starts to eliminate abundance in some mass chains. At \density{1}{11}, the extreme burst ashes shows a drastic reduction in $Q_{\rm imp}$ bringing it in line with the KEPLER ashes. This is due to the pycnonuclear fusion of oxygen produced via electron capture from the relatively large initial $^{20}$Ne abundance. In addition, compared to the KEPLER ashes, the extreme burst ashes has relatively smaller initial abundances of $^{24}$Mg and $^{28}$Si, causing a much larger impact on $Q_{\rm imp}$ once lighter nuclei from the abundance $^{20}$Ne start fusing. Between \density{2}{11} and \density{7}{11}, light element fusion and SEC chains lead to a steady reduction in $Q_{\rm imp}$ for all cases. 

Interestingly, all initial compositions converge to a comparable $Q_{\rm imp}=$7-11 between \density{8}{11} and \density{1.3}{12} due to shell effects that lock abundance in different locations. This includes even the pure $^{56}$Fe ashes, which turns into a multi-component composition beyond neutron drip due to the splitting of the reaction path discussed in section \ref{Sec:Fe56}. However, beyond \density{1.5}{12} material trapped at the $N=50$ spherical shell closure is destroyed and all compositions but the extreme burst ashes converge to a single nucleus and  $Q_{\rm imp}$ drops to less than one. This is in line with previous predictions \citep{Jones2005,Gupta2008,Steiner2012} that $Q_{\rm imp}$ is reduced to a value near one when transitioning from the outer to the inner crust, though we find that the transition is gradual and exhibits some variations. The exception is the extreme burst ashes, the only case where material is also locked in at the $N=82$ spherical shell closure due to the heavy $A \ge 106$ nuclei contained in the ashes. In this case, the conversion of $N=50$ nuclei into lighter species  together with the unchanged heavy $N=82$ nuclei, leads to the opposite behavior, an increase of $Q_{\rm imp}$ in the inner crust to values of around 20. 

Our theoretical predictions of $Q_{\rm imp}$ can be compared with constraints extracted from observed cooling curves of transiently accreting neutron stars using crust cooling models. For KS1731-260, the most recent analysis by \citet{Merritt2016} obtains $Q_{\rm imp}=4.4^{+2.2}_{-0.5}$, in agreement with earlier results from \citet{Brown2009} ($<4$). For MXB1659-29, \citet{Turlione2015} find $Q_{\rm imp}=$3.3--4 in agreement with earlier results from \citet{Brown2009}. These results are also in line with work by \citet{Page2013} who use models with different $Q_{\rm imp}$ values for the outer and the inner crust and find $Q_{\rm imp}=5$ and 3 for KS1731-260, $Q_{\rm imp}=10$ and 3 for MXB1659-29, and $Q_{\rm imp}=20$ and 4 for XTE J1701-462 for the outer and inner crust, respectively. A significantly higher $Q_{\rm imp}=40$ has been found in an analysis of EXO 0748-676 \citep{Degenaar2014}. \citet{Turlione2015} obtain a good fit of the data from this source with $Q_{\rm imp} \approx 1$ but do not include the most recent data points considered in \citet{Degenaar2014}. 

\citet{Roggero2016} performed Path Integral Monte Carlo calculations of the electron-ion scattering with impurities at \density{1}{10} and found that heat conductivity is reduced by about a factor of 2--4 compared to the simple approximation employed in current crust cooling models. If this result is indeed broadly applicable at higher densities, it would imply that the required impurity parameters to fit observations are reduced by about a factor of 2--4, and that thus an $Q_{\rm imp} \approx 1-2$ is needed to explain observations of KS1731-260, MXB1659-29, and XTE J1701-462. 

Most crust cooling models used to analyze observational data employ a single $Q_{\rm imp}$ for the entire crust. These $Q_{\rm imp}$ values can be compared to our predictions at $\rho > 10^{12}$~g/cm$^3$ where electron-impurity scattering is expected to dominate heat transport \citep{Brown2009}. Only \citet{Page2013} provide values for $Q_{\rm imp}$ in the outer crust. However, a comparison is difficult as we predict significant changes in $Q_{\rm imp}$ as a function of density, and as it is unclear how constraining the observational data are for the best fit $Q_{\rm imp}$ values given. 

The observational constraints of a small inner crust $Q_{\rm imp}$ in most sources overall agree with our predictions for KEPLER burst and superburst ashes, but clearly disagree with our prediction for extreme rp-process ashes. Our finding of a high $Q_{\rm imp}=20$ for an initial composition from extreme rp-process ashes could in principle explain the high $Q_{\rm imp}=40$ inferred for EXO 0748-676 \citep{Degenaar2014}. An extreme rp-process would require particularly long bursts with burst durations of the order of 200~s. Indeed EXO 0748-676 does exhibit such bursts \citep{Boirin2007} during its outburst phase. 

It has been suggested, that chemical separation effects during the freezing of the crust at the ocean crust boundary, which are not included in our model, lead to a reduction in $Q_{\rm imp}$ in the outer crust \citep{Horowitz2007,Horowitz2009,Medin2011,Medin2014,Mcinven2016}. This is due to the different freeze-out properties of light and heavy nuclei. However, in the absence of nuclear reactions, once steady state is achieved, the solid crust composition must match on average the composition of the ashes entering the crust. Either chemical separation effects lead to time dependent compositional changes in the ocean that counteract the chemical separation once steady state is achieved, or alternating layers of light and heavy nuclei form. In the latter case, $Q_{\rm imp}$ would indeed be significantly reduced, as each layer would have a higher purity than the average crust. However, if such layers form, what their thickness is, and what the impact of layer boundaries are remain open questions. 

\section{Conclusions}

We identify the typical nuclear reaction sequences in the crust of accreting neutron stars down to a depth where \density{2}{12} using a full reaction network for a range of initial compositions. Significant differences are found from calculations using the single nuclear species approximation or equilibrium considerations, including Urca cooling, already reported in
\citet{Schatz2014}; compositional changes in the outer crust due to early release and recapture of neutrons; the superthreshold electron capture cascade (SEC) mechanism found in \citet{Gupta2008} but proceeding sometimes in steps; branchings of the reaction path that lead to a more diverse composition; a much broader range of pycnonuclear fusion reactions including reactions among unlike species; and the formation of fusion-SEC cycles formed by fusion reactions and SECs. Because of the formation of fusion-SEC cycles, fusion does not lead to the build up of heavier nuclei. The "layer cake" structure of regions with high and low $Z$ that might lead to alternating liquid and solid layers \citep{Brown2000} therefore does not exist. 

The location of neutron drip and the buildup of a free neutron abundance depend on the initial composition. Traces of free neutrons (abundance $>10^{-18}$) begin to occur in the \density{3\textup{--}5}{11} range and continue to increase with depth.  Significant ($>1$~\%) free neutron abundances are established between \density{6}{11} and \density{1}{12}. \citet{Jones2005} pointed out that, with the occurrence of free neutrons at neutron drip, neutron captures and $\beta^-$ decay sequences in principle open up a possible pathway towards much heavier equilibrium crust composition. We find that based on our current understanding of nuclear physics such a reaction sequence does not occur. Rather, the composition remains out of equilibrium and evolves towards lighter nuclei, in line with findings from simplified approaches \citep{Haensel1990}.  

While the total nuclear heating to the depth explored in our work is similar to previous simpler estimates, and rather independent of initial composition, the deposited heat distribution can differ substantially. A significant shallow heat source is required to explain observations of cooling transients 
\citep{Brown2009,Degenaar2011,Page2013,Degenaar2013,Degenaar2015,Deibel2015,Turlione2015,Waterhouse2016,Merritt2016}. An open question was, whether at least in some cases this shallow heating could be of nuclear origin. We find that this is not the case, even when taking into account the fusion of $A \le 28$ nuclei in the initial composition as suggested by \citet{Horowitz2008}. With realistic burst ashes containing only a few \% of $A \le 28$ nuclei, fusion near \density{1}{11} contributes at most $0.05\,\MeVu$, far short of the $1-10\,\MeVu$ needed to explain observations. It would be interesting to explore whether enhanced $A \le 28$ production in hydrogen/helium burning is possible, for example when taking into account nuclear uncertainties or new burning regimes such as mixed stable and unstable burning \citep{Narayan2003,Keek2016}. While even such an enhancement is unlikely to explain shallow heating in all sources, it may have a significant impact on the constraints inferred from observations. 

Our calculations confirm previous predictions that large initial crust impurity is reduced when transitioning from the outer crust (shallower than neutron drip) to the inner (deeper than neutron drip) crust \citep{Jones2005,Gupta2008,Steiner2012} and show that this is a robust feature of accreted crusts. This reduction in $Q_{\rm imp}$ explains the low crust impurity in the inner crust inferred from observations of a broad range of systems. We follow this transition for the first time and find that it is a gradual process, starting prior to neutron drip at \density{1}{10}, greatly accelerating around \density{1}{11}, and being completed around \density{1}{12}. We also find that shell effects in very neutron-rich nuclei, as predicted by the FRDM mass model, or any other mass surface anomalies that raise electron capture thresholds locally, can still lead to non-negligible impurities in the inner crust. We find that regardless of initial composition, even for an initial single species composition, shell effects lead to a layer in the inner crust between \density{8}{11} and \density{1.3}{12} where $Q_{\rm imp}=7$--11, which following the arguments of \citet{Roggero2016} would correspond to  an effective $Q^*_{\rm imp} \approx 30$ in current cooling models. It would be interesting to explore the significance of such a layer on crust cooling. Because the feature is relatively robust, an observational signature would open the possibility of  constraining shell effects and nuclear masses of very neutron rich nuclei. 

We find $Q_{\rm imp}=20$ in the inner crust if the initial composition contains significant amounts of $A \ge 102$ nuclei (for example, our extreme rp-process ashes). Such a high $Q_{\rm imp}$ is incompatible with observational constraints for 
KS1731-260, MXB1659-29, and XTE J1701-462. Either shell effects vanish for extremely neutron rich nuclei, or these systems do not contain $A \ge 102$ nuclei in the outer crust. In general, shell effects are not expected to disappear on approaching the neutron drip line though. For example, while covariant density functional theory predicts the disappearance of the $N=50$ shell gap at the neutron drip line, it also 
predicts the emergence of a new $N=40$ shell gap and the persistence of a significant, though somewhat weakened, $N=82$ shell gap \citep{Afanasjev2015}. In addition, as the example of $^{40}$Mg in the FRDM shows, deformed nuclei away from the traditional spherical shell gaps can still exhibit "shell effects" in form of single particle energy gaps. 
For ashes without $A \ge 102$ nuclei we find a pure crust with $Q_{\rm imp}=0$ beyond \density{2}{12}.

The high $Q_{\rm imp}$ found for extreme rp-process ashes is of comparable order of magnitude to the analysis of \citet{Degenaar2014} for EXO 0748-676, who find $Q_{\rm imp}=40$. Indeed, 
EXO 0748-676 does exhibit during outburst, at least occasionally, long ($\approx 200$~s) mixed H/He bursts \citep{Boirin2007}  that are expected to produce $A \ge 106$ nuclei \citep{Schatz2001,Jose2010}. Interestingly,  EXO 0748-676 also exhibits a large number of so called double and even triple bursts, where a regular burst is followed by one or more weaker secondary bursts with very short recurrence times of the order of 10~min \citep{Keek2010}. The origin of these double and triple bursts is not fully understood \citep{Keek2010}. The secondary bursts must be powered by fuel left over from the previous burst, as the recurrence time is too short to accrete fresh fuel. If the secondary bursts occur in the ashes of the first burst, the resulting pulsed rp-process could enhance the production of heavy elements \citep{Hencheck1995} and even overcome the Sn-Sb-Te cycle that limits the rp-process to $A \le 108$ \citep{Schatz2001}. In connection with shell effects for neutron rich $N=82$ nuclei, this could be an intriguing explanation of the slow cooling observed in EXO 0748-676. Indeed \citet{Keek2017} found in a 1D burst model that opacity driven convection can mix hydrogen fuel into ashes produced by previous bursts and lead to secondary bursts that may explain short recurrence time bursts. However, they do not find an enhanced heavy element production. More work on burst models and burst nuclear physics is needed to clarify this question, as well as the contradictory predictions of $A \ge 106$ synthesis in X-ray burst models by different groups \citep{Woosley2004,Jose2010}. One conclusion from our work is that transiently accreting neutron stars with particularly long mixed H/He bursts during outburst are the best systems to probe the interplay of burst physics, thermal conductivity of dense matter, and shell effects in very neutron-rich $N=50,82$ nuclei during their cooling phase. 

To date, crust cooling models mostly employ a single $Q_{\rm imp}$ throughout the crust. It would be interesting to explore the impact of the more realistic $Q_{\rm imp}$ profile predicted in this work. Any sensitivity, for example to impurities in the deeper layers of the outer crust, would provide interesting constraints on the outer crust composition that could then be brought to bear on our understanding of the hydrogen and helium burning processes during prior outbursts. 

This work is a first step in identifying the critical nuclear physics inputs for models of accreted neutron star crusts. We delineate the types of nuclear reactions and the typical nuclear element and mass regions involved. Future work is needed to vary the nuclear physics input and determine the sensitivity of observables to nuclear physics. It is clear that nuclear masses play a critical role, in particular, the relative locations of contours of $S_{\rm n}$ and $Q_{\rm EC}$ across the chart of nuclides near the neutron drip line (and the location of the neutron drip line itself), and deviations from smooth trends in the mass surface, for example, shell effects, around the neutron drip line. In particular, the FRDM shell effects in neutron-rich $N \approx 50$ and $N \approx 82$ nuclei that cause neutron captures in the r-process to sweep abundance rapidly into the closed neutron shell resulting in the underproduction of nuclei below the $A=130$ and $A=195$ abundance peaks \citep{Kratz1993,Chen1995}, have the same effect at the transition from outer to inner crust and can result in rather large inner crust impurities and slower cooling neutron stars. It will be important to address the question of shell effects near the neutron drip line through experiments at upcoming radioactive beam facilities such as FRIB. In the mean time,  
alternative mass models and more realistic mass predictions, for example, based on modern density functional theory that also predicts uncertainties (e.g. \citet{Erler2012,Afanasjev2015}) should be explored. However, masses are also needed for nuclei beyond the neutron drip line that are stabilized against neutron decay by a degenerate neutron gas. 

Improved electron capture and $\beta$ decay strength functions are also needed -- key elements are the ground state to ground state strength that determines the degree of nuclear Urca cooling in the outer crust, and the location of the lowest lying strength that plays a role in early neutron emission and heat deposition. In this context, the role of forbidden transitions, which especially
near spherical shell closures can be the lowest-lying strength are important, and more work on models for forbidden transitions is desirable. 
One important question relates to the low lying $\beta^-$ and EC transitions in the \Urca{55}{Sc}{55}{Ti} Urca pair, which is the only strong Urca cooling pair found in this study for systems that exhibit superbursts. The treatment of EC and $\beta^-$ induced neutron emission should also be improved taking into account neutron decay strength distributions and the corresponding density dependent Pauli blocking through the degenerate free neutrons. 

The predicted rates of pycnonuclear fusion and neutron capture are highly uncertain. For example, while the assumptions of the underlying nuclear model used to calculate the S-factors have been confirmed experimentally \citep{Carnelli2014}, pycnonuclear fusion rate predictions still have estimated uncertainties of about 7 orders of magnitude \citep{Yakovlev2006}. However, because of the steep density dependence, this would for a typical reaction only result in a 2\% change of the density at which the reaction occurs. Similarly the establishment of neutron equilibrium along isotopic chains may mitigate the impact of neutron capture rate uncertainties. Nevertheless, the impact of rate uncertainties could still be substantial. Our work provides a starting point for future sensitivity studies to characterize the impact of these uncertainties, and to determine how accurately these reaction rates need to be predicted for applications in crust models. Results from such studies can then guide future work to improve the theoretical prediction of these quantities. 

It should also be noted that the initial composition, and therefore the nuclear physics of hydrogen and helium burning, either in X-ray bursts or in steady state burning, is important. \citet{Cyburt2016} recently determined the relevant nuclear reaction rate uncertainties in X-ray bursts, and \citet{Schatz2017} the critical nuclear masses in X-ray bursts. More work needs to be done to explore the critical nuclear physics for a broader range of burning regimes. Also, the impact of nuclear physics variations on particularly salient features of the hydrogen and helium burning ashes should be explored - for example the amount of $A \le 28$ nuclei, which determine heat deposition from shallow fusion reactions, the amount of $A \ge 102$ nuclei that may lead to particularly high impurities in the inner crust, the amount of $^{28}$Si and $56 < A  < 106$ nuclei that increase crust impurity by populating the $N=50$ spherical shell closure, and the amount of odd $A$ nuclei, which determine the degree of nuclear Urca cooling. 

In principle, our calculations can serve as a starting point for an iterative process to determine the crust temperature profile that is consistent with the calculated heating and cooling, similar to \citet{Gupta2007}. This would require use of simplified models for heat deposition at depths greater than our calculation.
However, recent observations indicate that thermal profiles vary greatly from system to system (see for example \citet{Homan2014}), with the key parameters likely being the strong shallow heat source of unknown nature, the initial composition from hydrogen and helium burning or superbursts, as well as the mass of the neutron star and accretion rate history. The shallow heat source has an especially strong influence on the thermal profile. As this heat source cannot be predicted, it needs to be determined from observational constraints. Therefore, determining a realistic thermal profile is only possible using a multi-parameter analysis for a specific observed system. This should be pursued in future work. However, we do not expect the results of this work to depend strongly on temperature (see section \ref{Sec:Model}). Therefore, our conclusions should be applicable to a broad range of systems. The one exception is the strength of the nuclear Urca cooling, which depends strongly on temperature \citep{Schatz2014}. Our analysis with a relatively high temperature allows us to identify the important Urca cooling pairs. These are the Urca pairs that would limit crustal heating should sufficiently strong heat sources be present in a particular system. 

\acknowledgments
We acknowledge stimulating discussions with L. Bildsten, D. Yakovlev, A. Cumming, and within the JINA-CEE crust working group. This work was supported in part by the US National Science Foundation under grant PHY-1430152 (JINA Center for the Evolution of the Elements).
Support by the US National Science Foundation is acknowledged by E.F.B. and A.D. under grant AST-1516969, by A.S. under grant PHY 1554876, and by H.S. under grant PHY-1102511. A.V.A is supported by the US Department of Energy, Office of Science, 
Office of Nuclear Physics under Award No. DE-SC0013037. The work of P.S.S. on (n, $\gamma$) reactions was supported by the Russian Science Foundation grant 14-32-00316.
\bibliography{hsref_v5}

\begin{thebibliography}{}
\expandafter\ifx\csname natexlab\endcsname\relax\def\natexlab#1{#1}\fi

\bibitem[{{Afanasjev} {et~al.}(2015){Afanasjev}, {Agbemava}, {Ray}, \&
  {Ring}}]{Afanasjev2015}
{Afanasjev}, A.~V., {Agbemava}, S.~E., {Ray}, D., \& {Ring}, P. 2015, \prc, 91,
  014324

\bibitem[{{Afanasjev} {et~al.}(2012){Afanasjev}, {Beard}, {Chugunov},
  {Wiescher}, \& {Yakovlev}}]{Afanasjev2012}
{Afanasjev}, A.~V., {Beard}, M., {Chugunov}, A.~I., {Wiescher}, M., \&
  {Yakovlev}, D.~G. 2012, \prc, 85, 054615

\bibitem[{{Altamirano} {et~al.}(2012){Altamirano}, {Keek}, {Cumming},
  {Sivakoff}, {Heinke}, {Wijnands}, {Degenaar}, {Homan}, \&
  {Pooley}}]{Altamirano2012}
{Altamirano}, D., {Keek}, L., {Cumming}, A., {et~al.} 2012, \mnras, 426, 927

\bibitem[{Bastin {et~al.}(2007)Bastin, Gr\'evy, Sohler, Sorlin, Dombr\'adi,
  Achouri, Ang\'elique, Azaiez, Baiborodin, Borcea, Bourgeois, Buta, B\"urger,
  Chapman, Dalouzy, Dlouhy, Drouard, Elekes, Franchoo, Iacob, Laurent, Lazar,
  Liang, Li\'enard, Mrazek, Nalpas, Negoita, Orr, Penionzhkevich, Podoly\'ak,
  Pougheon, Roussel-Chomaz, Saint-Laurent, Stanoiu, Stefan, Nowacki, \&
  Poves}]{Bastin2007}
Bastin, B., Gr\'evy, S., Sohler, D., {et~al.} 2007, Phys. Rev. Lett., 99,
  022503

\bibitem[{{Baumann} {et~al.}(2007){Baumann}, {Amthor}, {Bazin}, {Brown}, {},
  {Gade}, {Ginter}, {Hausmann}, {Mato{\v s}}, {Morrissey}, {Portillo},
  {Schiller}, {Sherrill}, {Stolz}, {Tarasov}, \& {Thoennessen}}]{Baumann2007}
{Baumann}, T., {Amthor}, A.~M., {Bazin}, D., {et~al.} 2007, \nat, 449, 1022

\bibitem[{{Beard} {et~al.}(2010){Beard}, {Afanasjev}, {Chamon}, {Gasques},
  {Wiescher}, \& {Yakovlev}}]{Beard2010}
{Beard}, M., {Afanasjev}, A.~V., {Chamon}, L.~C., {et~al.} 2010, At. Data Nucl.
  Data Tables, 96, 541

\bibitem[{{Becerril Reyes} {et~al.}(2006){Becerril Reyes}, {Gupta}, {Schatz},
  {Kratz}, \& {M{\"o}ller}}]{Becerril2006}
{Becerril Reyes}, A.~D., {Gupta}, S., {Schatz}, S., {Kratz}, K.~L., \&
  {M{\"o}ller}, P. 2006, in International Symposium on Nuclear Astrophysics -
  Nuclei in the Cosmos, 75.1

\bibitem[{{Bernardini} {et~al.}(2013){Bernardini}, {Cackett}, {Brown},
  {D'Angelo}, {Degenaar}, {Miller}, {Reynolds}, \& {Wijnands}}]{Bernardini2013}
{Bernardini}, F., {Cackett}, E.~M., {Brown}, E.~F., {et~al.} 2013, \mnras, 436,
  2465

\bibitem[{{Bildsten}(1998)}]{Bildsten1998grav}
{Bildsten}, L. 1998, \apjl, 501, L89

\bibitem[{Bisnovatyi-Kogan \& Chechetkin(1979)}]{bisnovatyui1979}
Bisnovatyi-Kogan, G., \& Chechetkin, V. 1979, Soviet Physics Uspekhi, 22, 89

\bibitem[{{Boirin} {et~al.}(2007){Boirin}, {Keek}, {M{\'e}ndez}, {Cumming},
  {in't Zand}, {Cottam}, {Paerels}, \& {Lewin}}]{Boirin2007}
{Boirin}, L., {Keek}, L., {M{\'e}ndez}, M., {et~al.} 2007, \aap, 465, 559

\bibitem[{{Brown}(2000)}]{Brown2000}
{Brown}, E.~F. 2000, \apj, 531, 988

\bibitem[{{Brown} {et~al.}(1998){Brown}, {Bildsten}, \& {Rutledge}}]{Brown1998}
{Brown}, E.~F., {Bildsten}, L., \& {Rutledge}, R.~E. 1998, \apjl, 504, L95

\bibitem[{{Brown} \& {Cumming}(2009)}]{Brown2009}
{Brown}, E.~F., \& {Cumming}, A. 2009, \apj, 698, 1020

\bibitem[{Brown {et~al.}(2018)Brown, Cumming, Fattoyev, Horowitz, Page, \&
  Reddy}]{Brown2018}
Brown, E.~F., Cumming, A., Fattoyev, F.~J., {et~al.} 2018, \prl, submitted

\bibitem[{{Cackett} {et~al.}(2006){Cackett}, {Wijnands}, {Linares}, {Miller},
  {Homan}, \& {Lewin}}]{Cackett2006}
{Cackett}, E.~M., {Wijnands}, R., {Linares}, M., {et~al.} 2006, \mnras, 372,
  479

\bibitem[{{Carnelli} {et~al.}(2014){Carnelli}, {Almaraz-Calderon}, {Rehm},
  {Albers}, {Alcorta}, {Bertone}, {Digiovine}, {Esbensen}, {Niello},
  {Henderson}, {Jiang}, {Lai}, {Marley}, {Nusair}, {Palchan-Hazan}, {Pardo},
  {Paul}, \& {Ugalde}}]{Carnelli2014}
{Carnelli}, P.~F.~F., {Almaraz-Calderon}, S., {Rehm}, K.~E., {et~al.} 2014,
  Phys. Rev. Lett., 112, 192701

\bibitem[{Chamel \& Haensel(2008)}]{Chamel2008}
Chamel, N., \& Haensel, P. 2008, Living Reviews in Relativity, 11, 10

\bibitem[{Chen {et~al.}(1995)Chen, Dobaczewski, Kratz, Langanke, Pfeiffer,
  Thielemann, \& Vogel}]{Chen1995}
Chen, B., Dobaczewski, J., Kratz, K.-L., {et~al.} 1995, Phys. Lett. B, 355, 37

\bibitem[{{Colpi} {et~al.}(2001){Colpi}, {Geppert}, {Page}, \&
  {Possenti}}]{Colpi2001}
{Colpi}, M., {Geppert}, U., {Page}, D., \& {Possenti}, A. 2001, \apjl, 548,
  L175

\bibitem[{Crawford {et~al.}(2010)Crawford, Janssens, Mantica, Berryman, Broda,
  Carpenter, Cieplicka, Fornal, Grinyer, Hoteling, Kay, Lauritsen, Minamisono,
  Stefanescu, Stoker, Walters, \& Zhu}]{Crawford2010}
Crawford, H.~L., Janssens, R. V.~F., Mantica, P.~F., {et~al.} 2010, \prc, 82,
  014311

\bibitem[{{Crawford} {et~al.}(2014){Crawford}, {Fallon}, {Macchiavelli},
  {Clark}, {Brown}, {Tostevin}, {Bazin}, {Aoi}, {Doornenbal}, {Matsushita},
  {Scheit}, {Steppenbeck}, {Takeuchi}, {Baba}, {Campbell}, {Cromaz},
  {Ideguchi}, {Kobayashi}, {Kondo}, {Lee}, {Lee}, {Lee}, {Li}, {Michimasa},
  {Motobayashi}, {Nakamura}, {Ota}, {Paschalis}, {Petri}, {Sako}, {Sakurai},
  {Shimoura}, {Takechi}, {Togano}, {Wang}, \& {Yoneda}}]{Crawford2014}
{Crawford}, H.~L., {Fallon}, P., {Macchiavelli}, A.~O., {et~al.} 2014, \prc,
  89, 041303

\bibitem[{{Cumming} {et~al.}(2017){Cumming}, {Brown}, {Fattoyev}, {Horowitz},
  {Page}, \& {Reddy}}]{Cumming2017}
{Cumming}, A., {Brown}, E.~F., {Fattoyev}, F.~J., {et~al.} 2017, \prc, 95,
  025806

\bibitem[{{Cumming} {et~al.}(2006){Cumming}, {Macbeth}, {in 't Zand}, \&
  {Page}}]{Cumming2006}
{Cumming}, A., {Macbeth}, J., {in 't Zand}, J.~J.~M., \& {Page}, D. 2006, \apj,
  646, 429

\bibitem[{{Cyburt} {et~al.}(2016){Cyburt}, {Amthor}, {Heger}, {Johnson},
  {Keek}, {Meisel}, {Schatz}, \& {Smith}}]{Cyburt2016}
{Cyburt}, R.~H., {Amthor}, A.~M., {Heger}, A., {et~al.} 2016, \apj, 830, 55

\bibitem[{{Degenaar} {et~al.}(2011){Degenaar}, {Brown}, \&
  {Wijnands}}]{Degenaar2011}
{Degenaar}, N., {Brown}, E.~F., \& {Wijnands}, R. 2011, \mnras, 418, L152

\bibitem[{{Degenaar} {et~al.}(2013){Degenaar}, {Wijnands}, {Brown},
  {Altamirano}, {Cackett}, {Fridriksson}, {Homan}, {Heinke}, {Miller},
  {Pooley}, \& {Sivakoff}}]{Degenaar2013}
{Degenaar}, N., {Wijnands}, R., {Brown}, E.~F., {et~al.} 2013, \apj, 775, 48

\bibitem[{{Degenaar} {et~al.}(2014){Degenaar}, {Medin}, {Cumming}, {Wijnands},
  {Wolff}, {Cackett}, {Miller}, {Jonker}, {Homan}, \& {Brown}}]{Degenaar2014}
{Degenaar}, N., {Medin}, Z., {Cumming}, A., {et~al.} 2014, \apj, 791, 47

\bibitem[{{Degenaar} {et~al.}(2015){Degenaar}, {Wijnands}, {Bahramian},
  {Sivakoff}, {Heinke}, {Brown}, {Fridriksson}, {Homan}, {Cackett}, {Cumming},
  {Miller}, {Altamirano}, \& {Pooley}}]{Degenaar2015}
{Degenaar}, N., {Wijnands}, R., {Bahramian}, A., {et~al.} 2015, \mnras, 451,
  2071

\bibitem[{{Deibel} {et~al.}(2015){Deibel}, {Cumming}, {Brown}, \&
  {Page}}]{Deibel2015}
{Deibel}, A., {Cumming}, A., {Brown}, E.~F., \& {Page}, D. 2015, \apjl, 809,
  L31

\bibitem[{{Deibel} {et~al.}(2017){Deibel}, {Cumming}, {Brown}, \&
  {Reddy}}]{Deibel2017}
{Deibel}, A., {Cumming}, A., {Brown}, E.~F., \& {Reddy}, S. 2017, \apj, 839, 95

\bibitem[{{Deibel} {et~al.}(2016){Deibel}, {Meisel}, {Schatz}, {Brown}, \&
  {Cumming}}]{Deibel2016}
{Deibel}, A., {Meisel}, Z., {Schatz}, H., {Brown}, E.~F., \& {Cumming}, A.
  2016, \apj, 831, 13

\bibitem[{{Dillmann} {et~al.}(2003){Dillmann}, {Kratz}, {W{\"o}hr}, {Arndt},
  {Brown}, {Hoff}, {Hjorth-Jensen}, {K{\"o}ster}, {Ostrowski}, {Pfeiffer},
  {Seweryniak}, {Shergur}, \& {Walters}}]{Dillmann2003}
{Dillmann}, I., {Kratz}, K.-L., {W{\"o}hr}, A., {et~al.} 2003, Phys. Rev.
  Lett., 91, 162503

\bibitem[{Erler {et~al.}(2012)Erler, Birge, Kortelainen, Nazarewicz, Olsen,
  Perhac, \& Stoitsov}]{Erler2012}
Erler, J., Birge, N., Kortelainen, M., {et~al.} 2012, Nature, 486, 509

\bibitem[{{Fisker} {et~al.}(2008){Fisker}, {Schatz}, \&
  {Thielemann}}]{Fisker2008}
{Fisker}, J.~L., {Schatz}, H., \& {Thielemann}, F.-K. 2008, \apj Suppl., 174,
  261

\bibitem[{Glasmacher {et~al.}(1997)Glasmacher, Brown, Chromik, Cottle,
  Fauerbach, Ibbotson, Kemper, Morrissey, Scheit, Sklenicka, \&
  Steiner}]{GLASMACHER1997}
Glasmacher, T., Brown, B., Chromik, M., {et~al.} 1997, Phys. Lett. B, 395, 163

\bibitem[{Guillemaud-Mueller {et~al.}(1984)Guillemaud-Mueller, Detraz,
  Langevin, Naulin, de~Saint-Simon, Thibault, Touchard, \&
  Epherre}]{GUILLEMAUD1984}
Guillemaud-Mueller, D., Detraz, C., Langevin, M., {et~al.} 1984, \npa, 426, 37

\bibitem[{{Gupta} {et~al.}(2007){Gupta}, {Brown}, {Schatz}, {M{\"o}ller}, \&
  {Kratz}}]{Gupta2007}
{Gupta}, S., {Brown}, E.~F., {Schatz}, H., {M{\"o}ller}, P., \& {Kratz}, K.-L.
  2007, \apj, 662, 1188

\bibitem[{{Gupta} {et~al.}(2008){Gupta}, {Kawano}, \& {M{\"o}ller}}]{Gupta2008}
{Gupta}, S.~S., {Kawano}, T., \& {M{\"o}ller}, P. 2008, Phys. Rev. Lett., 101,
  231101

\bibitem[{{Haensel} \& {Zdunik}(1990)}]{Haensel1990}
{Haensel}, P., \& {Zdunik}, J.~L. 1990, \AsAs, 227, 431

\bibitem[{{Haensel} \& {Zdunik}(2003)}]{Haensel2003}
---. 2003, \AsAsp, 404, L33

\bibitem[{Haensel \& Zdunik(2008)}]{Haensel2008}
Haensel, P., \& Zdunik, J.~L. 2008, \AsAs, 480, 459

\bibitem[{{Heger} {et~al.}(2007){Heger}, {Cumming}, {Galloway}, \&
  {Woosley}}]{Heger2007}
{Heger}, A., {Cumming}, A., {Galloway}, D.~K., \& {Woosley}, S.~E. 2007, \apjl,
  671, L141

\bibitem[{{Hencheck} {et~al.}(1995){Hencheck}, {Boyd}, {Meyer},
  {Hellstr{\"o}m}, {Morrissey}, {Balbes}, {Chloupek}, {Fauerbach}, {Mitchell},
  {Pfaff}, {Powell}, {Raimann}, {Sherrill}, {Steiner}, {Vandegriff}, \&
  {Yennello}}]{Hencheck1995}
{Hencheck}, M., {Boyd}, R.~N., {Meyer}, B.~S., {et~al.} 1995, in American
  Institute of Physics Conference Series, Vol. 327, Nuclei in the Cosmos III,
  ed. M.~{Busso}, C.~M. {Raiteri}, \& R.~{Gallino}, 331

\bibitem[{{Homan} {et~al.}(2014){Homan}, {Fridriksson}, {Wijnands}, {Cackett},
  {Degenaar}, {Linares}, {Lin}, \& {Remillard}}]{Homan2014}
{Homan}, J., {Fridriksson}, J.~K., {Wijnands}, R., {et~al.} 2014, \apj, 795,
  131

\bibitem[{{Horowitz} {et~al.}(2015){Horowitz}, {Berry}, {Briggs}, {Caplan},
  {Cumming}, \& {Schneider}}]{Horowitz2015}
{Horowitz}, C.~J., {Berry}, D.~K., {Briggs}, C.~M., {et~al.} 2015, Phys. Rev.
  Lett., 114, 031102

\bibitem[{{Horowitz} {et~al.}(2007){Horowitz}, {Berry}, \&
  {Brown}}]{Horowitz2007}
{Horowitz}, C.~J., {Berry}, D.~K., \& {Brown}, E.~F. 2007, \pre, 75, 066101

\bibitem[{{Horowitz} {et~al.}(2009){Horowitz}, {Caballero}, \&
  {Berry}}]{Horowitz2009}
{Horowitz}, C.~J., {Caballero}, O.~L., \& {Berry}, D.~K. 2009, \pre, 79, 026103

\bibitem[{{Horowitz} {et~al.}(2008){Horowitz}, {Dussan}, \&
  {Berry}}]{Horowitz2008}
{Horowitz}, C.~J., {Dussan}, H., \& {Berry}, D.~K. 2008, \prc, 77, 045807

\bibitem[{{Jones}(2005)}]{Jones2005}
{Jones}, P.~B. 2005, \prd, 72, 083006

\bibitem[{{Jos{\'e}} {et~al.}(2010){Jos{\'e}}, {Moreno}, {Parikh}, \&
  {Iliadis}}]{Jose2010}
{Jos{\'e}}, J., {Moreno}, F., {Parikh}, A., \& {Iliadis}, C. 2010, \apj Suppl.,
  189, 204

\bibitem[{{Keek} {et~al.}(2010){Keek}, {Galloway}, {in't Zand}, \&
  {Heger}}]{Keek2010}
{Keek}, L., {Galloway}, D.~K., {in't Zand}, J.~J.~M., \& {Heger}, A. 2010,
  \apj, 718, 292

\bibitem[{{Keek} \& {Heger}(2011)}]{Keek2011}
{Keek}, L., \& {Heger}, A. 2011, \apj, 743, 189

\bibitem[{{Keek} \& {Heger}(2016)}]{Keek2016}
---. 2016, \mnras, 456, L11

\bibitem[{{Keek} \& {Heger}(2017)}]{Keek2017}
---. 2017, \apj, 842, 113

\bibitem[{Keek {et~al.}(2012)Keek, Heger, \& in't Zand}]{Keek2012}
Keek, L., Heger, A., \& in't Zand, J. J.~M. 2012, \apj, 752, 150

\bibitem[{{Keek} \& {in't Zand}(2008)}]{Keek2008}
{Keek}, L., \& {in't Zand}, J.~J.~M. 2008, in Proceedings of the 7th INTEGRAL
  Workshop

\bibitem[{Klotz {et~al.}(1993)Klotz, Baumann, Bounajma, Huck, Knipper, Walter,
  Marguier, Richard-Serre, Poves, \& Retamosa}]{Klotz1993}
Klotz, G., Baumann, P., Bounajma, M., {et~al.} 1993, \prc, 47, 2502

\bibitem[{{Kratz} {et~al.}(1993){Kratz}, {Bitouzet}, {Thielemann}, {Moeller},
  \& {Pfeiffer}}]{Kratz1993}
{Kratz}, K.-L., {Bitouzet}, J.-P., {Thielemann}, F.-K., {Moeller}, P., \&
  {Pfeiffer}, B. 1993, \apj, 403, 216

\bibitem[{{Krumlinde} \& {M{\"o}ller}(1984)}]{Krumlinde1984}
{Krumlinde}, J., \& {M{\"o}ller}, P. 1984, \npa, 417, 419

\bibitem[{{Liu} {et~al.}(2007){Liu}, {van Paradijs}, \& {van den
  Heuvel}}]{Liu2007}
{Liu}, Q.~Z., {van Paradijs}, J., \& {van den Heuvel}, E.~P.~J. 2007, \aap,
  469, 807

\bibitem[{{Mackie} \& {Baym}(1977)}]{Mackie1977}
{Mackie}, F.~D., \& {Baym}, G. 1977, \npa, 285, 332

\bibitem[{Maierbeck {et~al.}(2009)Maierbeck, Gernh{\"a}user, Kr{\"u}cken,
  Kr{\"a}ll, Alvarez-Pol, Aksouh, Aumann, Behr, Benjamim, Benlliure, Bildstein,
  BÃ¶hmer, Boretzky, Borge, BrÃŒnle, BÃŒrger, CaamaÃ±o, Casarejos,
  Chatillon, Chulkov, Cortina-Gil, Enders, Eppinger, Faestermann, Friese,
  Fabbietti, GascÃ³n, Geissel, Gerl, Gorska, Hansen, Jonson, Kanungo,
  Kiselev, Kojouharov, Klimkiewicz, Kurtukian, Kurz, Larsson, Bleis, Mahata,
  Maier, Nilsson, Nociforo, Nyman, Pascual-Izarra, Perea, Perez, Prochazka,
  Rodriguez-Tajes, Rossi, Schaffner, Schrieder, Schwertel, Simon, Sitar,
  Stanoiu, SÃŒmmerer, Tengblad, Weick, Winkler, Brown, Otsuka, Tostevin, \&
  Rae}]{Maierbeck2009}
Maierbeck, P., Gernh{\"a}user, R., Kr{\"u}cken, R., {et~al.} 2009, Phys. Lett.
  B, 675, 22

\bibitem[{{Mayer}(1949)}]{Mayer1949}
{Mayer}, M.~G. 1949, Phys. Rev., 75, 1969

\bibitem[{{Mayer}(1950{\natexlab{a}})}]{Mayer1950a}
---. 1950{\natexlab{a}}, Phys. Rev., 78, 16

\bibitem[{{Mayer}(1950{\natexlab{b}})}]{Mayer1950b}
---. 1950{\natexlab{b}}, Phys. Rev., 78, 22

\bibitem[{{Mckinven} {et~al.}(2016){Mckinven}, {Cumming}, {Medin}, \&
  {Schatz}}]{Mcinven2016}
{Mckinven}, R., {Cumming}, A., {Medin}, Z., \& {Schatz}, H. 2016, \apj, 823,
  117

\bibitem[{{Medin} \& {Cumming}(2011)}]{Medin2011}
{Medin}, Z., \& {Cumming}, A. 2011, \apj, 730, 97

\bibitem[{{Medin} \& {Cumming}(2014)}]{Medin2014}
---. 2014, \apjl, 783, L3

\bibitem[{{Meisel} \& {Deibel}(2017)}]{Meisel2017}
{Meisel}, Z., \& {Deibel}, A. 2017, \apj, 837, 73

\bibitem[{{Meisel} {et~al.}(2015{\natexlab{a}}){Meisel}, {George}, {Ahn},
  {Bazin}, {Brown}, {Browne}, {Carpino}, {Chung}, {Cole}, {Cyburt},
  {Estrad{\'e}}, {Famiano}, {Gade}, {Langer}, {Mato{\v s}}, {Mittig}, {Montes},
  {Morrissey}, {Pereira}, {Schatz}, {Schatz}, {Scott}, {Shapira}, {Smith},
  {Stevens}, {Tan}, {Tarasov}, {Towers}, {Wimmer}, {Winkelbauer}, {Yurkon}, \&
  {Zegers}}]{Meisel2015}
{Meisel}, Z., {George}, S., {Ahn}, S., {et~al.} 2015{\natexlab{a}}, Phys. Rev.
  Lett., 115, 162501

\bibitem[{{Meisel} {et~al.}(2015{\natexlab{b}}){Meisel}, {George}, {Ahn},
  {Browne}, {Bazin}, {Brown}, {Carpino}, {Chung}, {Cyburt}, {Estrad{\'e}},
  {Famiano}, {Gade}, {Langer}, {Mato{\v s}}, {Mittig}, {Montes}, {Morrissey},
  {Pereira}, {Schatz}, {Schatz}, {Scott}, {Shapira}, {Smith}, {Stevens}, {Tan},
  {Tarasov}, {Towers}, {Wimmer}, {Winkelbauer}, {Yurkon}, \&
  {Zegers}}]{Meisel2015Ar}
---. 2015{\natexlab{b}}, Phys. Rev. Lett., 114, 022501

\bibitem[{{Merritt} {et~al.}(2016){Merritt}, {Cackett}, {Brown}, {Page},
  {Cumming}, {Degenaar}, {Deibel}, {Homan}, {Miller}, \&
  {Wijnands}}]{Merritt2016}
{Merritt}, R.~L., {Cackett}, E.~M., {Brown}, E.~F., {et~al.} 2016, \apj, 833,
  186

\bibitem[{{M{\"o}ller} {et~al.}(1997){M{\"o}ller}, {Nix}, \&
  {Kratz}}]{Moller1997}
{M{\"o}ller}, P., {Nix}, J.~R., \& {Kratz}, K.-L. 1997, At. Data Nucl. Data
  Tables, 66, 131

\bibitem[{{M{\"o}ller} {et~al.}(1995){M{\"o}ller}, {Nix}, {Myers}, \&
  {Swiatecki}}]{Moller1995}
{M{\"o}ller}, P., {Nix}, J.~R., {Myers}, W.~D., \& {Swiatecki}, W.~J. 1995, At.
  Data Nucl. Data Tables, 59, 185

\bibitem[{{M{\"o}ller} \& {Randrup}(1990)}]{Moller1990}
{M{\"o}ller}, P., \& {Randrup}, J. 1990, \npa, 514, 1

\bibitem[{{Mottelson} \& {Nilsson}(1959)}]{Mottelson1959}
{Mottelson}, B.~R., \& {Nilsson}, S.~G. 1959, Kgl. Danske Videnskab. Selskab.
  Mat.-Fys. Medd., 1, 1

\bibitem[{{Narayan} \& {Heyl}(2003)}]{Narayan2003}
{Narayan}, R., \& {Heyl}, J.~S. 2003, \apj, 599, 419

\bibitem[{{Negele} \& {Vautherin}(1973)}]{Negele1973}
{Negele}, J.~W., \& {Vautherin}, D. 1973, Nucl. Phys. A, 207, 298

\bibitem[{{Nilsson}(1955)}]{Nilsson1955}
{Nilsson}, S.~G. 1955, Kgl. Danske Videnskab. Selskab. Mat.-Fys. Medd., 29, 1

\bibitem[{{Ootes} {et~al.}(2016){Ootes}, {Page}, {Wijnands}, \&
  {Degenaar}}]{Ootes2016}
{Ootes}, L.~S., {Page}, D., {Wijnands}, R., \& {Degenaar}, N. 2016, \mnras,
  461, 4400

\bibitem[{{Page} \& {Reddy}(2013)}]{Page2013}
{Page}, D., \& {Reddy}, S. 2013, Phys. Rev. Lett., 111, 241102

\bibitem[{Parikh {et~al.}(2017)Parikh, Wijnands, Degenaar, Ootes, Page,
  Altamirano, Cackett, Deller, Gusinskaia, Hessels, Homan, Linares, Miller, \&
  Miller-Jones}]{Parikh2017}
Parikh, A.~S., Wijnands, R., Degenaar, N., {et~al.} 2017, \mnras, 466, 4074

\bibitem[{{Patruno} {et~al.}(2017){Patruno}, {Haskell}, \&
  {Andersson}}]{Patruno2017}
{Patruno}, A., {Haskell}, B., \& {Andersson}, N. 2017, \apj, 850, 106

\bibitem[{Roggero \& Reddy(2016)}]{Roggero2016}
Roggero, A., \& Reddy, S. 2016, \prc, 94, 015803

\bibitem[{{Rutledge} {et~al.}(2002){Rutledge}, {Bildsten}, {Brown}, {Pavlov},
  {Zavlin}, \& {Ushomirsky}}]{Rutledge2002}
{Rutledge}, R.~E., {Bildsten}, L., {Brown}, E.~F., {et~al.} 2002, \apj, 580,
  413

\bibitem[{{Sato}(1979)}]{Sato1979}
{Sato}, K. 1979, Prog. Theor. Phys., 62, 957

\bibitem[{{Schatz} {et~al.}(2003){Schatz}, {Bildsten}, \&
  {Cumming}}]{Schatz2003}
{Schatz}, H., {Bildsten}, L., \& {Cumming}, A. 2003, \apjl, 583, L87

\bibitem[{Schatz {et~al.}(1999)Schatz, Bildsten, Cumming, \&
  Wiescher}]{Schatz1999}
Schatz, H., Bildsten, L., Cumming, A., \& Wiescher, M. 1999, \apj, 524, 1014

\bibitem[{Schatz \& Ong(2017)}]{Schatz2017}
Schatz, H., \& Ong, W.-J. 2017, \apj, 844, 139

\bibitem[{{Schatz} \& {Rehm}(2006)}]{Schatz2006a}
{Schatz}, H., \& {Rehm}, K.~E. 2006, \npa, 777, 601

\bibitem[{{Schatz} {et~al.}(1998){Schatz}, {Aprahamian}, {Goerres}, {Wiescher},
  {Rauscher}, {Rembges}, {Thielemann}, {Pfeiffer}, {Moeller}, {Kratz},
  {Herndl}, {Brown}, \& {Rebel}}]{Schatz1998}
{Schatz}, H., {Aprahamian}, A., {Goerres}, J., {et~al.} 1998, Phys. Rep., 294,
  167

\bibitem[{Schatz {et~al.}(2001)Schatz, Aprahamian, Barnard, Bildsten, Cumming,
  Ouellette, Rauscher, Thielemann, \& Wiescher}]{Schatz2001}
Schatz, H., Aprahamian, A., Barnard, V., {et~al.} 2001, Phys. Rev. Lett., 86,
  3471

\bibitem[{{Schatz} {et~al.}(2014){Schatz}, {Gupta}, {M{\"o}ller}, {Beard},
  {Brown}, {Deibel}, {Gasques}, {Hix}, {Keek}, {Lau}, {Steiner}, \&
  {Wiescher}}]{Schatz2014}
{Schatz}, H., {Gupta}, S., {M{\"o}ller}, P., {et~al.} 2014, \nat, 505, 62

\bibitem[{{Shternin} {et~al.}(2012){Shternin}, {Beard}, {Wiescher}, \&
  {Yakovlev}}]{Shternin2012}
{Shternin}, P.~S., {Beard}, M., {Wiescher}, M., \& {Yakovlev}, D.~G. 2012,
  \prc, 86, 015808

\bibitem[{{Shternin} {et~al.}(2007){Shternin}, {Yakovlev}, {Haensel}, \&
  {Potekhin}}]{Shternin2007}
{Shternin}, P.~S., {Yakovlev}, D.~G., {Haensel}, P., \& {Potekhin}, A.~Y. 2007,
  \mnras, 382, L43

\bibitem[{{Sorlin} \& {Porquet}(2008)}]{Sorlin2008}
{Sorlin}, O., \& {Porquet}, M.-G. 2008, Prog. Part. Nucl. Phys., 61, 602

\bibitem[{{Steiner}(2012)}]{Steiner2012}
{Steiner}, A.~W. 2012, \prc, 85, 055804

\bibitem[{Taprogge {et~al.}(2014)Taprogge, Jungclaus, Grawe, Nishimura,
  Doornenbal, Lorusso, Simpson, S\"oderstr\"om, Sumikama, Xu, Baba, Browne,
  Fukuda, Gernh\"auser, Gey, Inabe, Isobe, Jung, Kameda, Kim, Kim, Kojouharov,
  Kubo, Kurz, Kwon, Li, Sakurai, Schaffner, Steiger, Suzuki, Takeda, Vajta,
  Watanabe, Wu, Yagi, Yoshinaga, Benzoni, B\"onig, Chae, Coraggio, Covello,
  Daugas, Drouet, Gadea, Gargano, Ilieva, Kondev, Kr\"oll, Lane,
  Montaner-Piz\'a, Moschner, M\"ucher, Naqvi, Niikura, Nishibata, Odahara,
  Orlandi, Patel, Podoly\'ak, \& Wendt}]{Taprogge2014}
Taprogge, J., Jungclaus, A., Grawe, H., {et~al.} 2014, Phys. Rev. Lett., 112,
  132501

\bibitem[{Tripathi {et~al.}(2008)Tripathi, Tabor, Mantica, Utsuno, Bender,
  Cook, Hoffman, Lee, Otsuka, Pereira, Perry, Pepper, Pinter, Stoker, Volya, \&
  Weisshaar}]{Tripathi2008}
Tripathi, V., Tabor, S.~L., Mantica, P.~F., {et~al.} 2008, Phys. Rev. Lett.,
  101, 142504

\bibitem[{{Turlione} {et~al.}(2015){Turlione}, {Aguilera}, \&
  {Pons}}]{Turlione2015}
{Turlione}, A., {Aguilera}, D.~N., \& {Pons}, J.~A. 2015, \aap, 577, A5

\bibitem[{{Ushomirsky} {et~al.}(2000){Ushomirsky}, {Cutler}, \&
  {Bildsten}}]{Ushomirsky2000}
{Ushomirsky}, G., {Cutler}, C., \& {Bildsten}, L. 2000, \mnras, 319, 902

\bibitem[{{Wallace} \& {Woosley}(1981)}]{Wallace1981}
{Wallace}, R.~K., \& {Woosley}, S.~E. 1981, \apj Suppl., 45, 389

\bibitem[{{Wang} {et~al.}(2012){Wang}, {Audi}, {Wapstra}, {Kondev},
  {MacCormick}, {Xu}, \& {Pfeiffer}}]{AME12}
{Wang}, M., {Audi}, G., {Wapstra}, A.~H., {et~al.} 2012, Chinese Physics C, 36,
  3

\bibitem[{Watanabe {et~al.}(2013)Watanabe, Lorusso, Nishimura, Xu, Sumikama,
  S\"oderstr\"om, Doornenbal, Browne, Gey, Jung, Taprogge, Vajta, Wu, Yagi,
  Baba, Benzoni, Chae, Crespi, Fukuda, Gernh\"auser, Inabe, Isobe, Jungclaus,
  Kameda, Kim, Kim, Kojouharov, Kondev, Kubo, Kurz, Kwon, Lane, Li, Moon,
  Montaner-Piz\'a, Moschner, Naqvi, Niikura, Nishibata, Nishimura, Odahara,
  Orlandi, Patel, Podoly\'ak, Sakurai, Schaffner, Simpson, Steiger, Suzuki,
  Takeda, Wendt, \& Yoshinaga}]{Watanabe13}
Watanabe, H., Lorusso, G., Nishimura, S., {et~al.} 2013, Phys. Rev. Lett., 111,
  152501

\bibitem[{{Waterhouse} {et~al.}(2016){Waterhouse}, {Degenaar}, {Wijnands},
  {Brown}, {Miller}, {Altamirano}, \& {Linares}}]{Waterhouse2016}
{Waterhouse}, A.~C., {Degenaar}, N., {Wijnands}, R., {et~al.} 2016, \mnras,
  456, 4001

\bibitem[{Woosley {et~al.}(2004)Woosley, Heger, Cumming, Hoffman, Pruet,
  Rauscher, Fisker, Schatz, Brown, \& Wiescher}]{Woosley2004a}
Woosley, S., Heger, A., Cumming, A., {et~al.} 2004, \apj Suppl., 151, 75

\bibitem[{{Woosley} {et~al.}(2004){Woosley}, {Heger}, {Cumming}, {Hoffman},
  {Pruet}, {Rauscher}, {Fisker}, {Schatz}, {Brown}, \&
  {Wiescher}}]{Woosley2004}
{Woosley}, S.~E., {Heger}, A., {Cumming}, A., {et~al.} 2004, \apj Suppl., 151,
  75

\bibitem[{{Xu} {et~al.}(2013){Xu}, {Goriely}, {Jorissen}, {Chen}, \&
  {Arnould}}]{Xu2013}
{Xu}, Y., {Goriely}, S., {Jorissen}, A., {Chen}, G.~L., \& {Arnould}, M. 2013,
  \aap, 549, A106

\bibitem[{Xu {et~al.}(2014)Xu, Nishimura, Lorusso, Browne, Doornenbal, Gey,
  Jung, Li, Niikura, S\"oderstr\"om, Sumikama, Taprogge, Vajta, Watanabe, Wu,
  Yagi, Yoshinaga, Baba, Franchoo, Isobe, John, Kojouharov, Kubono, Kurz,
  Matea, Matsui, Mengoni, Morfouace, Napoli, Naqvi, Nishibata, Odahara,
  \ifmmode~\mbox{\c{S}}\else \c{S}\fi{}ahin, Sakurai, Schaffner, Stefan,
  Suzuki, Taniuchi, \& Werner}]{Xu2014}
Xu, Z.~Y., Nishimura, S., Lorusso, G., {et~al.} 2014, Phys. Rev. Lett., 113,
  032505

\bibitem[{{Yakovlev} {et~al.}(2006){Yakovlev}, {Gasques}, {Afanasjev}, {Beard},
  \& {Wiescher}}]{Yakovlev2006}
{Yakovlev}, D.~G., {Gasques}, L.~R., {Afanasjev}, A.~V., {Beard}, M., \&
  {Wiescher}, M. 2006, \prc, 74, 035803

\bibitem[{Yordanov {et~al.}(2010)Yordanov, Blaum, De~Rydt, Kowalska, Neugart,
  Neyens, \& Hamamoto}]{Yordanov2010}
Yordanov, D.~T., Blaum, K., De~Rydt, M., {et~al.} 2010, Phys. Rev. Lett., 104,
  129201

\end{thebibliography}

\end{document}